\documentclass[titlepage,12pt,subeqn]{article}
\usepackage{amsmath,amssymb,amscd,amsxtra}
\usepackage[mathscr]{eucal}
\usepackage{jheppub}
\usepackage{bigints}
\usepackage{color,colortbl}
\usepackage{latexsym}
\usepackage{csquotes}
\usepackage{mathtools}   
\usepackage{amsfonts}
\usepackage{ifthen}
\usepackage[]{graphicx}
\usepackage[]{hyperref} 
\usepackage{bbm}
\usepackage{amsthm}
\usepackage{tikz}
\usetikzlibrary{shapes,snakes}
\usetikzlibrary{arrows,decorations}
\usepackage[]{booktabs,tabularx}
\definecolor{Gray}{gray}{0.9}
\newcolumntype{b}{X}
\newcolumntype{s}{>{\hsize=.5\hsize}X}
\usepackage{empheq}

\newcommand*\widefbox[1]{\fbox{\hspace{2em}#1\hspace{2em}}}

\title{Describing codimension two defects}
\author{Aswin Balasubramanian}
\affiliation{Theory Group\\ Department of Physics \\ University of Texas at Austin\\2515 Speedway Stop C1608\\
Austin, TX 78712-1197 }
\emailAdd{aswin@utexas.edu}

\abstract{Codimension two defects of the $(0,2)$ six dimensional theory $\mathscr{X}[\mathfrak{j}]$ have played an important role in understanding dualities for certain $\mathcal{N}=2$ SCFTs in four dimensions. These defects are typically understood by their behaviour under various dimensional reduction schemes. In their various guises, the defects admit  partial descriptions in terms of singularities of Hitchin systems, Nahm boundary conditions or Toda operators. Here, a uniform dictionary between these descriptions is given for a large class of such defects in  $\mathscr{X}[\mathfrak{j}], \mathfrak{j} \in A,D,E$.}
\begin{document}
\begin{titlepage}
\begin{flushright}
UTTG-37-13
\end{flushright}

\maketitle

\vskip 1cm

\end{titlepage}
\section{Introduction and Summary}

The study of defect operators in quantum field theories has a long history and has received closer attention in recent years. Apart from exposing deep connections to representation theory, such studies turn out to be useful in the understanding of various non-perturbative dualities.  A particular six dimensional $(0,2)$ SCFT has played a special in some of the recent developments along this theme. This SCFT is sometimes called theory $\mathscr{X}[\mathfrak{j}]$ to signify the fact that there is such a theory for every lie algebra $\mathfrak{j} \in A,D,E$. The theory lacks an intrinsic description in terms of classical fields, Lagrangians and action principles and thus precludes much direct investigation. Yet, under various dimensional reductions, this theory can be better understood. The specific objects that would be the focus of this paper are certain 1/2 BPS codimension two defects of theory $\mathscr{X}[\mathfrak{j}]$. The focus of this paper is on four dimensional $\mathcal{N}=2$ SCFTs (and their massive deformations) that can be built out of the codimension two defects\footnote{Henceforth, any invocation of the term `codimension two defect' should be taken to mean `codimension two defects of theory $\mathscr{X}[\mathfrak{j}]$'. }.
For a large class of regular (twisted or untwisted) codimension two defect of $\mathscr{X}[\mathfrak{j}]$, we have (following \cite{Chacaltana:2012zy} and the general lesson from \cite{Kanno:2009ga}),
\begin{itemize}
\item An associated nilpotent orbit in $\mathfrak{g}$ called the Nahm orbit ($\mathcal{O}_N$). This arises as a Nahm type boundary condition in 4d $\mathcal{N}=4$ SYM with gauge group $G$ \footnote{The gauge group $G$ is compact. But it turns out that the defects of concern are classified by nilpotent orbits in the complexified lie algebra $\mathfrak{g}_\mathbb{C}$, which will still denote by $\mathfrak{g}$ to simplify notation.} on a half space (or equivalently a boundary condition for 5d SYM with gauge group $G$ on a half space times a circle $S$),
\item An associated nilpotent orbit in Langlands/GNO dual $\mathfrak{g}^\vee$ called the Hitchin orbit ($\mathcal{O}_H$) with some further discrete data that can be captured by specifying a subgroup of $\overline{A}(\mathcal{O}_H)$, where $\overline{A}(\mathcal{O}_H)$ is Lusztig's quotient of the component group of the centralizer of the corresponding nilpotent element (identified upto $\mathfrak{g}^\vee$- conjugacy). This arises as a codimension two defect for 5d SYM with gauge group $G^\vee$ on a half space times a circle $\tilde{S}$,
\item A semi-degenerate primary  of the $\text{Toda}[\mathfrak{g}]$ theory that is given by the specification of a set of null vectors in the corresponding W-algebra Verma module.
\end{itemize}

Here, $\mathfrak{g}$ is an arbitrary simple lie algebra. For the untwisted defects, the lie algebra $\mathfrak{g}$ isomorphic to $\mathfrak{j}$ and thus simply laced. For the twisted sector defects, $\mathfrak{g}$ is a subalgebra of $\mathfrak{j}$ \footnote{The naming of lie algebras $\mathfrak{j}$ and $\mathfrak{g}$ in the current version of the paper is consistent with how they appear in \cite{Chacaltana:2012zy}.}. In particular, the twisted sector defects require the cases where $\mathfrak{g}$ is non-simply laced. This set of regular defects will be called the CDT class of defects in the rest of the paper.

The availability of these multiple descriptions is convenient since different aspects of the defects become manifest when expressed in each of these terms. However, one would expect that each one of these constitute a partial description of a given codimension two defect. This paper concerns the relationship between these three descriptions. A dictionary between the Hitchin data and the Nahm data has already been provided in \cite{Chacaltana:2012zy} for arbitrary $\mathfrak{g}$ and the discussion here hopes to complement the one provided in \cite{Chacaltana:2012zy}. Further, the relationship of this data to that of a Toda semi-degenerate primary is explained for a particular subset of defects that correspond to the Nahm data being a nilpotent orbit of principal Levi type. The relevant set of Toda operators were obtained in the work of \cite{Kanno:2009ga} for type $A$. In type $A$, all non-zero nilpotent orbits are principal Levi type. So, the setup here covers all of them. Outside of type $A$, there are nontrivial orbits that occur as non-principal orbits in Levi subalgebras. Extending the Toda part of the dictionary to such Nahm orbits would be an interesting problem.

The task that is accomplished here is modest if viewed in the larger scheme of things and the results only point to a need for more detailed investigations into the connections between geometric representation theory and the construction of class $\mathcal{S}$ theories. It should be mentioned here that almost all of the mathematical considerations in this paper arise from well known results and can be found in the existing literature. The one exception is a certain property that is discussed in Section \ref{setup} that places the `\textit{Higgs branch Springer invariant}' on a different footing from what one may call a `\textit{Coulomb branch Springer invariant}'. Further, it is hoped that the presentation of the known mathematical results is in a language that is friendly to physicists. The placing of these results in a physical framework yields some new insights into the physics and is also likely to motivate future investigations.

The plan of the paper is as follows. Section \ref{reductions} offers a review of some dimensional reduction schemes used in the study of codimension two defects. Section \ref{GWbc} reviews the set of boundary conditions studied by Gaiotto-Witten and action of S-duality on certain classes of these boundary conditions. Section \ref{dualitymaps} collects results from the mathematical literature on order reversing duality maps and the closely related representation theory of Weyl groups. In Section \ref{fourdimensional}, a way to relate the Hitchin and Nahm descriptions is provided using properties of the Higgs branch associated to the defect. This reproduces the setup of \cite{Chacaltana:2012zy} and provides a physical framework for some defining properties of the order reversing duality used in \cite{Chacaltana:2012zy}. Equivalently, this provides the S-duality map for the subset of boundary conditions in $\mathcal{N}=4$ SYM that correspond to the CDT class of codimension two defects. In Section \ref{Todapart}, a map is constructed between the set of codimension two defects and the set of semi-degenerate primary operators in Toda theory for the cases where the Nahm orbit is of principal Levi type. 

In Section \ref{setup}, the results in Section \ref{fourdimensional} and Section \ref{Todapart} are combined and the complete setup relating Toda, Nahm and Hitchin data is presented. Numerous realizations of this setup are collected in the tables in Section \ref{tables}. Sections \ref{fourdimensional},\ref{Todapart},\ref{setup},\ref{tables} form the core of the paper. It is worth emphasizing that much of the tight representation theoretic structures become obvious only with the compiling of detailed tables for various cases. The arguments in Sections 5-7 apply for all simple $\mathfrak{g}$. So, the tables include data for the non-simply laced $\mathfrak{g}$ as well. These are relevant for local properties of the twisted defects of the theory $\mathscr{X}[\mathfrak{j}]$, $\mathfrak{j} \in A,D,E$ and for S-duality of boundary conditions between $\mathcal{N}=4$ SYM with non-simply laced gauge groups $G$ and $G^\vee$, where $\mathfrak{g}$ is the subalgebra of $\mathfrak{j}$ that is invariant under the twist  \cite{Chacaltana:2012zy}. However, there is a feature of the setup in the non-simply laced cases that raises some puzzles about the case for arbitrary $\mathfrak{g}$. This is discussed in Section \ref{setup}. 

 Displaying information in the tables in a succinct way requires the introduction of some notation for nilpotent orbits and irreducible representations of Weyl groups. This is introduced in Appendices \ref{orbitsappendix}, \ref{repsappendix}. Also included are two appendices that provide a short summary of the Borel-de Seibenthal method (Appendix \ref{BoreldeSiebenthal}) to find all possible centralizers of semi-simple elements and the Macdonald-Lusztig-Spaltenstein induction method (Appendix \ref{jinduction}). A variation of the setup presented in Section \ref{setup} appeared in \cite{Balasubramanian:2013kva} for case of type $A$ theories. The discussion here is more detailed and is provided in a language that generalizes directly to the case of arbitrary $\mathfrak{j} \in A,D,E$.

\section{Codimension two defects under dimensional reductions}
\label{reductions}
Let us take the theory $\mathscr{X}[\mathfrak{j}]$ on various six manifolds $M_6$ with the required partial twists to preserve some of the supersymmetries. For the current purposes, it is helpful to recall a small subset of the various reduction schemes that are helpful while studying the supersymmetric defect operators in this theory.
 Each scheme will be summarized by a dot $(\cdot)$ and dash $(\leftrightarrow)$ table. Unless specified otherwise, the co-ordinate labels in such tables are in the obvious order implied by the notation for the manifold $M_6$.
\subsection{$\mathbb{R}^{3,1}\times C_{g,n}$}
\label{WittenGaiotto}
Consider the theory $\mathscr{X}[\mathfrak{j}]$ formulated on $\mathbb{R}^{3,1}\times C_{g,n}$ where $C_{g,n}$ is a Riemann surface of genus $g$ in the presence of $n$ codimension two defects $\mathcal{O}_i$. When the area of the Riemann surface tends to zero, an effectively four dimensional $\mathcal{N}=2$ field theory is obtained \cite{Witten:1997sc,Gaiotto:2009we}. 
\begin{center}
\begin{tabularx}{\textwidth}{sssssss}
\toprule
 & 1 & 2  & 3 & 4 & 5 & 6 \\ 
 \midrule 
 \rowcolor{Gray} $\mathcal{O}_i$  & $\leftrightarrow$ & $\leftrightarrow$ & $\leftrightarrow$ & $\leftrightarrow$ & $\cdot$  & $\cdot$\\  \bottomrule
 \end{tabularx}
 \end{center}
 
 The coupling constant moduli space of such theories is the moduli space of the Riemann surface with punctures. The low energy effective action of $\mathcal{N}=2$ theories in four dimensions is captured by the Seiberg-Witten solution. For these theories obtained from six dimensions, the SW solution is identified with an algebraic complex integrable system associated to the Riemann surface $C_{g,n}$ called the Hitchin system. In particular, the SW curve is identified with the spectral curve of the Hitchin system and the SW differentials are the conserved ``Hamiltonians" of the same.
 \subsection{$\mathbb{R}^{2,1}\times \mathbb{S}^1 \times C_{g,n}$}
 Following \cite{Gaiotto:2009hg}, one can seek a description of the codimension two defect in terms of a Hitchin system using a compactification on $\mathbb{R}^{2,1}\times \mathbb{S}^1 \times C_{g,n}$, with a codimension two defect wrapping the circle $\mathbb{S}^1$.
 \begin{center}
\begin{tabularx}{\textwidth}{sssssss}
\toprule
 & 1 & 2  & 3 & 4 & 5 & 6 \\ 
 \midrule 
  \rowcolor{Gray} $\mathcal{O}_1$  & $\leftrightarrow$ & $\leftrightarrow$ & $\leftrightarrow$ & $\leftrightarrow$ & $\cdot$  & $\cdot$ \\  \bottomrule
  \end{tabularx}
 \end{center}
The nature of the defect is captured by the singularity structure of the Higgs fields near the location of the defect on $C$. When the Higgs field has a simple pole, 
\begin{equation}
\phi(z) = \frac{\rho}{z} + \ldots,
\label{hitchin}
\end{equation}
it corresponds to the tamely ramified case and corresponding defects are called regular defects. For regular defects with no mass deformations, the residue at the simple pole ($\rho$) is a nilpotent element of the lie algebra $\mathfrak{j}$. The nature of the defect depends only the nilpotent orbit to which element $\rho$ belongs. While prescribing the behaviour in \ref{hitchin} is sufficient to identify a defect (upto perhaps some additional discrete data), we will momentarily see that pairs of nilpotent orbits are in some ways a more efficient description of a given codimension two defect. When the poles for the Higgs field occur at higher orders, it corresponds to the case of wild ramification and the corresponding defects are called irregular defects \cite{Witten:2007td,Gaiotto:2009hg}.
 
\subsection{$\mathbb{R}^{2,1} \times H \times \mathbb{S}^1  $}
To see that a pair of nilpotent orbits are relevant for the description of a single codimension two defect, follow \cite{Chacaltana:2012zy} and formulate $\mathscr{X}[\mathfrak{j}]$ on $\mathbb{R}^{2,1} \times H \times \mathbb{S}^1  $. Here, $H$ is a half-cigar which can be thought of as a circle ($\tilde{S}_1$) fibered over a semi-infinite line. Here again, consider the reduction with a single defect $\mathcal{O}_1$ (along with, maybe, a twist that allows for non-simple laced gauge groups to appear in five and four dimensions). The fifth co-ordinate refers to the co-ordinate along $\tilde{S}_1$.
\begin{center}
\begin{tabularx}{\textwidth}{sssssss}
\toprule
 & 1 & 2  & 3 & 4 & 5 & 6 \\ 
 \midrule 
 \rowcolor{Gray} $\mathcal{O}_1$  & $\leftrightarrow$ & $\leftrightarrow$ & $\leftrightarrow$ & $\cdot$ & $\leftrightarrow$  & $\cdot$ \\  \bottomrule
 \end{tabularx}
 \end{center}
Upon dimensional reduction in the fifth and six dimensions, this setup reduces to the one considered by Gaiotto-Witten \cite{Gaiotto:2008sa} in their analysis of supersymmetric boundary conditions in $\mathcal{N}=4$ SYM on a half-space. Performing a reduction first on $\mathbb{S}^1$ gives us 5d SYM with gauge group $G$ and a codimension one defect. Further reducing on $\tilde{\mathbb{S}}^1$ gives  4d SYM with gauge group $G$ on a half-space and 1/2 BPS boundary condition that is labeled by a triple $(\mathcal{O},H,\mathcal{B})$, where $\mathcal{O}$ is a nilpotent orbit, $H$ is a subgroup of the centralizer of the $\mathfrak{sl}_2$ triple associated to the nilpotent orbit $\mathcal{O}$ and $\mathcal{B}$ is a three dimensional boundary SCFT. Interchanging the order of dimensional reductions, one gets 4d SYM with gauge group $G^\vee$ on a half space with a dual boundary condition $(\mathcal{O}',H',\mathcal{B}')$. In the case of $\mathfrak{g}=A_{N-1}$, nilpotent orbits have a convenient characterization in terms of partitions of $N$. An order reversing duality on nilpotent orbits plays an important role in the description of the S-duality of boundary conditions. This duality acts as an involution only in the case of $A_{n-1}$ and fails to be an involution in the other cases. This failure to be an involution leads to a much richer and complex structure than the case for type $A$. This more general order reversing duality will hover around much of the considerations in the rest of the paper and will be discussed in greater detail in subsequent sections.

 \subsection{$\mathbb{R}^{1,1}\times \mathbb{R}^{2} \times \mathbb{T}^2$}
 \label{GWsurface}

\begin{center}
\begin{tabularx}{\textwidth}{sssssss}
\toprule
 & 1 & 2  & 3 & 4 & 5 & 6 \\ 
 \midrule 
\rowcolor{Gray} $\mathcal{O}_1$  & $\cdot$ & $\cdot$ & $\leftrightarrow$ & $\leftrightarrow$ & $\leftrightarrow$  & $\leftrightarrow$ \\  \bottomrule
 \end{tabularx}
 \end{center}
 
 Here, let us consider the reduction with a single defect $\mathcal{O}_1$ on $\mathbb{R}^{1,1}\times \mathbb{R}^{2} \times \mathbb{T}^2$ such that the defect wraps the $\mathbb{T}^2$ \cite{Chacaltana:2012zy} (again, possibly with a twist). The theory in four dimensions is now $\mathcal{N}=4$ SYM  with gauge group $G$ and a surface operator inserted along a surface $\mathbb{R}^2 \subset \mathbb{R}^{1,3}$. This is the kind of setup considered in \cite{Gukov:2006jk}. The S-dual configuration is then a surface operator in $\mathcal{N}=4$ SYM with gauge group $G^\vee$.

\subsection{Associating invariants to a defect}
Under various duality operations, it may turn out that the most obvious description of a given codimension two defect is quite different. So, it is helpful to associate certain invariants to a given defect which can be calculated independently in the various descriptions. If the defect comes associated with non-trivial moduli spaces of vacua, then a basic invariant is the dimension of these moduli spaces. For the codimension two defects in question, one can associate, in general, a Higgs branch dimension and a graded Coulomb branch dimension. These will correspond to the local contributions to the Higgs and Coulomb branch dimensions of a general class $\mathcal{S}$ theory built out of these defects. 

In the work of \cite{Chacaltana:2012zy}, the graded coulomb branch dimension played an important role in the interpretation of the role played by an order reversing duality that related the two descriptions of these four dimensional defects in their realizations as boundary conditions for $\mathcal{N}$=4 SYM. In this paper, a complementary discussion that relies crucially on properties of the Higgs branch will be provided. To this end, associate an invariant to the defect that will be called the \textit{Higgs branch Springer invariant}. This will be an irreducible representation of the Weyl group $W[\mathfrak{g}]( \simeq W[\mathfrak{g}^\vee])$ and can be calculated on both sides of the S-duality for boundary conditions in $\mathcal{N}=4$ SYM. This will turn out to be a more refined invariant than just the dimension of the Higgs branch. The discussion will also have the added advantage that it provides a physical setting for certain \textit{defining} properties of the order reversing duality map as formulated in \cite{sommers2001lusztig} (and used in \cite{Chacaltana:2012zy}). Associated to this invariant is a number that will be called the Sommers invariant $\tilde{b}$ highlighting the fact it plays a crucial role in \cite{sommers2001lusztig}. Its numerical value equals the quaternionic Higgs branch dimension. 
\subsubsection{An invariant via the Springer correspondence}
\label{springerCorr}
This invariant is attached to the defect by considering the Springer resolution of either the nilpotent cone $\mathcal{N}^\vee$ or $\mathcal{N}$ (depending on which side of the duality the invariant is being calculated). The discussion in this section will be somewhat generic and is meant to give an introduction to the Springer correspondence. The calculation of the invariant is deferred to a later section. For some expositions of the theory behind the Springer resolution, see \cite{humphreys2011conjugacy,chriss2009representation,de2009decomposition}. The explicit description of what is known as the Springer correspondence can be found in \cite{carter1985finite}.

Now, consider the nilpotent variety $\mathcal{N}$ and how the closures of other nilpotent orbits sit inside the nilpotent variety $\mathcal{N}$. This leads to a pattern of intricate singularities. For example, in the case of closure of the subregular orbit $\overline{\mathcal{O}}^{sr}$ inside $\mathcal{N}[\mathfrak{g}]$ for $\mathfrak{g} \in A,D,E$, we get the Kleinien singularities  $\mathbb{C}^2/\Gamma$ where $\Gamma$ is a finite subgroup of $SU(2)$. Such finite subgroups also have a similar A,D,E classification. A well known fact is that these singularities admit canonical resolutions. For types $B_n,C_n,G_2,F_4$, one can still obtain a very explicit description of these singularities by considering the $A_{2n-1},D_{n+1},D_4,E_6$ singularities with some additional twist data \cite{slodowy1980simple}. The deeper singularities of the nilpotent variety, however, do not have such a direct presentation. There is however a general construction due to Springer which is a simultaneous resolution of all the singularities of the Nilpotent variety. It enjoys many interesting properties and plays a crucial role in the study of the representation theory of $G_\mathbb{C}$. It is constructed in the following way. Consider pairs $(e,\mathfrak{b})$ where $e$ is a nilpotent element and $\mathfrak{b}$ is a Borel subalgebra containing $e$. This space of pairs is called the Springer variety $\tilde{\mathcal{N}}$. It is also canonically isomorphic to $T^*\mathcal{B}$, the co-tangent bundle to the Borel variety. The Borel variety $\mathcal{B}$ is the space of all Borel subalgebras in $\mathfrak{g}$ and is also called the flag manifold since elements of the Borel variety stabilize certain sequences of vector spaces of increasing dimension (`flags').  The condition that a non-zero nilpotent element $e$ should belong to $\mathfrak{b}$ leads to a smaller set of Borel subalgebras that will be denoted by $\mathcal{B}_e$. This is a subvariety of the full Borel variety. The subvariety so obtained depends only on the orbit to which $e$ belong. So, a more convenient notation is $\mathcal{B}_\mathcal{O}$, where $\mathcal{O}$ is a nilpotent orbit containing $e$.  Now, consider the map that just projects to one of the factors in the pair $\mu : (e,\mathfrak{b}) \rightarrow e$. When $e$ to allowed take values in arbitrary nilpotent orbits, the map $\mu : \tilde{\mathcal{N}} \rightarrow \mathcal{N}$ provides a simultaneous resolution of the singularities of $\mathcal{N}$. For $e$ being the zero element, the fiber over $e$, $\mu^{-1} (0)$ is the full Borel variety. And, $\text{dim} (\mathcal{B}) = \frac{1}{2} \text{dim} (\mathcal{N})$. For more general nilpotent elements, this dimension formula is modified to (see \cite{spaltenstein1982classes,carter1985finite})
\begin{equation}
\text{dim}(\mathcal{B}_{\mathcal{O}}) = \frac{1}{2} (\text{dim} (\mathcal{N}) - \text{dim}(\mathcal{O})).
\end{equation}

Resolutions in which the fibers obey the above relationship belong to a class of maps called semi-small resolutions. In other words, the Springer resolution of the nilpotent cone is a semi-small resolution \cite{borho1983partial}.  Apart from constructing the resolution, Springer also showed that the Weyl group acts on the cohomology ring of the fiber $\mathcal{B}_{\mathcal{O}}$. This action commutes with the action of the component group $A(\mathcal{O})$ which acts just by permuting the irreducible components of $\mathcal{B}_{\mathcal{O}}$.  In particular, the top dimensional cohomology $H^{2k}(\mathcal{B}_{\mathcal{O}},\mathbb{C})$ (with $k=\text{dim}_\mathbb{C}(\mathcal{B}_{\mathcal{O}})$) decomposes in the following way as a $W[\mathfrak{g}] \times A(\mathcal{O})$ module,
\begin{equation}
H^{2k}(\mathcal{B}_{\mathcal{O}},\mathbb{C}) = \bigoplus_{\chi \in Irr(A(\mathcal{O}))} V_{\mathcal{O},\chi} \otimes \chi
\label{springertop}
\end{equation}
where $\chi$ is an irreducible representation of the $A(\mathcal{O})$ and $V_{\mathcal{O},\chi}$ is an irreducible representation of the Weyl group. The component group $A(\mathcal{O})$ is defined as $C_G(e)/C_G(e)^0$, where $C_G(e)$ is the centralizer of the $e$ in group $G_{\mathbb{C}}$ and $C_G(e)^0$ is its connected component. The groups $A(\mathcal{O})$ are known for any nilpotent orbit $\mathcal{O}$ and can be obtained from the mathematical literature \cite{collingwood1993nilpotent,sommers1998generalization}. When the decomposition in \ref{springertop} involves nontrivial $\chi$, there are non-trivial local systems associated to the nilpotent orbit and $V_{\mathcal{O},\chi}$ corresponds to one of these local systems. In the classical cases, $A(\mathcal{O})$ is either trivial or the abelian group $(S_2)^n$ for some $n$. In type $A$, the component group is always trivial. In the exceptional cases, $A(\mathcal{O})$ belongs to the list $S_2,S_3,S_4,S_5$. While $S_2,S_3$  occur as component groups for numerous orbits in the exceptional cases, the groups $S_4$ and $S_5$ correspond to unique nilpotent orbits in $F_4$ and $E_8$ respectively. 

In most cases, all irreducible representations of $A(\mathcal{O})$ appear in the above direct sum (\ref{springertop}). In cases where this does not occur, the number of missing representations is always one and the pair $(\mathcal{O}, \chi)$ is called a \textit{cuspidal} pair. Such cuspidal pairs are classified and a generalization due to Lusztig incorporates these pairs as well into what is called the generalized Springer correspondence (see \cite{shoji1988geometry} for a review). One can further show that all irreps of $W[\mathfrak{g}]$ occur as part of the summands like \ref{springertop} for some unique pair $(\mathcal{O},\chi)$. The irreps of $W[\mathfrak{g}]$ which occur with the trivial representation of $A(\mathcal{O})$ (in other words, those that correspond to some pair $(\mathcal{O},1)$) are sometimes called the Orbit representations of $W[\mathfrak{g}]$ \footnote{This terminology however is not uniformly adopted. The name Springer representation is also used sometimes as an alternative.}.

Let $Irr(W)$ be the set of all irreducible representation of $W[\mathfrak{g}]$ and let $[\mathcal{O}]$ be the set of all nilpotent orbits in $\mathfrak{g}$ and $[\tilde{\mathcal{O}}]$ be the set of all pairs $(\mathcal{O},\chi)$, where $\chi$ is an irreducible representation of $A(\mathcal{O})$. The nature of the decomposition in \ref{springertop} defines an injective map,
\begin{equation}
Sp[\mathfrak{g}] : Irr(W) \rightarrow [\tilde{\mathcal{O}}].
\end{equation}
This injective map is called the Springer correspondence. A specific instance of this map will be denoted by $Sp[\mathfrak{g},r] : r \mapsto (\mathcal{O},\chi)$ for a unique pair $(\mathcal{O},\chi) \in [\tilde{\mathcal{O}}] $.

When the inverse exists, it will be denoted by $Sp^{-1}[\mathfrak{g},(\mathcal{O},\chi)]$ or (when $\chi=1$) $Sp^{-1}[\mathfrak{g},\mathcal{O}]$. The following two instances of the Springer map hold for all $\mathfrak{g}$. Let $\mathcal{O}^{pr}$ and $\mathcal{O}^0$ denote the principal orbit and the zero orbit respectively. Then,
\begin{eqnarray}
Sp^{-1}[\mathfrak{g},\mathcal{O}^{pr}] &=& \text{Id} \\
Sp^{-1}[\mathfrak{g},\mathcal{O}^0] &=& \epsilon,
\end{eqnarray}
where $\text{Id}, \epsilon$ refer (respectively) to the trivial and the sign representations of $W[\mathfrak{g}]$. This is the Springer correspondence in Lusztig's normalization. In \cite{carter1985finite}, the Springer correspondence is described in this normalization. Many geometric notions that one may associate with the theory of nilpotent orbits like partial orders, induction methods, duality transformations, special orbits, special pieces etc. have algebraic analogues in the world of Weyl group representations. The two worlds interact via the Springer correspondence.

In the context of understanding properties of codimension two defects, an interest in the Springer correspondence can be justified in the following way. For the class of defects under discussion, there is an associated Higgs branch moduli space which admits at least two different descriptions. One of them is as the space of solutions to Nahm equations with a certain boundary condition. This involves a nilpotent orbit in $\mathfrak{g}$ that will be called the Nahm orbit $\mathcal{O}_N$. The second realization is obtained as the Higgs branch of theory $T^\rho[G]$. In either case, an invariant to the defect can be assigned using the Springer correspondence. In the former case, the association is somewhat direct once the Nahm orbit $\mathcal{O}_N$ is known. In the latter case, this invariant will satisfy a non-trivial compatibility condition with properties of the Springer fiber over another nilpotent orbit $\mathcal{O}_H$ (the Hitchin orbit in $\mathfrak{g}^\vee$) that goes into the description of the Coulomb branch of $T^\rho[G]$. Requiring that this consistency condition hold for all defects will turn out to determine the pairs $(\mathcal{O}_N,\mathcal{O}_H)$ that can occur in the description of the defect. The ability to do so is completely independent of the availability of brane constructions and this allows one to understand the exceptional cases as well. Explaining how this can be done would be the main burden of the following two sections. It is also useful at this point to note that the bridge to representation theory of Weyl groups will also turn out be helpful in understanding the relationship to the Toda picture of codimension two defects which we will turn to in Section \ref{Todapart}.
\subsubsection{An invariant via the Kazhdan-Lusztig Map}

An alternative to using the Springer correspondence to define an invariant for a co-dimension two defect would be to consider the Kazhdan-Lusztig map which provides an injection from the set of nilpotent orbits in $\mathfrak{g}$ to the set of conjugacy classes in $W[\mathfrak{g}]$. This is, in a sense, a dual invariant to the one provided by considering the Springer correspondence. In the context of the four dimensional defects of the theory $\mathscr{X}[\mathfrak{j}]$, one could consider the compactification scheme of \ref{GWsurface}. The resulting four dimensional picture would involve $\mathcal{N}=4$ SYM with a surface operator, similar to the setup considered in \cite{Gukov:2008sn}. There, it was necessary to match the local behaviour of polar polynomials formed out of the Higgs field in an associated Hitchin system on the $G$ \& $G^\vee$ sides for the determination of the S-duality map. It was argued in \cite{Gukov:2008sn} that the KL map offered a compact way to implement this check. In this paper, this invariant will not play a central role. But, it will feature in a discussion of a possible extension of the setup provided in Section \ref{setup}.
\section{S-duality of Gaiotto-Witten boundary conditions}
\label{GWbc}

Recall that Gaiotto-Witten constructed a vast set of 1/2 BPS boundary conditions for $\mathcal{N} =4$ SYM on a half space \cite{Gaiotto:2008sa}. The most general boundary condition in this set can be described by a triple $(\mathcal{O}, H, \mathcal{B})$. Here, $\mathcal{O}$ is a nilpotent orbit. By the Jacobson-Morozov theorem, to every nilpotent orbit $\mathcal{O}$ is an associated $\mathfrak{sl}_2$ embedding $\rho_{\mathcal{O}} : \mathfrak{sl}_2 \rightarrow \mathfrak{g}$.  $H$ is a subgroup of the centralizer of $\mathfrak{sl}_2$ triple associated to $\mathcal{O}$ and $\mathcal{B}$ is a three dimensional SCFT living on the boundary that has a $H$ symmetry. This data is translated to a boundary condition as below,
\begin{itemize}
\item Impose a Nahm pole boundary condition that is of type $\rho_{\mathcal{O}}$,
\item At the boundary, impose Neumann boundary conditions for gauge fields valued in the subalgebra $\mathfrak{h} $ of $\mathfrak{g}$,
\item Gauge the $H$ symmetry of three dimensional boundary $\mathcal{B}$ and couple it to the corresponding four dimensional vector multiplets.
\end{itemize}

In talking about these boundary conditions, it is very helpful to always think of some special cases. Take  $\{\mathcal{O}^0,\mathcal{O}^{m}, \mathcal{O}^{sr},\mathcal{O}^{pr}\}$ to refer respectively to $\{$the zero orbit, the minimal orbit, the sub-regular orbit,the principal orbit $\}$. The principal orbit is sometimes called the regular orbit in the literature but in the discussions here, only the former name will appear. For the subgroup $H$, take $\{Id \}$ to denote the case where the gauge group is completely Higgsed at the boundary and $\{G \}$ to be case where it is not Higgsed. For the boundary field theory $\mathcal{B}$, the value $\varnothing$ corresponds to the case where there is no boundary field theory that is coupled to the bulk vector multiplets. A class of boundary theories named $T^\rho[G]$ played an important role in the discussion of S-dualities in \cite{Gaiotto:2008ak} and cases where $\mathcal{B} =  T^\rho[G]$ will turn out to be important in the current discussion as well.

The Higgs and Coulomb branches of these theories are certain sub-spaces \footnote{ \textit{strata} would, technically, be a more accurate term.} inside the Nilpotent cones $\mathcal{N}$ and $\mathcal{N}^{\vee}$. For much of what follows, various notions associated with the structure theory of nilpotent orbits in complex semi-simple Lie algebras will be routinely invoked. Accessible introductions to these aspects can be found in \cite{collingwood1993nilpotent,mcgovern2002adjoint}.

With these preliminaries established, one can now look at how S-dualities act on some of the simplest boundary conditions. For example, consider the triple $(\mathcal{O}^0,Id,\varnothing)$ that corresponds to the Dirichlet boundary conditions for the gauge fields and $(\mathcal{O}^0,G,\varnothing)$ corresponds to Neumann boundary conditions for the gauge fields. One of the important features of the GW set of boundary conditions is that it is \textit{closed} under S-duality. But, the simple minded boundary conditions recounted above get mapped to non-trivial boundary conditions. The S-dual of $(\mathcal{O}^0,Id,\varnothing)$ in a theory with gauge group $G$ is the boundary condition $(\mathcal{O}^0,G^\vee,T[G])$ in a theory with gauge group $G^\vee$. On the other hand, the dual of $(\mathcal{O}^0,G,\varnothing)$   is $(\mathcal{O}^{pr},Id,\varnothing)$. One strong evidence in favor of the identification of S-duality between these boundary conditions is the fact that dimensions of the vacuum moduli space of $\mathcal{N}=4$ SYM with these boundary conditions happen to match on both sides. In the two cases considered above, the moduli space is the nilpotent cone $\mathcal{N}$ in the first case and a point in the second case. These occurrences of the S-duality map \footnote{We are concerned here just with the $\mathbb{Z}_2$ subgroup of the full S-duality group that acts on the coupling constant as $\tau^\vee = -1/ n_r \tau$, where $n_r$ is the ratio of lengths of the longest root to the shortest root.} are listed in table \ref{Sdualitymap}.

\begin{table} [!h]
\centering 
\caption{S-duality of boundary conditions in $\mathcal{N}=4$ SYM} 
\begin{center}
\begin{tabular}{c|c|c}
\toprule
$\mathcal{N}=4$ SYM with gauge group $G$  & $\mathcal{N}=4$ SYM with gauge group $G^\vee$& Associated moduli space\\ \midrule
$(\mathcal{O}^0,G,\varnothing)$ & $(\mathcal{O}^{pr},Id,\varnothing)$ & $\cdot$ \\
$(\mathcal{O}^0,Id,\varnothing)$ & $(\mathcal{O}^0,G^\vee,T[G])$ & $\mathcal{N}$ \\
$(\mathcal{O}^\rho,Id,\varnothing)$ & $(\mathcal{O}^0,G^\vee,T^\rho[G]) $ & $\mathcal{S}^\rho \cap \mathcal{N}$\\ \bottomrule
\end{tabular}
\end{center}
\label{Sdualitymap}
\end{table}

We will not be needing the constructions of Gaiotto-Witten in their full generality. The cases that will be of direct relevance to discussions here correspond to the ones with a pure Nahm pole boundary condition and its S-dual case of a Neumann boundary condition along with a coupling to a three dimensional theory $T^{\rho}[G]$ and certain deformations thereof. In the rest of the section, we will look closely at duality between $(\mathcal{O}^\rho,Id,\varnothing)$ in the theory with gauge group $G$  and $(\mathcal{O}^0,G^\vee,T^{\rho}[G])$ in the theory with gauge group $G^\vee$. An important point to note here is that the specification of the boundary condition on the $G^\vee$ is incomplete without a description of how the theory $T^\rho[G]$ is coupled to boundary multiplets. In the adopted conventions, the Higgs branch of $T[G]$ will have a $G$ global symmetry, while the Coulomb branch has a $G^\vee$ global symmetry. So, the natural way to couple $T^\rho[G]$ would be to gauge the global symmetry on the Coulomb branch\footnote{The symmetries on the Coulomb branch are not obvious in any Lagrangian description of $T^\rho[G]$. So, a more practical way to describe this coupling is to use the description of this branch as the Higgs branch of the mirror theory $T_{\rho^\vee}[G]$. But, to simplify things, all statements in this paper are made with the theories $T^\rho[G]$.} and couple it to the boundary vector multiplets of the $G^\vee$ theory. The Higgs branch of $T^\rho[G]$ is now understood to be the vacuum moduli space of the full four dimensional theory with this boundary condition. As one may guess, understanding this instance of the duality map requires a careful study of the moduli spaces of Nahm equations under different pole boundary conditions and the theories $T^{\rho}[G]$ and their vacuum moduli spaces. Some of the main elements of such a study are outlined in the rest of the Section. 

\subsection{Moduli spaces of Nahm equations}

Various aspects of Nahm equations and their moduli space of solutions are reviewed in \cite{Gaiotto:2008sa}. For some other useful works which elucidate Nahm equation from different points of view, see \cite{Diaconescu:1996rk,atiyah2002nahm}.

In the setting of boundary conditions of $\mathcal{N}=4$ SYM \cite{Gaiotto:2008sa}, Nahm boundary conditions arise as a generalization of the usual Dirichlet boundary conditions. Recall that there are six real scalar fields in this theory. Let $\overrightarrow{X}$ be the triplet for which Nahm type boundary conditions conditions are imposed. Formulate the theory on $\mathbb{R}^3 \times \mathbb{R}^+$ and let $y$ be a co-ordinate along $\mathbb{R}^+$ with $y=0$ being the boundary. Let $\rho$ be a $\mathfrak{sl}_2$ embedding, $\rho : \mathfrak{sl}_2 \rightarrow \mathfrak{g}$. Then, the boundary conditions are of the form
\begin{eqnarray}
\frac{dX^i}{dy} &=& \epsilon_{ijk}[X^i,X^j] \\
X^i &=& \frac{t^i}{y} , y \rightarrow 0 \hspace{0.5 in} (i=1,2,3).
\label{Nahmpole}
\end{eqnarray}
with $t^i$ being a $\mathfrak{sl}_2$ triple associated to $\rho(e,f,h)$, $(e,f,h)$ being the standard triple. The first part is the usual Nahm equation while the second part of the boundary condition modifies it to a Nahm pole boundary condition. When $\rho$ is the zero embedding, this reduces to the case of a pure Dirichlet boundary condition. Following the works of Kronheimer \cite{kronheimer1990hyper}, it is known that solutions to \ref{Nahmpole} is a hyper-kahler manifold. Denote this by $\mathcal{M}_\rho(\overrightarrow{X}_\infty)$, where $\overrightarrow{X}_\infty$ are the values of $\overrightarrow{X}$ at $y \rightarrow \infty$. When $\overrightarrow{X}_\infty=0$,  $\mathcal{M}_{\rho}(\overrightarrow{X}_\infty)$ is a singular space. Some special cases are
\begin{itemize}
\item $\rho$ is the zero embedding. Here, $\mathcal{M}_\rho(0)$ is the nilpotent variety $\mathcal{N}$ of $G$.
\item $\rho$ is the sub-regular embedding. In this case, $\mathcal{M}_\rho(0)$ is a singularity of the form $\mathbb{C}^2/\Gamma$.
\item For $\rho$ being the principal embedding, $\mathcal{M}_\rho(0)$ is just a point.
\end{itemize}

In the more general cases, $\overrightarrow{X}_\infty$ is a non-zero semi-simple element and one obtains a resolution/deformation of the singular space. In this more general case,  $\overrightarrow{X}_\infty \in \mathfrak{t}^3/W$, where $W$ is the Weyl group.  Specializing to $\overrightarrow{X}_\infty=(i \tau, 0,0)$, one gets a resolution of the moduli space of solutions in one of the complex structures. It turns out that many of the ideas in the setup just reviewed play an important role in geometric representation theory. From a purely complex point of view, these moduli spaces have been studied in the works of Grothendieck-Brieskorn-Slodowy \cite{slodowy1980simple,slodowy1980four}. The general solution to Nahm pole boundary conditions is in fact best described as the intersection $\mathcal{S}^\rho \cap \mathcal{N}$ where $\mathcal{S}^\rho$ is the Slodowy slice that is transverse (in $\mathfrak{g}$) to the nilpotent orbit $\rho$. The realization of these spaces as solutions to Nahm equations gives a new hyper-kahler perspective. 

\subsubsection{Springer resolution of Slodowy slices}
Consider the Springer resolution $\mu$ discussed in Section \ref{springerCorr}. As already noted, this resolution is semi-small. Now, consider the preimage of $\mathcal{S}=\mathcal{S}^\rho \cap \mathcal{N}$ under $\mu$, given by $ \tilde{\mathcal{S}} = \mu^{-1}(\mathcal{S})$. It can be shown that $\text{dim}(\tilde{\mathcal{S}}) = \text{dim} (\mathcal{N}) - \text{dim}(\mathcal{O}_N)$ (all dimensions are complex dimensions unless stated otherwise). The Springer fiber $\mathcal{B}_N = \mu^{-1}(e)$, where $e$ is a representative of $\mathcal{O}_N$ is a space of dimension $\text{dim}(\mathcal{B}_N) = \frac{1}{2} ( \text{dim} (\mathcal{N}) - \text{dim}(\mathcal{O}_N) ) $. Further, $\mathcal{B}_N$ is a Lagrangian sub-manifold of $\tilde{\mathcal{S}}$ and can be obtained as a homotopy retract of $\tilde{\mathcal{S}}$  \cite{chriss2009representation,ginzburg2008harish}. In particular, $H^*(\tilde{\mathcal{S}}) = H^*(\mathcal{B}_N)$. Slodowy's construction naturally endows an action of the Weyl group on $H^*(\tilde{\mathcal{S}})$ as the monodromy representation. This then endows a Weyl group action on $H^*(\mathcal{B}_N)$. It is known that this action matches with the one from Springer's construction \cite{slodowy1980four} (in Lusztig's normalization). In particular, $H^\text{top}(\mathcal{B}_N)$ is a $W[\mathfrak{g}] \times A(\mathcal{O}_N)$ module. In light of the fact that the moduli space of solutions is actually a hyper-Kahler manifold, it is natural to associate to it a quaternionic dimension. Let $\text{dim}_{\mathbb{H}}(\mathcal{S}^\rho \cap \mathcal{N})$ be the quaternionic dimension. Then, the dimension formulas immediately imply 
\begin{equation}
\text{dim}_{\mathbb{H}}(\mathcal{S}^\rho \cap \mathcal{N}) = \text{dim}_{\mathbb{C}} (\mathcal{B}_N).
\end{equation}
It is convenient to note the above relation since $\text{dim}_{\mathbb{C}} (\mathcal{B}_N)$ is often readily available in the mathematical literature on Springer resolutions.
\subsection{Vacuum moduli spaces of $T^{\rho}[G]$}
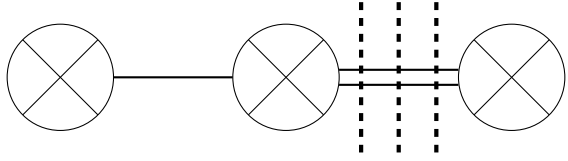
\begin{figure}
\begin{center}
\begin{tikzpicture}[node distance=3cm, auto]
    \node[matrix] (N1) {
        \draw (0,0) circle (0.707) ; 
        \draw (-0.5,-0.5) -- (0.5,0.5); 
        \draw (-0.5,0.5) -- (0.5,-0.5); \\
    };
    
    \node[matrix,right of=N1] (N2) 
    {
        \draw (0,0) circle (0.707) ; 
        \draw (-0.5,-0.5) -- (0.5,0.5); 
        \draw (-0.5,0.5) -- (0.5,-0.5); \\
    };
    \node[matrix,right of=N2] (N3) 
    {
        \draw (0,0) circle (0.707) ; 
        \draw (-0.5,-0.5) -- (0.5,0.5); 
        \draw (-0.5,0.5) -- (0.5,-0.5); \\
    };
	\draw [thick] (0.707,0) -- (3-0.707,0) ;
    \draw [thick] (3+0.707,0.1) -- (6-0.707,0.1) ;
     \draw [thick] (3+0.707,-0.1) -- (6-0.707,-0.1) ;
    \draw [dashed, ultra thick](4,1) -- (4,-1) ;
    \draw [dashed, ultra thick](5,1) -- (5,-1) ;
    \draw [dashed, ultra thick](4.5,1) -- (4.5,-1) ;
\end{tikzpicture}
\end{center}
\caption{Brane realization of $T[SU(3)]$. The D5 linking numbers are $l_i=(2,2,2)$ and the NS5 linking numbers are $\tilde{l_i}=(1,1,1)$}
\label{braneconstruction}
\end{figure}

The $T^\rho[G]$ theories are certain 3d $\mathcal{N}=4$ SCFTs that  play an important role in the description of S-duality of boundary conditions for $\mathcal{N}=4$ SYM. For $G$ classical, Gaiotto-Witten provide brane constructions in type IIB string theory (following the setup of \cite{Hanany:1996ie}) to describe the boundary conditions. In particular, their setup provides a brane construction of many of the three dimensional theories $T^\rho[G]$. An example of such a brane construction for $G=SU(N)$ is given in Fig \ref{braneconstruction}. For $G$ exceptional, the theories $T^\rho[G]$ exist although brane constructions are no longer available. There are however some general features that are expected to be shared by all $T^\rho[G]$. Most notable among this is the fact that the vacuum moduli spaces of these theories arise as certain subspaces of $\mathcal{N} \times \mathcal{N}^\vee$, where $\mathcal{N}$ is the nilpotent cone for the lie algebra $\mathfrak{g}$ while $\mathcal{N}^\vee$ is the nilpotent cone associated to the dual lie algebra $\mathfrak{g}^\vee$. More concretely \cite{Gaiotto:2008ak,Chacaltana:2012zy} let $(\mathcal{O}_N, \mathcal{O}_H)$ denote a pair of nilpotent orbits in $\mathfrak{g},\mathfrak{g}^\vee$.  The Higgs branch of $T^\rho[G]$ is a hyper-kahler manifold of complex dimension $\text{dim}(\mathcal{N})-\text{dim}(\mathcal{O}_N)$ and the Coulomb branch of $T^\rho[G]$ is another hyper-kahler manifold of dimension $\text{dim}(\mathcal{O}_H)$. It follows that for the corresponding four dimensional theory\footnote{Recall $T^\rho[G]$ is obtained by compactifying the four dimensional $\mathcal{N}=2$ codimension two defect theory on a circle and hence has a Higgs branch of the same dimension and a Coulomb branch that is twice the dimension of the 4d Coulomb branch.} on the co-dimension two defect, the dimensions of the Higgs branch and the Coulomb branch dimension are $\text{dim}(\mathcal{N})-\text{dim}(\mathcal{O}_N)$ and $\frac{1}{2}(\text{dim}(\mathcal{O}_H))$ respectively.
%

\subsubsection{Resolution of the Higgs branch}

Recall that under the conventions adopted, the theory $T^\rho[G]$ appears on the side of the duality with 4d SYM for gauge group $G^\vee$ and its Coulomb branch is a nilpotent orbit in $\mathfrak{g}^\vee$. Upon coupling to the boundary gauge fields, the Higgs branch of the theory is identified as the vacuum moduli space of the 4d theory with a boundary. The equivalence between this Higgs branch and the presentation of the space as $\mathcal{S}^\rho \cap \mathcal{N}$ is a highly non trivial assertion but one that can not be checked directly since an independent prescription for the Higgs branch does not exist for arbitrary $T^\rho[G]$. In this paper, it will be taken for granted that the S-dual boundary condition for a Nahm pole boundary condition should indeed involve one of the theories $T^\rho[G]$. Under this assumption, it will be possible to determine which of the $T^\rho[G]$ arise as part of the dual boundary condition to a particular Nahm boundary condition. Now, associated to the theory $T^\rho[G]$ are certain Fayet - Iliopoulos (FI) parameters $\overrightarrow{\zeta}$. The Springer resolution of the Higgs branch of $T^\rho[G]$ can be understood to arise from giving particular non-zero values to some of the FI parameters \cite{Gaiotto:2008ak}. Although an explicit description of this geometry is not available, one expects this to match the $\mathfrak{g}$ description where the resolution parameters entered the Nahm description as $\overrightarrow{X}_{\infty}$. The upshot of the argument here is that it makes sense to attach a Springer invariant to the resolved Higgs branch of $T^\rho[G]$. In Section \ref{fourdimensional}, it will be seen that requiring that the Springer invariant obtained from the $\mathfrak{g}$ and $\mathfrak{g}^\vee$ descriptions match is a strong constraint on the relationship between $\mathcal{O}_H$ and $\mathcal{O}_N$. The next section sets the ground by introducing several mathematical notions that are critical for Section \ref{fourdimensional}.

\section{Duality maps and Representations of Weyl groups}
\label{dualitymaps}
\subsection{Various duality maps}
Order reversing duality maps turn out to play an important role in understanding the physics of $T^\rho[G]$ theories and hence of the associated co-dimension two defects. But, there are different order reversing duality maps in the mathematical literature and it is helpful to know certain defining features of these maps to understand the nature of their relevance to the physical questions. To this end, here is a quick review of the available duality maps. Let us define the following. The set of all nilpotent orbits in $\mathfrak{g}$ will be denoted by $[\mathcal{O}]$. The set of all nilpotent orbits in $\mathfrak{g}^{\vee}$ will be denoted by $[\mathcal{O}^{\vee}]$. The special orbits within these two sets will be denoted by $[\mathcal{O}_{sp}]$, $[\mathcal{O}^{\vee}_{sp}]$. The notation $[\overline{\mathcal{O}}]$ refers to all pairs $(\mathcal{O},C)$ where $\mathcal{O} \in [\mathcal{O}]$ and $C$ is an conjugacy class of the group $\bar{A}(\mathcal{O})$. This group $\bar{A}(\mathcal{O})$ is a quotient (defined by Lusztig) of the component group $A(\mathcal{O})$ of the nilpotent orbit $\mathcal{O}$. The following order reversing duality maps have been constructed in the mathematical literature.
\begin{center}
\begin{tabularx}{\textwidth}{bb}
\toprule
The duality map & Its action \\ 
 \midrule 
  Lusztig-Spaltenstein  & $d_{LS} : [\mathcal{O}] \rightarrow [\mathcal{O}_{sp}] $ \\  
  Barbasch-Vogan  & $d_{BV} : [\mathcal{O}] \rightarrow [\mathcal{O}^{\vee}_{sp}] $ \\ 
  Sommers & $d_{S} : [\mathcal{O}] \rightarrow [\overline{\mathcal{O}^{\vee}}_{sp}] $ \\
  Achar & $d_{A} :  [\overline{\mathcal{O}}] \rightarrow [\overline{\mathcal{O}^{\vee}}] $ \\ \bottomrule
 \end{tabularx}
 \end{center}

Each of these maps invert the partial order on the set of nilpotent orbits. For example, the principal orbit is always mapped to the zero orbit and the zero orbit is always mapped to the principal orbit. The name `order-reversing duality' is meant to highlight this fact. The Lusztig-Spaltenstein map is explicitly detailed in \cite{spaltenstein1982classes} and is the only order-reversing duality map that strictly stays within $\mathfrak{g}$ and does not pass to the dual lie algebra. In this sense, it occupies a different position from the other three maps. The order reversing map of Sommers \cite{sommers2001lusztig} (further elaborated upon in \cite{achar2002local} and extended by Achar in  \cite{achar2003order}) is defined \footnote{One could equivalently view the Sommers map as being defined in the opposite direction, $d_S : [\overline{\mathcal{O}^{\vee}}_{sp}] \rightarrow [\mathcal{O}]$. The way it is written here is the direction in which it is invoked in \cite{Chacaltana:2012zy}.} by combining the duality construction due to Lusztig-Spaltenstein \cite{spaltenstein1982classes} and a map constructed by Lusztig in \cite{lusztig1984characters}. The duality map of Barbasch-Vogan \cite{barbasch1985unipotent} arises from the study of primitive ideals in universal enveloping algebras (equivalently of Harish-Chandra modules) and can be thought of as a special case of the duality maps due to Sommers and Achar.

Everytime an order reversing duality map is used in this paper, it will be explicitly one of the maps summarized in the table above. The order reversing duality that is used in \cite{Chacaltana:2012zy} is the Sommers duality map $d_S$. If one forgets the additional discrete data associated to the special orbit that arises on the $\mathfrak{g}^\vee$ side, this reduces to the duality map of Barbasch-Vogan, $d_{BV}$. In \cite{Chacaltana:2012zy}, the name \textit{Spaltenstein dual} is used for describing a duality map that passes to the dual lie algebra. This terminology is potentially confusing if one wants to compare with the mathematical literature and will not be adopted here. All of these maps are easiest to describe when their domain is restricted to just the special orbits. It is an important property of the maps that they act as involutions on the special orbits. Considering the case of special orbits in $\mathfrak{g}=\mathfrak{so}_8$, $\mathfrak{g}^\vee=\mathfrak{so}_8$. In this case, all the above maps coincide and their action is best seen as the unique order reversing involution acting on the closure diagram for special orbits.

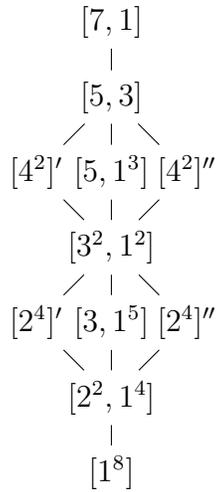
\begin{figure} [!h]
\begin{center}
\begin{tikzpicture}
  \node (max) at (0,7) {$[7,1]$};
  \node (a) at (0,6) {$[5,3]$};
  \node (b) at (0,5) {$[5,1^3]$};
  \node (c) at (-1,5) {$[4^2]'$};
  \node (d) at (1,5) {$[4^2]''$};
  \node (e) at (0,4) {$[3^2,1^2]$};
  \node (f) at (-1,3) {$[2^4]'$};
  \node (g) at (1,3) {$[2^4]''$};
  \node (h) at (0,3) {$[3,1^5]$};
  \node (i) at (0,2) {$[2^2,1^4]$};
  \node (j) at (0,1){$[1^8]$};
  \draw[preaction={draw=white, -,line width=6pt}] (max) -- (a)--(b)--(e)--(h)--(i)--(j);
  \draw[preaction={draw=white, -,line width=6pt}] (d)--(e);
  \draw[preaction={draw=white, -,line width=6pt}] (e) -- (g);
  \draw[preaction={draw=white, -,line width=6pt}] (e) --(c) --(a) --(d);
  \draw[preaction={draw=white, -,line width=6pt}] (e) -- (f) --(i) --(g);
 \end{tikzpicture}
\end{center}
\caption{Hasse diagram describing the closure ordering for special nilpotent orbits in $\mathfrak{so}_8$.}
\label{hassesl6}
\end{figure}

As one further remark, let us note here a particular subtlety. Even in scenarios where $d_{LS}$ and $d_{BV}$ have identical domain and image, they could disagree. For example, in the case of $\mathfrak{g}=F_4$, $\mathfrak{g}^\vee=F_4$. So, the domain and the image for $d_{LS}$ are identical to that for $d_{BV}$. But, $d_{LS}$ and $d_{BV}$ disagree for certain nilpotent orbits (see the Hasse diagram for $F_4$ in \cite{Chacaltana:2012zy}).

An important feature of all the duality maps is their close interaction with the Springer correspondence and consequently with the representation theory of Weyl groups. In fact, some of the maps are defined using the Springer correspondence. So, any attempt to gain a deeper understanding of how the duality maps work is aided greatly by a study of the representation theory of Weyl groups. In the rest of the section, some of the elements of this theory are recounted.

\subsection{Families, Special representations and Special orbits}
Let $Irr(W)$ denote the set of irreducible representation of the Weyl group $W$. There is a distinguished subset of $Irr(W)$ called special representations that are well behaved under a procedure known as truncated induction (or $j$ induction, see Appendix \ref{jinduction}) and duality. To explain this, denote the set of special representations by $S_W$. Now, let $s_p$ be a special representation of a parabolic subgroup $W_p$. Requiring that the identity representation be special and considering all parabolic subgroups of a Weyl group and proceeding inductively, define $s$ to be special if $s=j_{W_p}^W(s_p)$ for some parabolic subgroup $W_p$ and additionally $s'= i(s)$ is also special. Here, $i(s)$ refers to Lusztig's duality which in almost all cases acts as tensoring by the sign representation. The exceptions are certain cases in $E_7$ and $E_8$ which will be discussed at a later point (See Section \ref{exceptionalorbits}). Proceeding in this fashion, Lusztig determined the set of all special representations in an arbitrary Weyl group in \cite{lusztig1979class}. 

Another important notion that is defined inductively is that of a cell module\footnote{An equivalent term is that of a `constructible representation' but the term cell module will be preferred in this paper.}. This is a not-necessarily irreducible module of $W$ that, again, has some very nice properties under induction and duality. The trivial representation $Id$ is defined to be a cell module by itself. One arrives at the other cell modules in the following way. Let $c$ be a cell module of $Irr(W)$ and $c_p$ be a cell module of a parabolic subgroup $W_p$ of $W$. Consider their behaviour under two operations for arbitrary subgroups $W_p$,
\begin{eqnarray}
c' &=& \epsilon \otimes c,\\
c'' &=& Ind_{W_p}^W (c_p),
\end{eqnarray}
where $Ind$ is the usual induction (in the sense of Frobenius) from a parabolic subgroup. Requiring that the above two operations always yield another cell module determines all the cell modules in $W[\mathfrak{g}]$ for every $\mathfrak{g}$. The structure of these cell modules has what may seem like a surprising property. Each cell module has a \textit{unique} special representation as one of its irreducible summands. Additionally, the representations that occur as part of a cell module that contains a special representation $s$ occur \textit{only} in the cell modules that contain $s$ as the special representation. This structure suggests a certain partitioning of $Irr(W)$ \cite{lusztig1982class}. It is of the following form \footnote{There is an equivalent partitioning of Weyl group representations using the idea of a two-cell of the finite Weyl group. In this paper, the term family will be used uniformly.},
\begin{equation}
Irr(W) = \coprod_{s} f_s
\label{families}
\end{equation}
where $s$ is a special representation. An irrep $r$ occurs in the family $f_s$ if and only if it occurs in a cell module along with the special representation $s$.
In type $A$, all representations are special and hence the above partitioning reduces to the statement that each irreducible representation of $W(A_n)$ belongs to a separate family in which it is the only constituent. This simple structure however does not hold for Weyl groups outside of type $A$. The general case includes non-special representations which occur as constituents of some of the families $f_s$. So, a typical family contains a unique special representation (which can be used to index the family as in \ref{families}) and a few non-special representations. Associated to each family are the cell modules in which the representation $s$ occurs as the special summand. As an example of a family with more than one constituent, consider the unique non-trivial family in $D_4$ (see Appendix \ref{typeD} for the notation adopted),
\begin{equation}
f_{([2,1],[1])} = \{([2,1],[1]),([2^2],-) ,([2],[1^2])\}.
\end{equation}
The special representation in this family is given by $([2,1],[1])$ and the cell modules that belong to this family are
\begin{eqnarray}
c_1 &=& ([2,1],[1]) \oplus ([2^2],-), \\
c_2 &=& ([2,1],[1]) \oplus ([2],[1^2]).
\end{eqnarray}
To every irreducible representation of a Weyl group, Lusztig assigns a certain invariant such that it is constant within a family and unique to it. Its value is equal to the dimension of the Springer fiber associated to the special element in a given family. For the family in the example discussed above, the $a$ value is  3 and it is the unique family in $W(D_4)$ that has $a=3$. Here, it is appropriate to also note that one of the earliest characterizations of \textit{special orbits} was via the Springer correspondence. A nilpotent orbit $\mathcal{O}$ in $\mathfrak{g}$ is special if and only if $Sp^{-1}[\mathfrak{g},\mathcal{O}]$ is a special representation of the Weyl group. Alternatively, a non-special orbit $\mathcal{O}$ is the one for which  $Sp^{-1}[\mathfrak{g},\mathcal{O}]$ yields a non-special irrep of $W$. Note that some irreps correspond under the Springer correspondence to non-trivial local systems on $\mathcal{O}$. So, not every non-special representation is associated to a non-special orbit. For example, in $D_4$,
\begin{eqnarray}
Sp[D_4,([2^2],-)] &=& ([3,2^2,1],1) \\
Sp[D_4,([2],[1^2])] &=& ([3^2,1^2],\psi_2),
\end{eqnarray}
where $\psi_2$ is the sign representation of $S_2$, the component group of $[3^2,1^2]$. In the first case above, the Springer correspondence assigns a non-special representation to a non-special orbit while in the second case, it assigns a non-special representation a non-trivial local system on a special orbit. The structure of the cell modules can now be seen as
\begin{eqnarray}
c_1 &=& \text{special orbit rep} \oplus \text{non-special orbit rep} \\ \nonumber
c_2 &=& \text{special orbit rep} \oplus \text{non-orbit rep}.
\label{threerepfamily}
\end{eqnarray}
For all families with three irreducible representations, the cell structure follows an identical pattern to the one just discussed. The special orbit together with all the non-special orbits to which the Springer correspondence assigns (when the orbits are taken with the trivial representation of the component groups) Weyl group irreps that are in the same family as that of the special representation (assigned to the special orbit by $Sp^{-1}$) form what is called a \textit{special piece} \cite{lusztig1997notes}. Geometrically, it is the set of all orbits which are contained in the closure of the special orbit $\mathcal{O}$ but are not contained in the closure of any other special orbit $\mathcal{O}'$ that obeys $\mathcal{O}' < \mathcal{O}$ in the closure ordering on special orbits. Note that in the example above, there is a cell module which contains all the Orbit representations corresponding to the special piece. The tables in the paper show, explicitly, that this pattern persists for every special piece in low rank classical cases and all the exceptional cases. That this pattern actually persists for every special piece can be shown using certain results in \cite{lusztig1984characters} (the summary of results at the end of pg. xiii and the beginning of pg. xv are most pertinent here)\footnote{ I thank G. Lusztig for correspondence on these matters.}. Further, the relevant results in \cite{lusztig1984characters} also imply that the number of orbits in the special piece is equal to the number of irreducible representations of the finite group $\bar{A}(\mathcal{O}^\vee)$ for some special orbit $\mathcal{O}^\vee$ in the dual lie algebra. A weaker statement that the Orbit representations of a special piece belong to the same family is available in \cite{lusztig1997notes}. 

For larger families, the overall structure of cell modules is a lot more complicated than \ref{threerepfamily}. For example, consider the family in $W(E_8)$ that contains the special representation $\phi_{4480,16}$ \cite{carter1985finite},

\begin{eqnarray*}
f_{\phi_{4480,16}} = \{ \phi_{4480,16},\phi_{7168,17},\phi_{3150,18},\phi_{4200,18},\phi_{4536,18},\phi_{5670,18}, \\ \nonumber \phi_{1344,19},\phi_{2016,19},\phi_{5600,19},\phi_{2688,20},\phi_{420,20},\phi_{1134,20},\\ \nonumber  \phi_{1400,20},\phi_{1680,22},\phi_{168,24},\phi_{448,25},\phi_{70,32}  \}.
\end{eqnarray*}
This family has $a=16$ and has a total of 17 irreps which organize themselves into the following seven cell modules,
\begin{eqnarray}
c_1 &=& \phi_{4480,16} \oplus \phi_{7168,17} \oplus \phi_{3150,18} \oplus \phi_{4200,18} \oplus \phi_{1344,19} \oplus \phi_{2016,19} \oplus \phi_{420,20} \\ \nonumber
c_2 &=& \phi_{4480,16} \oplus \phi_{7168,17} \oplus \phi_{3150,18} \oplus \phi_{4200,18} \oplus \phi_{5670,18} \oplus \phi_{1344,19} \oplus \phi_{5600,19} \oplus \phi_{1134,20}\\ \nonumber
c_3 &=& \phi_{4480,16} \oplus \phi_{7168,17} \oplus 2 \phi_{4200,18} \oplus \phi_{4536,18} \oplus \phi_{5670,18} \oplus \phi_{1344,19} \oplus \phi_{5600,19} \oplus \phi_{1400,20} \oplus \phi_{168,24} \\ \nonumber
c_4 &=& \phi_{4480,16} \oplus \phi_{7168,17} \oplus \phi_{3150,18} \oplus \phi_{4536,18} \oplus 2 \phi_{5670,18} \oplus 2 \phi_{5600,19} \oplus \phi_{1134,20} \oplus \phi_{1680,22} \oplus \phi_{448,25} \\ \nonumber
c_5 &=& \phi_{4480,16} \oplus \phi_{7168,17} \oplus 3 \phi_{4536,18} \oplus 3 \phi_{5670,18}  \oplus 2 \phi_{5600,19} \oplus 2 \phi_{1400,20} \oplus 3 \phi_{1680,22} \oplus \phi_{448,25} \oplus \phi_{70,32}\\ \nonumber
c_6 &=& \phi_{4480,16} \oplus 2 \phi_{7168,17} \oplus \phi_{3150,18} \oplus \phi_{4200,18} \oplus \phi_{4536,18} \oplus \phi_{5670,18} \oplus \phi_{2016,19} \oplus \phi_{5600,19} \oplus \phi_{2688,20}\\ \nonumber 
c_7 &=&\phi_{4480,16} \oplus 2 \phi_{7168,17} \oplus \phi_{4200,18} \oplus 2 \phi_{4536,18} \oplus 2 \phi_{5670,18} \oplus 2 \phi_{5600,19} \oplus \phi_{2688,20} \oplus \phi_{1400,20} \oplus \phi_{1680,22}.
\end{eqnarray}
Here again, $c_1$ is the collection of all Orbit representations in the family and the corresponding orbits form a special piece (see the table for $E_8$ in \ref{e8tablens} ). The patterns in the other cell modules for this family are not very obvious. 

In the following sections, the various notions introduced in this section will play an important role. For a more detailed exposition of the theory of Weyl group representations, see \cite{lusztig1984characters,carter1985finite}.

\section{Physical implications of duality maps}
\label{fourdimensional}

\subsection{CDT class of defects via matching of the Springer invariant}

Recall from the discussion of S-duality of 1/2 BPS boundary conditions in $\mathcal{N}=4$ SYM that the vacuum moduli space of the theory on a half space has two different realizations. One is its realization in the $G$ description and the other is its realization in the $G^\vee$ description. For the examples considered, the former was as a solution to Nahm equations with certain pole boundary conditions. The solution is in general of the form $\mathcal{S}^\rho\cap \mathcal{N}$, where $\rho$ is a nilpotent orbit in $\mathfrak{g}$. On the $G^\vee$ side, this space is realized as the Higgs branch of theory $T^\rho[G]$. Recall that the Higgs branch is a (singular) hyper-kahler space. So, the above statement in particular means that the metric on the moduli space is the same in both realizations. There is, at present, no known way to check this equality for arbitrary cases. However, there is strong evidence that the above identification holds for all $\mathcal{O}^\rho$ in any simple $\mathfrak{g}$. 

The S-duality map however would be incomplete if one could not say something about what the Coulomb branch of $T^\rho[G]$ should be. It is the Coulomb branch of $T^\rho[G]$ that is gauged and coupled to the boundary gauge fields on the $G^\vee$ side.  In \cite{Gaiotto:2008ak}, in the case of type $A_n$, it is shown that the Coulomb branch of $T^\rho[G]$ is given by a nilpotent orbit in $\mathfrak{g}^\vee=A_n$ whose partition type is $P^T$, the transpose of the partition type $P$ of the orbit $\rho$. Geometrically, transposition on the partition type acts as an order reversing duality on the set of nilpotent orbits taken with the partial order provided by their closure ordering\cite{collingwood1993nilpotent}. So, in the more general cases, one can guess that something similar to the case of $A_n$ prevails and description of the Coulomb branch of $T^\rho[G]$ will involve an order reversing duality between the data on the $\mathfrak{g}$ and the $\mathfrak{g}^\vee$ sides. Before the more general case is discussed, consider the case of $\mathfrak{g}=su(N)$ and a hypothetical scenario where one did not know that the right S-duality map between boundary conditions picks out the $T^\rho[SU(N)]$ that has a Coulomb branch given by a dual nilpotent orbit as the correct theory to couple at the boundary in the description of the S-dual of Nahm pole boundary condition of type partition type $P$. If, however, one is convinced that the boundary condition on the $G^\vee$ side should involve one of the $T^\rho[G]$ theories, then there is a unique theory whose Higgs branch matches the dimension of $\mathcal{S}^\rho \cap \mathcal{N}$. This theory would be the obvious candidate for the boundary theory on the $G^\vee$ side. And this theory has as its Coulomb branch the nilpotent orbit $P^T$.  One could call this argument \textit{dimension matching}, for merely requiring that the dimensions of the moduli space in its two realizations match turns out to completely specify the duality map. Outside of type $A$, the above argument can't be carried out directly for there are different $T^\rho[G]$ that have Higgs branches of the same dimension. 

Additionally, for certain $G$ in the classical types, the quivers that describe $T^\rho[G]$ turn out to be `bad' in the sense of \cite{Gaiotto:2008ak}. This complicates the description of the IR limit of the associated brane configurations.  Moreover, when $G$ is of exceptional type, a quiver description of the three dimensional theory is no longer available. In this context, it is convenient to use a more refined invariant which will be called the \textit{Higgs branch Springer invariant}. It has the advantage of being calculable for all $G$ and can distinguish $T^\rho[G]$ that have Higgs branches of the same dimension.   The point of view pursued here is that once the interaction between the representation theory and the vacuum moduli spaces of $T^\rho[G]$ is understood for $G$ classical (where brane constructions are available), then the available results from representation theory can be used to understand cases for which there is no brane construction available. Such a point of view is additionally supported by the fact that the corresponding representation theoretic results are highly constrained and enjoy a degree of uniqueness. This is also the point of view adopted in \cite{Chacaltana:2012zy} whose setup is what we are seeking to arrive at, albeit by a different route. 

Let us now proceed to associate a Higgs branch Springer invariant on both sides of the S-duality map and require that they match.  The irrep that occurs in this matching will be called $\bar{r}$. It seems suitable to call this check for the S-duality map as \textit{Higgs branch Springer invariant matching}, or $\bar{r}$-\textit{matching} for short. This invariant $\bar{r}$ is calculated on the $\mathfrak{g}$ in a straightforward manner,
\begin{equation}
\bar{r} = Sp^{-1}[\mathfrak{sl}_N,\mathcal{O}_N].
\label{Sp1}
\end{equation}
 From the brane constructions, we know that nilpotent orbits that enter the description of the Higgs and Coulomb branches of $T^\rho[SU(N)]$ are related by an order reversing duality between the nilpotent orbits. The analogue of an order reversing duality at the level of Weyl group representations is tensoring by the sign representation $\epsilon$. And, indeed, one sees that the $\bar{r}$ obtained as in \ref{Sp1} above obeys
\begin{equation}
\bar{r} = \epsilon \otimes Sp^{-1}[\mathfrak{sl}_N,\mathcal{O}_H].
\end{equation}
Alternatively, one can \textit{require} that
\begin{equation}
Sp^{-1}[\mathfrak{sl}_N,\mathcal{O}_N] = \epsilon \otimes  Sp^{-1}[\mathfrak{sl}_N,\mathcal{O}_H]
\label{typeAmatching}
\end{equation}
and this, in turn, determines $\mathcal{O}_N$ for a given $\mathcal{O}_H$.

Now, it is natural to try and generalize this for other $\mathfrak{g}$. For arbitrary $\mathfrak{g}$, the Springer correspondences in $\mathfrak{g}^\vee$ and $\mathfrak{g}$ would give irreps of $W[\mathfrak{g}^\vee]$ and $W[\mathfrak{g}]$. Since there is a canonical isomorphism between the two, it is natural to parameterize the irreps of the two Weyl groups in a common fashion (see Appendix \ref{repsappendix} and \cite{carter1985finite}). This would also allow one to formulate a `matching' argument along the lines of \ref{typeAmatching}. This does turn out to be hugely helpful as this simple-minded generalization specifies the duality map in numerous cases. Let us for a moment consider case where Hitchin data is $(\mathcal{O}_H,1)$.  Merely requiring that
\begin{equation}
Sp^{-1}[\mathfrak{g},\mathcal{O}_N] = \epsilon \otimes Sp^{-1}[\mathfrak{g}^\vee,\mathcal{O}_H] ,
\label{GmatchingSpecial}
\end{equation}
one can obtain the order reversing duality map for all $\mathcal{O}_N$ special except for the cases discussed in Section \ref{exceptionalorbits}. One can handle all the cases uniformly by replacing the RHS in \ref{GmatchingSpecial} with the unique  special representation in the family of $\epsilon \otimes Sp^{-1}[\mathfrak{g}^\vee,\mathcal{O}_H]$. This version of the duality operation that implements a fix for the `exceptional' (in the sense of Section \ref{exceptionalorbits} ) cases is due to Lusztig. In the discussion below, the duality operation will continue to the represented as tensoring by sign with the understanding that, if needed, the above fix can always be applied to the definition.

Now, consider the following equivalent formulation of Eq \ref{GmatchingSpecial},

\begin{equation}
\boxed{Sp^{-1}[\mathfrak{g},\mathcal{O}_N] =  Sp^{-1}[\mathfrak{g},d_{LS}(\mathcal{O}_H)]},
\label{LSSpringerMatching}
\end{equation}

where $d_{LS}$ is the Lusztig-Spaltenstein order reversing duality map that stays within the lie algebra $\mathfrak{g}$. The equivalence of the above formulation to Eq \ref{GmatchingSpecial} follows from a property of the map $d_{LS}$ when acting on special orbits,
\begin{equation}
Sp^{-1}[\mathfrak{g},d_{LS}(\mathcal{O})]  = \epsilon \otimes  Sp^{-1}[\mathfrak{g},\mathcal{O}].
\end{equation}

From \ref{LSSpringerMatching}, we get the order reversing duality for the cases where $\mathcal{O}_N$ is \textit{special}. For the other cases, one has to formulate a more sophisticated argument. Before we get to that, let us try to understand how the Springer invariant can be calculated when we allow for a particular symmetry breaking deformation in the bulk on the $\mathfrak{g}^\vee$ side.

The boundary condition on the $\mathfrak{g}^\vee$ side involves $\mathcal{N}=4$ SYM on a half space with a coupling to a three dimensional theory $T^\rho[G]$ that lives on the boundary. Now, deform this boundary condition by giving a vev to the adjoint scalars of the bulk theory. Let this vev be some semi-simple element $m \in T^\vee$. Now, in the $m \rightarrow \infty $ limit, the bulk symmetry is broken from $G^\vee$ to $L^\vee$, where $l^\vee$ is a subalgebra that arises as the centralizer $Z_{\mathfrak{g}^\vee}(m)$. Pick $m$ such that a representative $e^\vee$ of the Coulomb branch orbit $\mathcal{O}_H$ is a distinguished nilpotent element in $\mathfrak{l}^\vee$.  Taking the $m \rightarrow \infty $ limit gives a boundary condition in $\mathcal{N}=4$ SYM with gauge group $L^\vee$ with the theory at the boundary being $T^{\tilde{\rho}}[L]$. Let us call such a deformation of the boundary condition on the $G^\vee$ side a \textit{distinguished symmetry breaking},
\begin{equation}
(\mathcal{O}^0,G^\vee,T^\rho[G]) \longrightarrow_{d.s.b} (\mathcal{O}^0,L^\vee,T^{\tilde{\rho}}[L]).
\end{equation} 
The above deformation can be done for any boundary condition of the form $(\mathcal{O}^0,G^\vee,T^\rho[G])$ in $\mathcal{N}=4$ SYM. When $\mathfrak{l}^\vee$ is a Levi subalgebra, this procedure, in a sense, reproduces the Bala-Carter classification of nilpotent orbits in $\mathfrak{g}^\vee$ (see Appendix \ref{orbitsappendix} and \cite{carter1985finite}). Let us briefly restrict to the case where $\mathfrak{l}^\vee$ is indeed a Levi subalgebra. In what follow, it is helpful to note that every distinguished orbit is special and $d_{LS}$ always acts as an involution on special orbits. Now, associate an irrep of $W[\mathfrak{l}^\vee]$ to the Coulomb branch of $T^{\tilde{\rho}}[L]$ in the following way,
\begin{equation}
s = Sp^{-1}[\mathfrak{l^\vee},d_{LS}(\mathcal{O}_H^{\mathfrak{l}^\vee})],
\end{equation}
where $d_{LS}$ is the duality map that stays within $\mathfrak{l}^\vee$. Now, it turns out that the following is always true,
\begin{equation}
\bar{r} = j_{W[\mathfrak{l}^\vee]} ^{W[\mathfrak{g}^\vee]} (s),
\label{Jinduced}
\end{equation}
where $\bar{r}$ is \textit{Higgs branch Springer invariant} defined earlier and the operation $j_{W[\mathfrak{l}^\vee]} ^{W[\mathfrak{g}^\vee]}$ refers to Macdonald-Lusztig-Spaltenstein induction from irreps of the Weyl subgroup 
$W[\mathfrak{l}^\vee]$ to the parent Weyl group $W[\mathfrak{g}^\vee]$ (See Appendix \ref{jinduction}). The $j$ induction procedure is sometimes also called truncated induction. It plays a critical role in the interaction of Springer theory with induction within the Weyl group and especially in isolating how the $W[\mathfrak{g}^\vee]$ module structure of $H^\text{top}(\mathcal{B})$ can be induced from a $W[\mathfrak{l}^\vee]$ module structure. More generally, the cohomology in lower degrees also obey certain induction theorems (see, for example \cite{lusztig2004induction,treumann2009topological}). For the current purposes (associating a Springer invariant to the defect), only the structure of  $H^\text{top}(\mathcal{B})$ is relevant and hence \ref{Jinduced} is sufficient.

Now, \ref{Jinduced} allows us to rewrite the matching condition \ref{LSSpringerMatching} as
\begin{subequations}
\begin{empheq}[box=\widefbox]{align}
 s = Sp^{-1}[\mathfrak{l^\vee},d_{LS}(\mathcal{O}_H^{\mathfrak{l}^\vee})] \\
Sp^{-1}[\mathfrak{g},\mathcal{O}_N] =  j_{W[\mathfrak{l}^\vee]} ^{W[\mathfrak{g}^\vee]} (s)
\end{empheq}
\label{Specialmatching}
\end{subequations}

The above matching condition \textit{determines} the pairs $\mathcal{O}_N,\mathcal{O}_H$ for $\mathcal{O}_N$ being a special orbit. Different $\mathcal{O}_N$ arise on the $\mathfrak{g}$ side when the various non-conjugate Levi subalgebras $\mathfrak{l}^\vee$ are considered on the $\mathfrak{g}^\vee$ side. 

 Apart from this highly constraining structure, the matching condition \ref{Specialmatching} additionally enjoys the following beautiful feature. In order to extend the domain of the duality map to include cases where $\mathcal{O}_N$ is non-special, all that one has to do is to allow for $\mathfrak{l}^\vee$ to be an arbitrary centralizer and not just a Levi subalgebra. These more general centralizers are what are called pseudo-Levi subalgebras in \cite{sommers2001lusztig}. So, by allowing $\mathfrak{l}^\vee$ to a pseudo-Levi subalgebra in which a representative $e^\vee$ of the Hitchin orbit $O_H$ is distinguished, one obtains an order reversing duality map that recovers the entire CDT class of defects. By Sommers' extension of the Bala-Carter theorem \cite{sommers1998generalization}, this more refined data on the Hitchin side is actually equivalent to specifying $(\mathcal{O}_H,C)$ where $C$ is a conjugacy class in $\bar{A}(\mathcal{O}_H)$. $\bar{A}(\mathcal{O}_H)$ is always a Coxeter group. Within this Coxeter group, there is a well defined way to translate data of the form $(\mathcal{O}_H,C)$ to something of the form $(\mathcal{O}_H,\mathcal{C})$ \cite{achar2002local}, where $\mathcal{C}$ is the Sommers-Achar subgroup of $\bar{A}(\mathcal{O}_H)$ (in the notation and terminology of \cite{Chacaltana:2012zy}). For non-special Nahm orbits, this subgroup $\mathcal{C}$ enters the description of the Coulomb branch data in a crucial way as explained in \cite{Chacaltana:2012zy}. One also observes that the map between Hitchin and Nahm data offers the following distinction between special and non-special Nahm orbits in the language of boundary conditions for $\mathcal{N}=4$ SYM. When $\mathcal{O}_N$ is special, the distinguished symmetry breaking deformation on the $G^\vee$ side produces a theory on the boundary whose Coulomb branch is a distinguished orbit in a Levi subalgebra $\mathfrak{l}^\vee$. On the other hand, when $\mathcal{O}_N$ is non-special, the distinguished symmetry breaking deformation on the $G^\vee$ side produces a theory on the boundary whose Coulomb branch is a distinguished orbit in a pseudo-Levi subalgebra $\mathfrak{l}^\vee$ that is not a Levi subalgebra. The description given here is the exact definition of the map in \cite{sommers2001lusztig} \footnote{To avoid confusion, it is useful to note that in the notation adopted here, nontrivial local systems appear on the $\mathfrak{g}^\vee$ side, while they appear on the $\mathfrak{g}$ side in Sommers' notation.}. Here, the definition is placed in a physical context.

\subsection{Implications for four dimensional constructions}

Once the dictionary between the Nahm/Hitchin data is established, one has the following immediate consequences for some of the local properties of the codimension two defects \cite{Chacaltana:2012zy}, 
\begin{eqnarray}
\text{dim}_{\mathbb{H}} (\text{Higgs branch }) &=& \frac{1}{2} \bigg(\text{dim} (\mathcal{N}) - \text{dim} (\mathcal{O}_N) \bigg), \\
\text{dim}_{\mathbb{C}} (\text{Coulomb branch}) &=& \frac{1}{2} \text{dim}(\mathcal{O}_H).
\end{eqnarray}
Further, the contributions to the trace anomalies $a,c$ and the flavor central charge $k$ can also be determined as outlined nicely in \cite{Chacaltana:2012zy}. Before turning to the Toda description, here are some further comments which future work can presumably clarify.

In the discussion in the early part of this Section, a particular symmetry breaking deformation is applied to the four dimensional theory that was called distinguished symmetry breaking. One is able to retrieve the Springer invariant for the undeformed theory by an induction procedure from the Springer invariant for the deformed theory. In fact, outside of type $A$, this was a crucial part of the matching constraint on the duality map that enabled one to completely specify it.  But, it would be useful understand the physical underpinnings of the induction procedure and its potential applicability outside of the setup considered here. 

In particular, it would be interesting to explore the relationship between other calculable observables of these theories. In this direction, it is notable that there have been recent advances in the understanding of the Hilbert Series and $\mathbb{S}^3$ partition functions of 3d $\mathcal{N}=4$ theories  (see, for example \cite{Kapustin:2010xq,Hanany:2011db,Cremonesi:2013lqa,Dey:2013fea,Cremonesi:2014kwa} ).

\section{The part about Toda}
\label{Todapart}
In light of the observations of AGT-W \cite{Alday:2009aq,Wyllard:2009hg}, it is expected that the sphere partition function of a theory of class $\mathcal{S}$ (built using codimension two defects of $\mathscr{X}[\mathfrak{j}]$  as in \ref{WittenGaiotto}) can be expressed as a correlator in a two dimensional Toda CFT of type $\mathfrak{g}$. Let us briefly recall some facts about Toda CFTs. They are described by the following Lagrangian on a disc with a curvature insertion at infinity,
\begin{equation}
S_T = \frac{1}{2 \pi} \int \sqrt{\hat{g}} d^2 z \bigg ( \frac{1}{2} \hat{g}^{ab}\partial_a \phi \partial_b \phi + \sum_{i=1}^{\text{rank}(\mathfrak{g})} 2 \pi \Lambda e^{2b (e_i,\phi)} \bigg) + \frac{1}{\pi} \int (Q,\phi) d \theta + (\ldots),
\end{equation}
where $e_i \in \mathfrak{h}^*$ are the simple roots of the root system associated to $\mathfrak{g}$, $\phi \in \mathfrak{h}$ is the Toda field and $Q=b+b^{-1}$. A special case of Toda[$\mathfrak{g}$] is Liouville CFT. It corresponds to the case $\mathfrak{g}=A_1$. Recall that the chiral algebra of Liouville CFT is the Virasoro algebra. The chiral algebra of the more general Toda[$\mathfrak{g}$] theories are certain affine $\mathcal{W}$ algebras. These theories have conserved currents $\mathcal{W}^k(z)$ of integer spins $k$. The spectrum of values $\{k-1\}$ in a particular Toda[$\mathfrak{g}$] theory is equal to the set of exponents of the lie algebra $\mathfrak{g}$. The unique spin 2 conserved current in this set is the stress tensor $\mathcal{W}^2(z)=T(z)$. 

The $\mathcal{W}$-algebras that arise in such theories have the additional property that they can be obtained by a Hamiltonian reduction procedure from affine Lie algebras which arise as the chiral algebras of non-compact WZW models. This procedure admits a generalization for every $\sigma : \mathfrak{sl}_2 \rightarrow\mathfrak{g}$ and this allows one construct other W algebras. When $\sigma$ is taken to be principal, then one obtains the usual Toda[$\mathfrak{g}$] theories. It is only the Toda[$\mathfrak{g}$] theories that will concern us in what follows since this is the setting for the direct generalizations of \cite{Alday:2009aq,Wyllard:2009hg} to arbitrary theories of class $\mathcal{S}$.  While Toda theories exist for both simply laced and non-simply laced $\mathfrak{g}$, the discussion that follows will be confined to the case $\mathfrak{g} (\cong \mathfrak{j}) \in A,D,E$. If one were to consider the twisted defects and seek a Toda interpretation for them, an adaptation of much of the arguments below for $\mathfrak{g} \in B,C,F_4,G_2$ would likely be relevant. 
 
When trying to build an understanding of the AGT conjecture for an arbitrary theory of class $\mathcal{S}$, a good starting point is to have the following local-global setup in mind,
\begin{itemize}
\item \textit{Local aspects of the AGT conjecture} : This is the claim that the regular codimension two defects of the $\mathscr{X}[\mathfrak{g}]$ admit a description in terms of certain primary operators of the principal Toda theory of type $\mathfrak{g}$. Let us call this part of the AGT dictionary the \textit{primary map} $\wp$. This map is a bijection from the set of defects to the set of semi-degenerate states (borrowing terminology from \cite{Kanno:2009ga}) in the Toda theory and concerns data that is local to the codimension two defect insertion on the Riemann surface $C_{g,n}$ and does not involve the Riemann surface in any way. 
\item \textit{Global aspects of the AGT conjecture} : If the description of the four dimensional theory involves compactification of $\mathscr{X}[\mathfrak{g}]$ on $C_{g,n}$, then the sphere partition function (including non-perturbative contributions) of this theory is obtained by a Toda correlator on $C_{g,n}$ with insertions of the corresponding primary operators of Toda theory at the $n$ punctures. The identification of the corresponding Toda primary is done according to the map $\wp$. The identification of the conformal block with the instanton partition function is a crucial ingredient in the global AGT conjecture. Checks of the conjecture for the sphere partition function in cases of arbitrary $\mathfrak{g}$ are available in specific corners of the coupling constant moduli space where Lagrangian descriptions become available for the four dimensional theories\cite{Alday:2009aq,Wyllard:2009hg}.
\end{itemize}

In the discussion above, a choice was made to restrict to four dimensional SCFTs obtained by the compactification from six dimensions involving just the regular defects. But, it is interesting to note that the formalism associated to the AGT conjecture can also be extended to the cases where SCFTs are built out of irregular defects\footnote{The terminology of regular and irregular defects is from \cite{Witten:2007td,Gaiotto:2009hg}.} as in \cite{Bonelli:2011aa,Gaiotto:2012sf,Kanno:2013vi} and certain aspects extend to the case of asymptotically free theories (See, for example \cite{Gaiotto:2009ma,Keller:2011ek}). There exist generalizations which involve partition functions in the presence of supersymmetric loop and surface operators of the 4d theory (See, for example \cite{Alday:2009fs,Drukker:2009id,Drukker:2010jp} and \cite{Alday:2010vg}). Some of the mathematical implications that follow from the observations of AGT have been explored in \cite{2011CMaPh.308..457B,2011arXiv1107.5073N,2012arXiv1202.2756S,2012arXiv1211.1287M}. For a more complete review of the literature, consult \cite{TachikawaLecs}.

 The global AGT conjecture suggests that the OPE of codimension two defects of the six dimensional theory is controlled by the $\mathcal{W}$-algebra symmetry of the Toda theory. While this is powerful as an organizing idea, it is particularly hard to proceed in practice as the non-linear nature of $\mathcal{W}$ algebras complicates their representation theory. In the discussion that follows, the goal is only to establish the \textit{primary map} for as many defects as possible in arbitrary $\mathfrak{g}$. In particular, global aspects of the AGT conjecture or any of its generalizations are not analyzed (except for a discussion about scale factors).

\subsection{The primary map $\wp$}
In the original work of AGT, this map was obtained for the case of $A_1$. There is just a single nontrivial codimension two defect \footnote{The trivial defect (the defect corresponding to the principal Nahm pole) is always mapped to the identity operator on the 2d CFT side.} in this case. So, the map is particularly straightforward to describe. After setting the radius of the four sphere to be unity (see \cite{Balasubramanian:2013kva} for how the radius dependence on the overall partition function can be analyzed), this map can be described as
\begin{equation}
\wp : [1^2]_N \rightarrow e^{2 \alpha \phi} \mid \alpha = Q/2 + im,
\end{equation}
where $\phi$ is the Liouville field. In the map above, the Nahm orbit is used to identify the defect operator. The defect could have alternatively been identified by the Hitchin orbit associated to it, namely the orbit $[2]_H$. But, it will turn out that the Nahm orbit is the one that is convenient for obtaining the generalization of this for arbitrary $\mathfrak{g}$. So, it is convenient to use it to tag a particular codimension two defect. Two important aspects of the above map are
\begin{itemize}
\item A precise identification of $\Re{(\alpha)}$
\item An identification of $\Im{(\alpha)}$ with $im$ where $m$ is a mass deformation parameter.
\end{itemize}

An identification similar to the one above for the mass parameter $m$ exists for the Coulomb branch modulus $a$. In both of these cases, a distinguished real subspace of the $\mathcal{N}=2$ theory's parameters is picked out in writing the map to the corresponding Liouville primary. 

To extend these argument to higher rank cases, a natural thing to try and obtain is a generalization of the primary map $\wp$ that is in the same form. Say,
\begin{equation}
\wp : \mathcal{O}_N \rightarrow e^{(\alpha,\phi)} \mid \alpha = \Re(\alpha) + \Im(\alpha),
\end{equation}
with some prescribed conditions on $\Re(\alpha) $ and $\Im(\alpha)$ that depend on $\mathcal{O}_N$. Here, $\phi \in \mathfrak{h}$ is the Toda field and it is a $r$-dimensional vector of scalar fields where $r$ is the rank of $\mathfrak{g}$ and $\alpha \in \mathfrak{h}^*$ is the Toda momentum.  The relevant primaries for the case of $A_n$ were identified in \cite{Kanno:2009ga} (a precise formulation in terms of the Nahm orbit data can be found in \cite{Balasubramanian:2013kva} and is explained in greater detail below). The general picture is that $\wp$ maps the zero Nahm orbit to the maximal puncture while the other Nahm orbits are mapped to certain semi-degenerate primary operators in the corresponding Toda theory. The principal Nahm orbit is mapped to the identity operator. The semi-degenerate primaries of \cite{Kanno:2009ga} contain null vectors at level-1 with the exact number and nature of these null vectors depending on the associated Nahm orbit. Combinatorially, specifying the level-1 null vectors amounts to specifying a certain subset of the simple roots in the root system associated to $A_n$. One gets the relationship to the Nahm orbit by noticing a very natural connection between subsets of simple roots and nilpotent orbits in $A_n$. This connection is offered by the Bala-Carter classification of nilpotent orbits in $\mathfrak{g}$. For a quick summary of the work of Bala-Carter, see Appendix \ref{orbitsappendix} and for a more detailed account, see \cite{collingwood1993nilpotent,carter1985finite,binegar}. For the current purposes, the important fact will be that the Bala-Carter classification amounts to specifying a pair $(\mathfrak{a},e)$ where $\mathfrak{a}$ is a Levi subalgebra of $\mathfrak{g}$ and $e$ is a distinguished nilpotent element in that Levi subalgebra.\footnote{The Levi subalgebra $\mathfrak{a}$ should not be confused with the Levi subalgebra $\mathfrak{l}^\vee$. The former is a subalgebra of $\mathfrak{g}$ and arises as part of the Nahm data while the latter is a subalgebra of $\mathfrak{g}^\vee$ and is part of the Hitchin data.} 

Levi subalgebra are naturally classified by non-conjugate subsets of the set of simple roots. When $e$ is principal nilpotent in a Levi subalgebra, the corresponding orbit is called principal Levi type \footnote{Interestingly, certain finite $\mathcal{W}$ algebras associated to nilpotent orbits of principal Levi type also play an important role in the mathematical approach to a variant of the original setup of AGT \cite{2011CMaPh.308..457B}, extended to arbitrary $\mathfrak{g}$. }. It turns out that all the non-zero orbits in type $A$ are principal Levi type. In particular, the combinatorial data associated to a Nahm orbit by the Bala-Carter theory is precisely the subset of simple roots corresponding to the Levi subalgebra $\mathfrak{a}$. Once the combinatorial data is placed in the setting of nilpotent orbits, a reasonable generalization would be to consider all principal Levi type orbits in arbitrary $\mathfrak{g}$. The combinatorial data assigned to such orbits is always a subset of the simple roots of the root system associated to $\mathfrak{g}$. Additionally, let $F$ denote the reductive part of the connected component of the centralizer of a nilpotent representative $e$ of the Nahm orbit. This is the \textit{global symmetry} associated to the Higgs branch of the codimension two defect, or equivalently of $T^\rho[G]$ \cite{Chacaltana:2012zy}. Now, the mass deformation parameters of $T^\rho[G]$ (and hence of the defect) are valued in a Cartan subalgebra of $\mathfrak{f}$. In particular, the number of such linearly independent parameters is equal to $\text{rank}(\mathfrak{f})$. For any non-zero orbit of principal Levi type, this quantity is necessarily non-zero. It is a general property that
\begin{equation}
\text{rank} (\mathfrak{f}) = \text{rank}(\mathfrak{g}) - \text{rank}(\mathfrak{a}).
\end{equation}
Now, consider a Toda primary with momentum $\alpha \in \Lambda^+$ that obeys
\begin{eqnarray}
\label{Todaprimary}
(\Re({\alpha}),e_i)&=& 0,\\ \nonumber
 0 \leq \Re({\alpha}) &\leq& Q \rho ,\\ \nonumber
\Im(\alpha) &=& 0 ,
\end{eqnarray}
where $e_i$ is any simple root of the Levi subalgebra $\mathfrak{a}$ and $\rho$ is the Weyl vector of $\mathfrak{g}$ and the relation $\leq$ is in the partial order on the set of dominant weights $\Lambda^+$.  Imposing the above conditions would also mean, in particular, that $(\alpha,\rho_\mathfrak{a})=0$, where $\rho_{\mathfrak{a}}$ is the Weyl vector of the subalgebra $\mathfrak{a}$. When the Nahm orbit associated to codimension two defect is principal Levi type, I argue that (\ref{Todaprimary}) provides the right Toda primary in the massless limit. A piece of evidence that supports such a statement is the following. Let us write $\Re({\alpha})$ as a combination of the fundamental weights of $\mathfrak{g}$
\begin{equation}
\Re{(\alpha)} = a_i \omega_i,
\label{realpart}
\end{equation}
where $a_i \neq 0$ and $\{ \omega_i \}$ is some subset of the fundamental weights. Now, deform the Toda momentum such that it acquires an imaginary part given by
\begin{equation}
\Im{(\alpha)} = m_i \omega_i,
\end{equation}
so that $(\alpha,e_i)=0$ holds for all $e_i$ being simple roots of $\mathfrak{a}$. The $m_i$ introduced above are the mass parameters that one would associate with the codimension two defect. And the total number of such linearly independent parameters will equal the number of fundamental weights occurring in \ref{realpart} and this is equal to precisely $\text{rank}(\mathfrak{f})$, as expected.  For type A, the above procedure reproduces the semi-degenerate primaries considered in \cite{Kanno:2009ga} \footnote{This point was also made in \cite{Balasubramanian:2013kva} using the Dynkin weight $h$ of the Nahm orbit.}. For non-zero orbits that are not principal Levi type, one natural guess is that the level-1 null vectors that are imposed are still given by the set of simple roots that one associates to the Bala-Carter Levi. In these cases, a nilpotent representative will correspond to a non-principal distinguished nilpotent orbit in $\mathfrak{a}$. This corresponds to picking a further subset of the simple roots of $\mathfrak{a}$. This additional combinatorial data may presumably be translated to null vector conditions at higher level, but this needs to be made precise. The case of non-principal Levi type orbits for which $\text{rank}(\mathfrak{f})$ is zero would be particularly interesting since the mere existence of such cases challenges the wisdom that $\Im({\alpha})$ should give rise to an associated mass deformation. In $\mathfrak{g}=E_8$, for example, all orbits that are distinguished in $\mathfrak{a}=E_8$ have $\text{rank}(\mathfrak{f})=0$. To give some idea about how many of the nilpotent orbits in $\mathfrak{g}$ tend to be of principal Levi type, the data for certain low rank $\mathfrak{g}$ is displayed in Table \ref{principalLevitype}.

It should be mentioned here that one can device some local checks of the map $\wp$ that are sensitive to the Coulomb branch data. In \cite{Kanno:2009ga}, it was checked that the behaviour of the Seiberg-Witten curve near the punctures is reproduced in a `semi-classical' limit of the Toda correlators together with insertions of the currents $\mathcal{W}^k(z)$. This is really a direct check on the local contribution to the Coulomb branch from a Toda perspective. Here, the map between the Nahm and Hitchin data obtained in the previous section already provides a candidate for the local contribution to the Coulomb branch from a Toda primary whose Nahm orbit is principal Levi type. But, a direct check of this assertion would be more pleasing.

\begin{table} [!h]
\centering 
\caption{Nilpotent orbits of principal Levi type in certain Lie algebras} 
\begin{center}
\begin{tabular}{c|c|c}
\toprule
$\mathfrak{g}$ & $\#$ of Nilpotent orbits & $\#$ of principal Levi orbits  \\ 
 \midrule 
 $A_4$ & 7 & 7   \\ 
 $B_4$ & 13 &  10 \\ 
 $C_4$ & 14 &  10  \\
 $D_4$ & 12 &  9  \\ 
 $E_6$ & 21 & 17  \\ 
 $E_7$ & 45 & 32  \\  
 $E_8$ & 70 & 41  \\ 
 $F_4$ & 16 & 12   \\ 
 $G_2$ & 5 &  4 \\ 
 \bottomrule
\end{tabular}
\end{center}
\label{principalLevitype}
\end{table}

\subsection{Local contributions to Higgs and Coulomb branch dimensions}
As just discussed, once the relation between the Nahm data and the Toda primary is known, one can use the dictionary between the Nahm/Hitchin data to associate a Hitchin orbit to a Toda primary. With this, the effective contribution to the local Higgs branch and the local Coulomb branch from a particular Toda primary can be inferred. From the tinkertoy constructions \cite{Chacaltana:2012zy}, the following expressions are known for $n_h -n_v$ (the total quaternionic Higgs branch dimension) and $d$ (the total Coulomb branch dimension) in terms of the Nahm and Hitchin orbit data for each defect $(\mathcal{O}_H^i,\mathcal{O}_N^i)$,
\begin{eqnarray}
(n_h - n_v) &=& \sum (n_h - n_v)^i + (n_h - n_v)^\text{global} \\
d &=& \sum_i d^i + d^\text{global}
\end{eqnarray}
with
\begin{eqnarray}
(n_h - n_v)^i &=& \frac{1}{2}\bigg(\text{dim}(\mathcal{N})-\text{dim}(\mathcal{O}_N^i)\bigg) = \text{dim}(\mathcal{B}_N^i) \\
d^i &=& \frac{1}{2} \text{dim} (\mathcal{O}_H^i)
\end{eqnarray}
and 
\begin{eqnarray}
(n_h  - n_v)^{\text{global}} &=& (1-g) \text{rank}(\mathfrak{g})  \\
d^{\text{global}} &=&  (g-1) \text{dim} (\mathfrak{g})
\end{eqnarray}
 
\subsection{Scale factors in Toda theories}
As a simple illustration of the local-global interplay, one can consider how the scale factor in the sphere partition function that captures the Euler anomaly of the four dimensional theory is calculated. From a purely four dimensional perspective, the Euler anomaly is very well understood in the tinkertoy constructions. In \cite{Balasubramanian:2013kva}, the radius dependent factor in the sphere partition function that encodes the Euler anomaly was made explicit and the relation to a corresponding scale factor in the two dimensional CFT was pointed out. The scale factor in question should be calculated for a (canonically defined) stripped version of the Toda correlator. In certain simple cases like correlators corresponding to free theories, this scale factor directly encodes the number of polar divisors. In the more complicated cases, it provides an interesting constraint on the analytical structure of the correlator and its factorizing limits. For Toda correlators corresponding to a subset of the class $\mathcal{S}$ theories, this scale factor can be directly calculated starting from a purely 2d perspective. For other cases, one still expects the scale factor for the stripped correlators to be such that it reproduces the Euler anomaly accurately. A conjecture to this effect was formulated in \cite{Balasubramanian:2013kva}. The work in this paper provides an extension of the framework for the conjecture outside of type $A$ for cases where the Nahm orbit is principal Levi type.
\newpage

\section{The setup}
\label{setup}

\subsection*{Notation}
All the relevant notation for the subsequent sections of the paper are collected here for convenience. 

\begin{table}[!h]
\begin{tabular}{ll}
$\{\mathcal{O}_N \}$ & Set of nilpotent orbits in $\mathfrak{g}$. \\
$\{\mathcal{O}_H \}$ & Set of special nilpotent orbits in $\mathfrak{g}^{\vee}$. \\
$\mathfrak{l}^\vee$ & A pseudo-Levi subalgebra of $\mathfrak{g}^\vee$ \\
$\mathfrak{l}$&  Langlands dual of $\mathfrak{l}^\vee$. May not be a subalgebra of $\mathfrak{g}$.\\
$\mathfrak{a}$ &  Levi subalgebra of $\mathfrak{g}$ that arises from Bala-Carter label for $\mathcal{O}_N$.\\
 $A(\mathcal{O}_H)$ & Component group of the Hitchin nilpotent orbit. \\ 
 $\bar{A}(\mathcal{O}_H)$ & Lusztig's quotient of the component group. \\
 $\psi_H$ & Irrep of  $\bar{A}(\mathcal{O}_H)$. \\
 $\mathcal{C}_H$ & Sommers-Achar subgroup of $\bar{A}(\mathcal{O}_H)$. It is such that $j_{\mathcal{C}_H}^{\bar{A}(\mathcal{O}_H)}(\text{sign})=\psi_H$.\\
$Irr(W)$ & Set of irreducible representations of the Weyl group $W$ of $\mathfrak{g}$. \\
$Irr(W^{\vee})$ & Set of irreducible representations of the Weyl group $W^{\vee}$ of $\mathfrak{g}^{\vee}$. \\
 $\bar{r}$ & An irreducible representation of the Weyl group ${W[\mathfrak{g}]}$. \\ 
 $r$ & The irrep $\bar{r}$ tensored with the sign representation.  \\ 
   $f_r$ & The family to which the irrep $r$ belongs. \\
$Sp[\mathfrak{g}]$ & Springer's injection from $Irr(W)$ to pairs $(\mathcal{O},\psi)$, \\ & where $\mathcal{O}$ is a nilpotent orbit in $\mathfrak{g}$ and $\psi$ is a representation of its component group $A(\mathcal{O})$. \\
$Sp^{-1}[\mathfrak{g}]$ & Inverse of Springer's injection. Acts only on the subset of $(\mathcal{O},\psi)$ \\ & which occurs in the image of $Sp[\mathfrak{g}]$. \\
$j_{W'}^{W}(r_{W'})$ &  The truncated induction procedure of Macdonald-Lusztig-Spaltenstein. \\
$n_h$ & Contribution to effective number of hypermultiplets. \\
  $n_v$ & Contribution to effective number of vector multiplets. \\
  $d$ & Contribution to the total Coulomb branch dimension. \\
  $\mathcal{B}_N$ & Springer fiber associated to the Nahm orbit. \\
  $\mathcal{B}_H$ & Springer fiber associated to the Hitchin orbit. \\
  $a(f_r)$ & Lusztig's invariant. Its value is the same for any irrep in a given family.  \\ & This equals $\text{dim}_{\mathbb{C}} \mathcal(\mathcal{B}_H)$ for the special orbit $\mathcal{O}_H$. \\ 
  $\tilde{b}(\bar{r})$ & Sommers' invariant. This equals $\text{dim}_{\mathbb{C}} \mathcal(\mathcal{B}_N)$. \\ 

\end{tabular}
\end{table}
\newpage

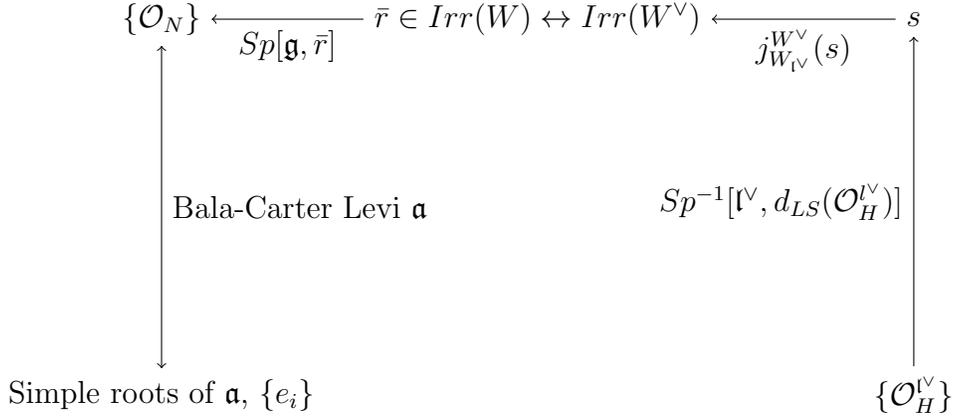
\begin{figure}[!h]
\centering
\caption{The setup}
\begin{center}
\begin{tikzpicture}[node distance=5cm, auto]
  \node (P) {$\bar{r} \in Irr(W) \leftrightarrow Irr(W^{\vee})$};
    \node (BB) [left of=P] {$\{\mathcal{O}_N \}$};
  	\node (A) [right of=P] {$s$ };
  	\node (T) [below of=BB] {Simple roots of $\mathfrak{a}$, $\{e_i\}$};
  	\node (H) [below of=A] {$\{\mathcal{O}_H^{\mathfrak{l}^\vee}\}$};
  \draw[->] (A) to node {$j_{W_{\mathfrak{l}^\vee}}^{W^{\vee}}(s)$} (P);
  \draw[->] (P) to node {$Sp[\mathfrak{g},\bar{r}]$} (BB);
  \draw[<->] (BB) to node {Bala-Carter Levi $\mathfrak{a}$} (T) ;
  \draw[->] (H) to node {$Sp^{-1}[\mathfrak{l^\vee},d_{LS}(\mathcal{O}_H^{l^\vee})]$}(A);
 \label{THEsetup}
\end{tikzpicture}
\end{center}
\end{figure}

%

As a useful summary, the constructions of Sections 5 and 6 have been summarized in the Fig \ref{THEsetup}. Some of the interesting physical quantities can be obtained from the above figure in the following way,
\begin{eqnarray}
 \text{simple roots for  } \mathfrak{a}, \{ e_i \}  &\implies& \{\text{level 1 null vectors for a Toda primary} \} , \\ 
 (n_h - n_v) &=& \frac{1}{2} \bigg( \text{dim} (\mathcal{N}) - \text{dim} (\mathcal{O}_N) \bigg) = \tilde{b}(\bar{r}), \\
 d &=& \frac{1}{2} \text{dim} (\mathcal{O}_H) = \mid \Lambda^+ \mid - a(f_r).
 \label{physical}
 \end{eqnarray}
The identification of the Toda primary in (7.1) is taken to be for just the cases where $\mathcal{O}_N$ is principal Levi type. The other two sets of relations in (7.2), (7.3) that give the local contributions to the Higgs and Coulomb branch dimensions hold for all $\mathcal{O}_N$. These quantities enter the description of the four dimensional theory (obtained via the class $\mathcal{S}$ constructions) and its partition function on a four sphere. 

Note the asymmetric nature of the setup. The asymmetry arises from the fact that in the CDT description of these defects, in cases outside type A, the Hitchin side involves only special orbits in $\mathfrak{g}^{\vee}$ with an additional datum involving subgroups of their component groups while the Nahm side involves all possible nilpotent orbits in $\mathfrak{g}$ along with the trivial representation of their component groups.\footnote{An expanded set of regular defects might allow one to think about the $\mathfrak{g}$ and $\mathfrak{g}^{\vee}$ descriptions of the defect in a more symmetric way. However, that possibility is not explored in this paper.}

Also included in the tables is the representation $r$ obtained by tensoring $\bar{r}$ with the sign representation and the value of Lusztig's invariant $a(f_r)$ for the family containing the irrep $r$. For the defects whose Nahm data is a special orbit, the irrep $r$ is the Orbit representation associated to the corresponding Hitchin orbit. For defects with non-special orbits as Nahm data, the irrep $Sp^{-1}[(\mathcal{O}_H,\psi_H]$ (when it exists) turns out to be a different non-special irrep belonging to the same family as $r$. It is notable that in these cases, the irrep $r$ is not one of the Springer reps associated to non-trivial local systems on the Hitchin orbit. The general pattern for a non-special $\mathcal{O}_N$ (observed by calculations in classical lie algebras of low rank and all exceptional cases) is that there exists a cell module $c_1' (= \epsilon \otimes c_1)$ belonging to the family that contains $r$ and the unique special representation in the family together with other such $r$ ($=\epsilon \times \bar{r}$) arising from all the non-special orbits in the same special piece.\footnote{It is interesting that in recent work \cite{2012arXiv1202.6097L}, finite $\mathcal{W}$-algebra methods are used to study certain properties of cell modules in a given family/two-cell.}  Further, the representations associated to the non-trivial local systems on $\mathcal{O}_H$ occur as summands in cell modules that are strictly \textit{different} from $c_1'$. This does not seem to have been recorded in the mathematical literature. It would be interesting to know if there is a proof of such a statement for arbitrary $\mathfrak{g}$. In any case, the physical consequence is the following. A matching argument for what one may call the Coulomb branch Springer invariant ($r$) is out of reach except for the cases where $\mathcal{O}_N$ is special.  However, intuitively, one expects that the Coulomb branch considerations in  \cite{Chacaltana:2012zy} and the Higgs branch $\bar{r}$ matching argument provided here should be part of one unified setup. In this context, associating certain other invariants like the conjugacy class of the Weyl group to the Coulomb branch data might be helpful. Achieving this would also seem relevant to developing a direct Coulomb branch check for the Toda primary for arbitrary $\mathfrak{g}$.

Every step in Toda-Nahm-Hitchin dictionary outlined in Fig \ref{THEsetup} remains perfectly applicable when $\mathfrak{g}$ and $\mathfrak{g}^\vee$ are non simply laced and thus one expects the dictionary to extend, as stated, to these cases as well.  As discussed earlier, these are the cases with relevance for the twisted defects of the six dimensional theory and for S-duality of boundary conditions in $\mathcal{N}=4$ SYM with non-simply laced gauge groups. But, there is a new feature in these cases that is worth pointing out. When $\mathfrak{g}^\vee$ is non-simply laced, the Langlands dual of the pseudo-Levi subalgebra $\mathfrak{l}^\vee$ which is denoted by $\mathfrak{l}$ is no longer guaranteed to be a subalgebra of $\mathfrak{g}$. The general procedure to find all possible centralizers of semi-simple elements in a complex lie algebra is to follow the Borel-de Seibenthal algorithm. Following this algorithm, one immediately recognizes the inevitability of the situation where $\mathfrak{l} \nsubseteq \mathfrak{g}$ (See Appendix \ref{BoreldeSiebenthal}). When such $\mathfrak{l}$ occur, the scenario is sometimes termed `elliptic-endoscopic'. More concretely, the corresponding group $L_{\mathbb{C}}$ would be an elliptic endoscopic group for $G_{\mathbb{C}}$. Such scenarios play an important role in the framework of geometric endoscopy explored in \cite{Frenkel:2007tx}. 

The occurrence of such data in the framework of Fig \ref{THEsetup} suggests the following question for $\mathfrak{g}$ arbitrary. Let $d_{BV}(O_H^{\mathfrak{l}^\vee})$ be the Barbasch-Vogan dual orbit in $\mathfrak{l}$. Is there a relationship between $d_{BV}(O_H^{\mathfrak{l}^\vee})$ and the orbit $\mathcal{O}_N$ (in $\mathfrak{g}$) that can be described in terms of the physics of Nahm boundary conditions in $\mathcal{N}=4$ SYM and/or the 3d $T^\rho[G]$ theories in a $\mathfrak{g}$ intrinsic way ?

\section{Tables}
\label{tables}
These detailed tables are included so that the reader can get some appreciation for the details of how the order reversing duality map works. The reader is especially encouraged to check these tables by following the map from one side to the other for a few scattered examples from the simply laced and non-simply laced cases.
 
Some of the calculations involved in compiling the tables were done using the CHEVIE package for the GAP system \cite{geck1996chevie,michel2013development}. Consulting the standard tables in Carter's book is also essential. The partitioning of the Weyl group representations into families is provided in Carter \cite{carter1985finite}. The Cartan type of the pseudo-Levi subalgebra $\mathfrak{l}^\vee$ that arises on the $\mathfrak{g}^\vee$ side is included as part of the tables for some simple cases. For the exceptional cases, it can be obtained from \cite{sommers2001lusztig}. The data collected in the tables is available in the mathematical literature often very explicitly or perhaps implicitly. It is hoped that the details help those who are not familiar with this literature. What is new is the physical interpretation of some defining features of the order reversing duality map. 

In the tables for $F_4, E_6, E_7,E_8$, the duality map for special orbits is detailed first and then separate tables are devoted for the non-trivial special pieces. The special orbits that are part of non-trivial special pieces thus occur in both tables.

In the non-simply laced cases, the number $d$ corresponds to a part of the local contribution to the Coulomb branch dimension. There is an additional contribution that comes from the fact that the nilpotent orbits for $G$ non-simply laced arise actually from the twisted defects of the six dimensional theory \cite{Chacaltana:2012zy}.

The tables themselves were generated in the following way. The data for the columns $\mathcal{O}_N, \tilde{b}, \bar{r}, (\mathcal{O}_H, C_H)$ follows directly from the data that is used in the description of the $\bar{r}$-matching. The irrep $r$ is obtained by tensoring $\bar{r}$ by the sign representation. The column $a(f_r)$ is Lusztig's invariant attached to the family to which the representation $r$ belongs. It is equal to the dimension of the Springer fiber associated to the Hitchin orbit.

The notation used in the tables is reviewed in the various Appendices. Appendix \ref{orbitsappendix} reviews the notation used for nilpotent orbits. This is relevant for the columns $\mathcal{O}_N,(\mathcal{O}_H,C_H)$. Appendix \ref{repsappendix} reviews the notation used for irreducible representations of Weyl groups and is relevant for columns $\bar{r}, r$.
\subsection{Simply laced cases}
\subsubsection{$A_3$}
$\mid \Lambda^+ \mid = 6$\\
\begin{table}[!h]
\centering 
\caption{Order reversing duality for $A_3=\mathfrak{su}(4)$} \label{a3}
\begin{center}
\begin{tabularx}{\textwidth}{bsbbssbc}
\toprule
($\mathcal{O}_N$) & $\tilde{b}$ & $\bar{r}$ & $r$ & $a(f_r)$ & $d$  & $(\mathcal{O}_H,C_H)$ & $\mathfrak{l}^\vee$  \\ 
 \midrule
 \rowcolor{Gray} $[1^4]$  & 6 & $[1^4]$ & $[4]$ & 0 & 6  & $[4]$ & $A_3$ \\ 
  $[2,1^2]$  & 3  & $[2,1^2]$ & $[3,1]$ & 1 & 5  & $[3,1]$  & $A_2$ \\ 
 \rowcolor{Gray} $[2,2]$   & 2 & $[2,2]$ & $[2,2]$ & 2 & 4  & $[2,2]$ & $A_1 + A_1$ \\ 
  $[3,1]$  & 1 & $[3,1]$ & $[2,1^2]$ & 3   & 3  & $[2,1^2]$ & $A_1$\\ 
 \rowcolor{Gray} $[4]$  & 0 & $[4]$ & $[1^4]$ & 6 & 0  & $[1^4]$ & $\varnothing$ \\ \bottomrule
 \end{tabularx}
 \end{center}
 \label{a3table}
\end{table}
\subsubsection*{Families with multiple irreps}
\begin{tabular}{ll}
& None
\end{tabular}
\newpage

\subsubsection{$D_4$}
$\mid \Lambda^+ \mid = 12$
\begin{table}[!h]
\centering 
\caption{Order reversing duality for $D_4=\mathfrak{so}_8$} \label{d4}
\begin{center}
\begin{tabularx}{\textwidth}{bsbbssbc}
\toprule
($\mathcal{O}_N$) & $\tilde{b}$ & $\bar{r}$ & $r$ & $a(f_r)$ & $d$  & $(\mathcal{O}_H,C_H)$ & $\mathfrak{l}^\vee$ \\ 
 \midrule
 \rowcolor{Gray} $[1^8]$  & 12 & $[1^4].-$ & $[4].-$ & 0 & 12  &$[7,1]$ & $D_4$ \\ 
  $[2^2 ,1^4]$  & 7 & $[1^3].[1]$ & $[3].[1]$ & 1 & 11 &$[5,3]$ & $D_4$ \\ 
\rowcolor{Gray}  $[2^4]^I$  &  6 & $([1^2].[1^2])'$ & $([2].[2])'$ & 2 & 10  &$[4^2]^I$ & $A_3$ \\ 
   $[2^4]^{II}$  & 6 & $([1^2].[1^2])''$ & $([2].[2])''$ & 2 & 10  &$[4^2]^{II}$ & $A_3$ \\ 
 \rowcolor{Gray}  $[3,1^5]$  &  6 & $[2,1^2].-$ & $([3,1].-)$ & 2 & 10  &$[5,1^3]$ & $A_3$\\ 
 $[3,2^2,1]$  & 4 & $[2^2].-$ & $[2^2].-$ & 3 & 9  & $[3^2,1^2],S_2$ &  $4 A_1$\\ 
 \rowcolor{Gray}  $[3^2,1^2]$  &  3 & $[2,1].[1]$ & $[2,1].[1]$ & 3 & 9  &  $[3^2,1^2]$ & $A_2$ \\ 
  $[5,1^3]$ & 2 & $[3,1].-$ & $[2,1^2].-$& 6 & 6   & $[3,1^5]$ & $2 A_1$\\ 
 \rowcolor{Gray}   $[4^2]^I$  & 2 & $([2].[2])'$ & $([1^2].[1^2])'$ & 6 & 6  & $[2^4]^I$ & $2 A_1 $  \\ 
 $[4^2]^{II}$  & 2 & $([2].[2])''$ & $([1^2].[1^2])''$ &  6 & 6   & $[2^4]^{II}$ & $2 A_1$ \\ 
   \rowcolor{Gray}  $[5,3]$  & 1 & $[3].[1]$ & $[1^3].[1]$ &  7 & 5  & $[2^2,1^4]$ & $A_1$ \\ 
   $[7,1]$  & 0 & $[4].-$ & $[1^4].-$ &  12 & 0  & $[1^8]$  & $\varnothing$ \\ \bottomrule
\end{tabularx} 
\end{center}
\label{d4table}
\end{table}

The Nahm orbits $[3,2^2,1]$ and $[3^2,1^2]$ are part of the only non-trivial special piece for $D_4$.

\subsubsection*{Families with multiple irreps}
\begin{tabularx}{\textwidth}{bs}
\toprule
Family $f$ & $a(f)$ \\
\midrule
 $\{([2,1],[1]),([2^2],-) ,([2],[1^2])\}$ & 3 \\ \bottomrule
\end{tabularx}

\newpage

\subsubsection{$E_6$}
$\mid \Lambda^+ \mid = 36$\\
\begin{table} [!h]
\centering 
\caption{Order reversing duality for special orbits in $E_6$} \label{e6s}
\end{table}
\begin{center}
\begin{tabularx}{\textwidth}{bssssssb}
\toprule
$(\mathcal{O}_N)$ & $\tilde{b}$ & $\bar{r}$ & $r$ & $a(f_r)$ & $d$  & $(\mathcal{O}_H)$ \\ 
 \midrule 
 \rowcolor{Gray}  $0$ & 36 & $\phi_{1,36}$  & $\phi_{1,0}$ & 0 & 36   & $E_6$ \\ 
  $A_1$ & 25 & $\phi_{6,25}$ & $\phi_{6,1}$ & 1 &  35 &  $E_6(a_1)$\\
  \rowcolor{Gray}  $2A_1$ &  20 & $\phi_{20,20}$ & $\phi_{20,2}$  & 2 & 34   & $D_5$\\ 
 $A_2$ & 15 & $\phi_{30,15}$ & $\phi_{30,3}$ & 3 & 33 & $E_6(a_3)$ \\
  \rowcolor{Gray} $A_2+A_1$  & 13 & $\phi_{64,13}$ & $\phi_{64,4}$ & 4 &  32  & $D_5(a_1)$ \\ 
 $A_2+2A_1$  & 11 & $\phi_{60,11}$ & $\phi_{60,5}$ & 5 & 31 & $A_4+A_1$ \\
  \rowcolor{Gray} $2A_2$ & 12 & $\phi_{24,12}$ & $\phi_{24,6}$ & 6 & 30 & $D_4$ \\ 
  $A_3$ & 10 & $\phi_{81,10}$ & $\phi_{81,6}$ & 6 & 30   & $A_4$ \\ 
  \rowcolor{Gray} $D_4(a_1)$ & 7 & $\phi_{80,7}$ & $\phi_{80,7}$ & 7 & 29   &  $D_4(a_1)$ \\ 
 $A_4$  & 6 & $\phi_{81,6}$ & $\phi_{81,10}$ & 10 & 24 & $A_3$ \\
  \rowcolor{Gray} $D_4$  & 6 & $\phi_{24,6}$ & $\phi_{24,12}$ & 12 &  26  & $2A_2$ \\ 
  $A_4+A_1$ & 5 & $\phi_{60,5}$ & $\phi_{60,11}$ & 11 & 25 & $A_2+2A_1$ \\
  \rowcolor{Gray}  $D_5(a_1)$ & 4 & $\phi_{64,4}$ & $\phi_{64,13}$ & 13 & 23   & $A_2+A_1$ \\ 
 $E_6(a_3)$  & 3 & $\phi_{30,3}$& $\phi_{30,15}$& 15 & 21 & $A_2$ \\
  \rowcolor{Gray}  $D_5$ & 2 & $\phi_{20,2}$ & $\phi_{20,20}$ & 20 & 16   & $2A_1$\\ 
 $E_6(a_1)$  & 1 & $\phi_{6,1}$ & $\phi_{6,25}$ & 25 & 11 &  $A_1$ \\
  \rowcolor{Gray} $E_6$ & 0 & $\phi_{1,0}$ & $\phi_{1,36}$ & 36 & 0   & $0$ \\ \bottomrule
 \end{tabularx}
\end{center}
\newpage

\begin{table}[!h]
\centering 
\caption{Order reversing duality for nontrivial special pieces in $E_6$} \label{e6}
\begin{center}
\begin{tabularx}{\textwidth}{bsssssb}
\toprule
($\mathcal{O}_N$) & $\tilde{b}$ & $\bar{r}$ & $r$ & $a(f_r)$ & $d$  & $(\mathcal{O}_H$,$C_H)$ \\ 
 \midrule 
 \rowcolor{Gray} 	$3A_1$  & 16 & $\phi_{15,16}$ & $\phi_{15,4}$ &  3 & 33  & $E_6(a_3),S_2$  \\ 
  					$A_2$   & 15 & $\phi_{30,15}$ & $\phi_{13,3}$ & 3 & 33  & $E_6(a_3)$\\ 
 \midrule
 \rowcolor{Gray} 	$2A_2 + A_1$ & 9 & $\phi_{10,9}$ & $\phi_{10,9}$ & 7  & 29 & $D_4(a_1),S_3$\\  
 					$A_3 + A_1 $  & 8 & $\phi_{60,8}$ & $\phi_{60,8}$ & 7 & 29  & $D_4(a_1),S_2$\\ 
 \rowcolor{Gray} 	$D_4(a_1)$  & 7 & $\phi_{80,7}$ & $\phi_{80,7}$ & 7 & 29 & $D_4(a_1)$ \\ \midrule
	$A_5$   & 4 & $\phi_{15,4}$ & $\phi_{15,16}$ & 15 & 21  & $A_2,S_2$\\ 
   \rowcolor{Gray} $E_6(a_3)$  & 3 & $\phi_{30,3}$ & $\phi_{30,15}$ &  15 & 21  & $A_2$  \\ 
 \bottomrule
\end{tabularx}
\end{center}
\end{table}
\subsubsection*{Families with multiple irreps}
\begin{tabularx}{\textwidth}{bs}
\toprule
Family $f$ & $a(f)$ \\
\midrule
$\{ \phi_{30,3},\phi_{15,4},\phi_{15,5} \}$  &  15 \\
$\{\phi_{80,7},\phi_{60,8},\phi_{90,8},\phi_{10,9},\phi_{20,10} \}$ & 7 \\
$\{ \phi_{30,15},\phi_{15,16},\phi_{15,17} \}$ & 3 \\ \bottomrule
\end{tabularx}
\newpage
\subsubsection{$E_7$}
$\mid \Lambda^+ \mid =63$ \\
\begin{table} [!h]
\centering 
\caption{Order reversing duality for special orbits in $E_7$} \label{e7s}
\end{table}
\begin{center}
\begin{tabularx}{\textwidth}{cssssssc}
\toprule
 $(\mathcal{O}_N)$ & $\tilde{b}$ & $\bar{r}$ & $r$ & $a(f_r)$ & $d$  & $(\mathcal{O}_H)$ \\ 
 \midrule 
 \rowcolor{Gray}  $0$  & 63 & $\phi_{1,63}$ & $\phi_{1,0}$  & 0 & 63   & $E_7$ \\ 
 $A_1$ & 46 & $\phi_{7,46}$ & $\phi_{7,1}$ & 1 & 62 & $E_7(a_1)$ \\
  \rowcolor{Gray}  $2A_1$ & 37 & $\phi_{27,37}$  & $\phi_{27,2}$ & 2 &  61  & $E_7(a_2)$ \\ 
  $A_2$ & 30 & $\phi_{56,30}$ & $\phi_{56,3}$ & 3 & 60 &  $E_7(a_3)$\\
  \rowcolor{Gray}  $(3A_1)''$ & 36 & $\phi_{21,36}$  & $\phi_{21,3}$ & 3 &  60  & $E_6$ \\ 
  $A_2+A_1$ & 25 & $\phi_{120,25}$ & $\phi_{120,4}$&  4 & 59 &  $E_6(a_1)$\\
  \rowcolor{Gray} $A_2+2A_1$  & 22 & $\phi_{189,22}$ & $\phi_{189,5}$ & 5 &  58  & $E_7(a_4)$  \\ 
  $A_2+3A_1$  & 21 & $\phi_{105,21}$ & $\phi_{105,6}$ & 6 & 57 & $A_6$\\
  \rowcolor{Gray} $A_3$ & 21 & $\phi_{210,21}$  & $\phi_{210,6}$ & 6 & 57 &  $D_6(a_1)$   \\ 
   $2A_2$ & 21 & $\phi_{168,21}$ & $\phi_{168,6}$& 6 & 57 & $D_5+A_1$\\
  \rowcolor{Gray}  $D_4(a_1)$ &  16 & $\phi_{315,16}$ & $\phi_{315,7}$ & 7 &  56  & $E_7(a_5)$ \\ 
 $(A_3+A_1)''$  &  20 & $\phi_{189,20}$& $\phi_{189,7}$& 7&  56 &  $D_5$ \\
  \rowcolor{Gray} $D_4(a_1)+A_1$  & 15 & $\phi_{405,15}$  & $\phi_{405,8}$ & 8 &  51  & $E_6(a_3)$  \\ 
$A_3+A_2$ &  14 & $\phi_{378,14}$& $\phi_{378,9}$& 9 &  54 & $D_5(a_1)+A_1$ \\
  \rowcolor{Gray} $D_4$ & 15 & $\phi_{105,15}$ & $\phi_{105,12}$ & 12 &  51  & $A_5''$  \\ 
  $A_3+A_2+A_1$  & 13 &$\phi_{210,13}$ &$\phi_{210,10}$ & 10 & 53 & $A_4+A_2$\\
  \rowcolor{Gray}  $A_4$ & 13 &$\phi_{420,13}$  & $\phi_{420,10}$ & 10 & 53   & $D_5(a_1)$  \\ 
   $^\spadesuit$ $A_4+A_1$ & 11 & $\phi_{510,11}$&$\phi_{510,12}$ & 12 & 51 & $A_4+A_1$\\
  \rowcolor{Gray}  $D_5(a_1)$ & 10 & $\phi_{420,10}$ & $\phi_{420,13}$ & 13 &  50  & $A_4$  \\ 
 $A_4+A_2$  & 10 & $\phi_{210,10}$ & $\phi_{210,13}$& 13 & 50 &  $A_3+A_2+A_1$ \\
\end{tabularx}
\end{center}
\newpage
\begin{center}
\begin{tabularx}{\textwidth}{csssssc}
	$A_5''$ & 12 & $\phi_{105,12}$ & $\phi_{105,15}$ & 15 & 48 & $D_4$ \\
  \rowcolor{Gray}  $D_5(a_1)+A_1$ & 9 & $\phi_{378,9}$ & $\phi_{378,14}$ & 14  &  49  & $A_3+A_2$  \\ 
  $E_6(a_3)$  & 8 & $\phi_{405,8}$& $\phi_{405,15}$ & 15 & 48 & $D_4(a_1)+A_1$\\
  \rowcolor{Gray}  $D_5$ & 7 & $\phi_{189,7}$ & $\phi_{189,20}$ & 20 &  43  & $(A_3+A_1)''$ \\ 
  $E_7(a_5)$  & 7 & $\phi_{315,7}$ & $\phi_{315,16}$ & 16 & 47 & $D_4(a_1)$\\
  \rowcolor{Gray}  $D_5+A_1$ & 6 & $\phi_{168,6}$ & $\phi_{168,21}$ & 21 & 42   & $2A_2$ \\ 
   $D_6(a_1)$ & 6 & $\phi_{210,6}$ & $\phi_{210,21}$ & 21 & 42 & $A_3$\\
  \rowcolor{Gray}  $A_6$ &  6 & $\phi_{105,6}$ & $\phi_{105,21}$  & 21 &  42  & $A_2+3A_1$ \\ 
   $E_7(a_4)$ & 5 & $\phi_{189,5}$ & $\phi_{189,22}$ &  22& 41 &  $A_2+2A_1$\\
  \rowcolor{Gray} $E_6(a_1)$ & 4 & $\phi_{120,4}$ & $\phi_{120,25}$ & 25 & 38   &  $A_2+A_1$  \\ 
   $E_6$ & 3 & $\phi_{21,3}$ & $\phi_{21,36}$ & 36 & 27 & $(3A_1)''$\\
  \rowcolor{Gray}  $E_7(a_3)$ & 3 & $\phi_{56,3}$ & $\phi_{56,30}$ & 30  & 33   & $A_2$   \\ 
  $E_7(a_2)$ & 2 & $\phi_{27,2}$ & $\phi_{27,37}$ & 37 & 26 &  $2A_1$ \\
  \rowcolor{Gray} $E_7(a_1)$ & 1 & $\phi_{7,1}$ & $\phi_{7,46}$ & 46 & 17  &  $A_1$  \\ 
  $E_7$  & 0 & $\phi_{1,0}$ & $\phi_{1,63}$ & 63 & 0 & $0$ \\
 \bottomrule
\end{tabularx}
\end{center}
\newpage
\begin{table}[!h]
\centering 
\caption{Order reversing duality for nontrivial special pieces in $E_7$} \label{e7}
\end{table}
\begin{center}
\begin{tabularx}{\textwidth}{bsssssb}
\toprule
($\mathcal{O}_N$) & $\tilde{b}$ & $\bar{r}$ & $r$ & $a(f_r)$ & $d$  & $(\mathcal{O}_H$,$C_H)$ \\ 
 \midrule 
 \rowcolor{Gray} $3A_1'$  & 31 & $\phi_{35,31}$ & $\phi_{35,4}$ & 3 & 60 & $E_7(a_3),S_2$\\ 
 $A_2$  & 30 & $\phi_{56,30}$ & $\phi_{56,3}$ & 3 &  60  & $E_7(a_3)$\\ 
 \midrule
  \rowcolor{Gray} $4A_1$  & 28 & $\phi_{15,28}$ & $\phi_{15,7}$ & 4 &  59  & $E_6(a_1),S_2$\\ 
 $A_2+A_1$  & 25 & $\phi_{120,25}$ & $\phi_{120,4}$ & 4 &  59  & $E_6(a_1)$\\
  \midrule
   \rowcolor{Gray} $A_3+2A_1$  & 16 & $\phi_{216,16}$ & $\phi_{216,9}$ & 8 &  55  & $E_6(a_3),S_2$\\ 
 $D_4 (a_1)+A_1$  & 15 & $\phi_{405,15}$ & $\phi_{405,8}$ & 8 &  55 & $E_6(a_3)$\\
  \midrule
   \rowcolor{Gray} $D_4+A_1$  & 12 & $\phi_{84,12}$ & $\phi_{84,15}$ & 13 &  50  & $A_4,S_2$\\ 
 $D_5 (a_1)$  & 10 & $\phi_{420,10}$ & $\phi_{420,13}$ & 13 &  50  & $A_4$\\
  \midrule
   \rowcolor{Gray} $(A_5)'$  & 9 & $\phi_{216,9}$ & $\phi_{216,19}$ & 15 &  48  & $D_4(a_1)+A_1,S_2$\\ 
 $E_6 (a_3)$  & 8 & $\phi_{405,8}$ & $\phi_{405,15}$ & 15 &  48  & $D_4(a_1)+A_1$\\
  \midrule
  \rowcolor{Gray} $D_6$  & 4 & $\phi_{35,4}$ & $\phi_{35,31}$ & 30 &  33  & $A_2,S_2$\\ 
 $E_7 (a_3)$  & 3 & $\phi_{56,3}$ & $\phi_{56,30}$ & 30 & 33  & $A_2$\\
  \midrule
  \end{tabularx}
(..contd)  
   \newpage
  \begin{tabularx}{\textwidth}{bsbssbb}
  
 \rowcolor{Gray} $2A_2 + A_1$  & 18 & $\phi_{70,18}$ & $\phi_{70,9}$  & 7 &  56  & $E_7(a_5),S_3$\\ 
  $(A_3 +A_1)'$  & 17 & $\phi_{280,17}$ & $\phi_{280,8}$ & 7 &  56  & $E_7(a_5),S_2$ \\
  \rowcolor{Gray} $D_4(a_1)$  & 16 & $\phi_{315,16}$ & $\phi_{315,7}$ & 7 &  56 & $E_7(a_5)$ \\ 
   \midrule
    \rowcolor{Gray} $A_5 + A_1$  & 9 & $\phi_{70,9}$ & $\phi_{70,18}$  & 16 &  47  & $D_4(a_1),S_3$\\ 
  $D_6(a_2)$  & 8 & $\phi_{280,8}$ & $\phi_{280,17}$ &  16 & 47   & $D_4(a_1),S_2$ \\
  \rowcolor{Gray} $E_7(a_5)$  & 7 & $\phi_{315,7}$ & $\phi_{315,16}$ & 16 &  47  & $D_4(a_1)$ \\ 
   \bottomrule  
 \end{tabularx}
\end{center}
\subsubsection*{Families with multiple irreps}
\begin{tabularx}{\textwidth}{bs}
\toprule
Family $f$ & $a(f)$ \\
\midrule
 $\{ \phi_{56,3},\phi_{35,4},\phi_{21,6} \}$ &  3\\
 $\{ \phi_{120,4},\phi_{105,5},\phi_{15,7} \}$ & 4\\
 $\{ \phi_{405,8},\phi_{216,9},\phi_{189,10} \}$ & 8\\
 $\{ \phi_{420,10},\phi_{336,11},\phi_{84,12} \}$ & 10\\
 $^\spadesuit \{ \phi_{512,11},\phi_{512,12} \}$ & 11\\
 $\{ \phi_{420,13},\phi_{336,14},\phi_{84,15} \}$ & 13\\
 $\{ \phi_{405,15},\phi_{216,16},\phi_{189,17} \}$ & 15\\
 $\{ \phi_{120,25},\phi_{105,26},\phi_{15,28} \}$ & 25\\
 $\{ \phi_{56,30},\phi_{35,31},\phi_{21,33} \}$ & 30\\
 $\{ \phi_{315,7},\phi_{280,8},\phi_{70,9},\phi_{280,9},\phi_{35,13} \}$ & 7\\
 $\{ \phi_{315,16},\phi_{280,17},\phi_{70,18},\phi_{280,18},\phi_{35,22} \}$ &  16\\ \bottomrule
\end{tabularx}

\newpage
\subsubsection{$E_8$}
$\mid \Lambda^+ \mid = 120$ \\
\begin{table} [!h]
\centering 
\caption{Order reversing duality for special orbits in $E_8$} \label{e8s}
\end{table}
\begin{center}
\begin{tabularx}{\textwidth}{bssssssb}
\toprule
 $\mathcal{O}_N$ & $\tilde{b}$ & $\bar{r}$ & $r$ & $a(f_r)$ & $d$  & $\mathcal{O}_H$ \\ 
 \midrule 
 \rowcolor{Gray} $0$ & 120 & $\phi_{1,120}$ &  $\phi_{1,0}$ & 0 & 120   &  $E_8$ \\ 
  $A_1$ & 91 & $\phi_{8,91}$ & $\phi_{8,1}$ & 1 &  119 & $E_8(a_1)$\\
  \rowcolor{Gray}$2A_1$  & 74& $\phi_{35,74}$ & $\phi_{35,2}$ & 2 & 118   &  $E_8(a_2)$ \\ 
  $A_2$ & 63 & $\phi_{112,63}$& $\phi_{112,3}$& 3 & 117 & $E_8(a_3)$ \\
  \rowcolor{Gray} $A_2+A_1$  & 52 & $\phi_{210,52}$ & $\phi_{210,4}$ & 4 &  116  & $E_8(a_4)$  \\ 
   $A_2+2A_1$ & 47 &$\phi_{560,47}$ &$\phi_{560,5}$ & 5 & 115  & $E_8(b_4)$\\
  \rowcolor{Gray}  $A_3$ & 46 & $\phi_{567,46}$ & $\phi_{567,6}$ & 6 & 114   & $E_7(a_1)$  \\ 
   $2A_2$ & 42 & $\phi_{700,42}$ & $\phi_{700,6}$& 6 & 114 & $E_8(a_5)$\\
  \rowcolor{Gray} $D_4(a_1)$ & 37 & $\phi_{1400,37}$ & $\phi_{1400,7}$ &  7 &  113  &  $E_8(b_5)$ \\ 
 $D_4(a_1)+A_1$ & 32 & $\phi_{1400,32}$ & $\phi_{1400,8}$ & 8 & 112 &  $E_8(a_6)$ \\
  \rowcolor{Gray} $A_3+A_2$  & 31 & $\phi_{3240,31}$ & $\phi_{3240,9}$ & 9 & 111   & $D_7(a_1)$  \\ 
 \rowcolor{Gray}  $D_4(a_1)+A_2$ & 28 & $\phi_{2240,28}$ & $\phi_{2240,10}$ & 10 & 110 & $E_8(b_6)$      \\ 
 $A_4$ & 30 & $\phi_{2268,30}$ & $\phi_{2268,10}$ & 10 & 110 & $E_7(a_3)$   \\
  \rowcolor{Gray} $D_4$ & 36 & $\phi_{525,36}$ & $\phi_{525,12}$ & 12 & 108   &  $E_6$ \\ 
$^\spadesuit A_4+A_1$  & 26 & $\phi_{4096,26}$ & $\phi_{4096,12}$ & 11 & 109 &  $E_6(a_1)+A_1$ \\
  \rowcolor{Gray} $A_4+2A_1$ & 24 & $\phi_{4200,24}$ & $\phi_{4200,12}$ & 12  & 108   &  $D_7(a_2)$  \\ 
  $A_4+A_2$ & 23 & $\phi_{4536,23}$ & $\phi_{4536,13}$ & 13 & 107 & $D_5+A_2$ \\
  \rowcolor{Gray} $D_5(a_1)$ & 25 &$\phi_{2800,25}$  & $\phi_{2800,13}$ & 13 & 107   &  $E_6(a_1)$   \\ 
   $A_4+A_2+A_1$ & 22  &  $\phi_{2835,22}$& $\phi_{2835,14}$ & 14 & 106 & $A_6+A_1$\\
  \rowcolor{Gray}  $D_4+A_2$ & 21 & $\phi_{4200,21}$ & $\phi_{4200,15}$ & 15 &  105  & $A_6$ \\ 
 $D_5(a_1)+A_1$ & 22 &  $\phi_{6075,22}$& $\phi_{6075,14}$& 14 & 106 &  $E_7(a_4)$ \\
 \end{tabularx}
\end{center}
\newpage
\begin{center}
\begin{tabularx}{\textwidth}{bsssssb}
 \rowcolor{Gray}$E_6(a_3)$  & 21 & $\phi_{5600,21}$ & $\phi_{5600,15}$ & 15 &  105  &  $D_6(a_1)$ \\ 
  $D_5$ &20 & $\phi_{2100,20}$& $\phi_{2100,20}$ & 20 & 100 &  $D_5$\\
  \rowcolor{Gray} $E_8(a_7)$  & 16 & $\phi_{4480,16}$ & $\phi_{4480,16}$ & 16  & 104   & $E_8(a_7)$ \\ 
  $D_6(a_1)$ & 15 & $\phi_{5600,15}$& $\phi_{5600,21}$ & 21 & 99 &  $E_6(a_3)$\\
  \rowcolor{Gray}  $E_7(a_4)$ &  14& $\phi_{6075,14}$ & $\phi_{6075,22}$ & 22 &  98  &  $D_5(a_1)+A_1$ \\ 
   $A_6$ & 15 & $\phi_{4200,15}$& $\phi_{4200,21}$ & 21 & 99 & $D_4+A_2$\\
  \rowcolor{Gray}$A_6+A_1$  & 14& $\phi_{2835,14}$ & $\phi_{2835,22}$ &  22 & 98   & $A_4+A_2+A_1$ \\ 
  $E_6(a_1)$& 13 & $\phi_{2800,13}$ & $\phi_{2800,25}$ & 25 & 95  &$D_5(a_1)$  \\
  \rowcolor{Gray} $D_5+A_2$ & 13 & $\phi_{4536,13}$ & $\phi_{4536,23}$ & 23 & 97   &  $A_4+A_2$ \\ 
 $D_7(a_2)$  &12 & $\phi_{4200,12}$ & $\phi_{4200,24}$ & 24 & 96 & $A_4+2A_1$ \\
  \rowcolor{Gray}  $^\spadesuit  E_6(a_1)+A_1$ & 11 & $\phi_{4096,11}$  & $\phi_{4096,27}$ & 26 & 94   & $A_4+A_1$ \\ 
   $E_6$ & 12 & $\phi_{525,12}$ & $\phi_{525,36}$ & 36 & 84 & $D_4$\\
  \rowcolor{Gray}  $E_7(a_3)$ & 10 & $\phi_{2268,10}$ & $\phi_{2268,30}$ & 30 &  90  & $A_4$ \\ 
  $E_8(b_6)$  &  10&  $\phi_{2240,10}$& $\phi_{2240,28}$& 28 & 92  &  $D_4(a_1)+A_2$\\
  \rowcolor{Gray} $D_7(a_1)$ &  9 & $\phi_{3240,9}$ & $\phi_{3240,31}$ & 31 & 89   & $A_3+A_2$  \\ 
  $E_8(a_6)$ & 8 & $\phi_{1400,8}$ & $\phi_{1400,32}$ & 32 & 88 & $D_4(a_1)+A_1$ \\
    \rowcolor{Gray}$E_8(b_5)$  & 7  & $\phi_{1400,7}$ & $\phi_{1400,37}$ & 37 & 83   &  $D_4(a_1)$  \\ 
   $E_8(a_5)$ &  6 & $\phi_{700,6}$& $\phi_{700,42}$& 42 & 78 &  $2A_2$\\
  \rowcolor{Gray}  $E_7(a_1)$ & 6 & $\phi_{567,6}$ & $\phi_{567,46}$ & 46 & 74   & $A_3$  \\ 
 $E_8(b_4)$ & 5 & $\phi_{560,5}$ & $\phi_{560,47}$ & 47 & 73 & $A_2+2A_1$ \\
  \rowcolor{Gray} $E_8(a_4)$  & 4 & $\phi_{210,4}$ & $\phi_{210,52}$  &  52 & 68   &  $A_2+A_1$ \\ 
  $E_8(a_3)$ & 3 & $\phi_{112,3}$ & $\phi_{112,63}$& 63 &  57& $A_2$\\
  \rowcolor{Gray} $E_8(a_2)$ & 2 & $\phi_{35,2}$ & $\phi_{35,74}$ & 74 & 46   &  $2A_1$ \\ 
   $E_8(a_1)$ & 1 & $\phi_{8,1}$ & $\phi_{8,91}$ & 91 & 29 & $A_1$\\
    \rowcolor{Gray}  $E_8$ & 0 & $\phi_{1,0}$ & $\phi_{1,120}$ & 120 & 0   & $0$ \\ 
 \bottomrule
\end{tabularx}
\end{center}
\newpage
\begin{table} [!h]
\centering 
\caption{Order reversing duality for nontrivial special pieces in $E_8$} \label{e8}
\end{table}
\begin{center}
\begin{tabularx}{\textwidth}{bsssssb}
\toprule
($\mathcal{O}_N$) & $\tilde{b}$ & $\bar{r}$ & $r$ & $a(f_r)$ & $d$  & $(\mathcal{O}_H$,$C_H)$ \\  
 \midrule 
\rowcolor{Gray} $3A_1$  & 64 & $\phi_{84,64}$ & $\phi_{84,4}$ & 3 &  117  & $E_8(a_3),S_2$\\ 
 $A_2$  & 63 & $\phi_{112,63}$ & $\phi_{112,3}$ & 3 &  117  & $E_8(a_3)$\\
  \midrule
   \rowcolor{Gray} $4A_1$  & 56 & $\phi_{50,56}$ & $\phi_{50,8}$ & 4 &  116  & $E_8(a_4),S_2$\\ 
 $A_2+A_1$  & 52 & $\phi_{210,52}$ & $\phi_{210,4}$ & 4 &  116  & $E_8(a_4)$\\
  \midrule
   \rowcolor{Gray} $A_2+3A_1$  & 43 & $\phi_{400,43}$ & $\phi_{400,7}$ & 6 &  114  & $E_8(a_5),S_2$\\ 
 $2A_2$  & 42 & $\phi_{700,42}$ & $\phi_{700,6}$ & 6 &  114  & $E_8(a_5)$\\
  \midrule
   \rowcolor{Gray} $D_4+A_1$  & 28 & $\phi_{700,28}$ & $\phi_{700,16}$ & 13 &  107  & $E_6(a_1),S_2$\\ 
 $D_5(a_1)$  & 25 & $\phi_{2800,25}$ & $\phi_{2800,13}$ & 13 &  107  & $E_6(a_1)$\\
  \midrule
   \rowcolor{Gray} $2A_3$  & 26 & $\phi_{840,26}$ & $\phi_{840,14}$ & 12 &  108  & $D_7(a_2),S_2$\\ 
 $A_4+2A_1$  & 24 & $\phi_{4200,24}$ & $\phi_{4200,12}$ & 12 &  108  & $D_7(a_2)$\\
  \midrule 
   \rowcolor{Gray} $A_5$  & 22 & $\phi_{3200,22}$ & $\phi_{3200,16}$ & 15 &  105  & $D_6(a_1),S_2$\\ 
 $E_6(a_3)$  & 21 & $\phi_{5600,21}$ & $\phi_{5600,15}$ & 15 &  105  & $D_6(a_1)$\\
  \midrule
   \rowcolor{Gray} $D_5+A_1$  & 16 & $\phi_{3200,16}$ & $\phi_{3200,22}$ & 25 &  95  & $E_6(a_3),S_2$\\ 
 $D_6(a_1)$  & 15 & $\phi_{5600,15}$ & $\phi_{5600,21}$ & 25 &  95  & $E_6(a_3)$\\
  \midrule
  \end{tabularx} \newpage
  \begin{tabularx}{\textwidth}{bsssssb}
   \rowcolor{Gray} $D_6$  & 12 & $\phi_{972,12}$ & $\phi_{972,32}$ & 30 &  90  & $A_4,S_2$\\ 
 $E_7(a_3)$  & 10 & $\phi_{2268,10}$ & $\phi_{2268,30}$ & 30 &  90  & $A_4$\\
  \midrule 
   \rowcolor{Gray} $A_7$  & 11 & $\phi_{1400,11}$ & $\phi_{1400,29}$ & 28 &  92  & $D_4(a_1)+A_2,S_2$\\ 
 $E_8(b_6)$  & 10 & $\phi_{2240,10}$ & $\phi_{2240,28}$ & 28 &  92  & $D_4(a_1)+A_2$\\
  \midrule
  \rowcolor{Gray} $D_7$  & 7 & $\phi_{400,7}$ & $\phi_{400,43}$ & 42 &  78 & $E_8(a_5),S_2$\\ 
 $E_8(a_5)$  & 6 & $\phi_{700,6}$ & $\phi_{700,42}$ & 42 &  78  & $E_8(a_5)$\\
  \midrule
   \rowcolor{Gray} $E_7$  & 4 & $\phi_{84,4}$ & $\phi_{84,64}$ & 63 &  57 & $A_2,S_2$\\ 
 $E_8(a_3)$  & 3 & $\phi_{112,3}$ & $\phi_{112,63}$ & 63 &  57  & $A_2$\\
  \midrule 
    \rowcolor{Gray} $A_3+A_2+A_1$  & 29 & $\phi_{1400,29}$ & $\phi_{1400,11}$  & 10 &  110  & $E_8(b_6),S_2$\\ 
   $D_4(a_1)+A_2$  & 28 & $\phi_{2240,28}$ & $\phi_{2240,10}$ & 10 &  100 & $E_8(b_6)$ \\ 
 \midrule 
 \rowcolor{Gray} $2A_2+A_1$  & 39 & $\phi_{448,39}$ & $\phi_{448,9}$  & 7 &  113  & $E_8(b_5),S_3$\\ 
  $A_3+ A_1$  & 38 & $\phi_{1344,38}$ & $\phi_{1344,38}$ &  7 & 113   & $E_8(b_5),S_2$ \\
  \rowcolor{Gray} $D_4(a_1)$  & 37 & $\phi_{1400,37}$ & $\phi_{1400,8}$ & 7 &  113  & $E_8(b_5)$ \\ 
 \midrule 
  \rowcolor{Gray} $2A_2+2A_1$  & 36 & $\phi_{175,36}$ & $\phi_{175,12}$  & 8 &  112  & $E_8(a_6),S_3$\\ 
  $A_3+2A_1$  & 34 & $\phi_{1050,34}$ & $\phi_{1050,10}$ &  8 & 112   & $E_8(a_6),S_2$ \\
  \rowcolor{Gray} $D_4(a_1)+A_1$  & 32 & $\phi_{1400,32}$ & $\phi_{1400,8}$ & 8 &  112  & $E_8(a_6)$ \\ 
 \midrule 
  \rowcolor{Gray} $E_6+A_1$  & 9 & $\phi_{448,9}$ & $\phi_{448,39}$  & 37 &  83  & $D_4(a_1),S_3$\\ 
  $E_7(a_2)$  & 8 & $\phi_{1344,8}$ & $\phi_{1344,38}$ &  37 & 83   & $D_4(a_1),S_2$ \\
  \rowcolor{Gray} $E_8(b_5)$  & 7 & $\phi_{1400,7}$ & $\phi_{1400,37}$ & 37 &  83  & $D_4(a_1)$ \\ 
 \midrule 
 \end{tabularx} \newpage
  \begin{tabularx}{\textwidth}{bsssssb}
  \rowcolor{Gray} $A_4+A_3$  & 20 & $\phi_{420,20}$ & $\phi_{420,20}$  & 16 &  104  & $E_8(a_7),S_5$\\ 
  $D_5(a_1)+A_2$  & 19 & $\phi_{1344,19}$ & $\phi_{1344,19}$  & 16 &  104  & $E_8(a_7),S_4$\\  
   \rowcolor{Gray} $A_5+A_1$  & 19 & $\phi_{2016,19}$ & $\phi_{2016,19}$ &  16 & 104   & $E_8(a_7),S_3 \times S_2$ \\
  $E_6(a_3)+A_1$  & 18 & $\phi_{3150,18}$ & $\phi_{3150,18}$ & 16 &  104  & $E_8(a_7),S_3$ \\ 
  \rowcolor{Gray} $D_6(a_2)$  & 18 & $\phi_{4200,18}$ & $\phi_{4200,18}$ &  16 & 104   & $E_8(a_7),S_2 \times S_2$ \\ 
   $E_7(a_5)$  & 17 & $\phi_{7168,17}$ & $\phi_{7168,17}$ &  16 & 104   & $E_8(a_7),S_2$ \\
  \rowcolor{Gray} $E_8(a_7)$  & 16 & $\phi_{4480,16}$ & $\phi_{4480,16}$ & 16 &  104  & $E_8(a_7)$ \\ 
 \bottomrule
\end{tabularx}
\label{e8tablens}
\end{center}
\subsubsection*{Families with multiple irreps}
\begin{center}
\begin{tabularx}{\textwidth}{bs}
\toprule
Family $f$ & $a(f)$ \\ \midrule
 $\{ \phi_{112,3},\phi_{84,4},\phi_{28,8} \}$ & 3\\
 $\{ \phi_{210,4},\phi_{160,7},\phi_{50,8} \}$ & 4\\
 $\{\phi_{700,8},\phi_{400,7},\phi_{300,8}   \}$ & 8\\
 $\{\phi_{2268,10},\phi_{972,12},\phi_{1296,13}   \}$ & 10 \\
 $\{ \phi_{2240,10},\phi_{1400,11},\phi_{840,13}  \}$ & 10\\
 $ ^\spadesuit \{ \phi_{4096,11},\phi_{4096,12} \}$ & 11\\
 $\{\phi_{4200,12},\phi_{3360,13},\phi_{840,14}   \}$ & 13\\
 $\{ \phi_{2800,13},\phi_{700,16},\phi_{2100,16}  \}$ & 16\\
 $\{ \phi_{5600,15},\phi_{3200,16},\phi_{2400,17}  \}$ & 16\\
 $\{ \phi_{5600,21},\phi_{3200,22},\phi_{2400,23}  \}$ & 22\\
 $\{\phi_{4200,24},\phi_{3360,25},\phi_{840,31}   \}$ & 25\\
 $\{ \phi_{2800,25},\phi_{700,28},\phi_{2100,28}  \}$ &28 \\
 $^\spadesuit \{\phi_{4096,26},\phi_{4096,27}\}$ & 26\\
 $\{\phi_{2240,28},\phi_{1400,29},\phi_{840,31}   \}$ & 29\\
 $\{\phi_{2268,30},\phi_{972,32},\phi_{1296,33}   \}$& 32\\
 $\{\phi_{700,42},\phi_{400,43},\phi_{300,44}   \}$ & 43\\
 $\{ \phi_{210,52},\phi_{160,55},\phi_{50,56}  \}$ & 55\\
 $\{ \phi_{112,63},\phi_{84,64},\phi_{28,68}  \}$ & 64\\
 $\{\phi_{1400,7},\phi_{1344,8},\phi_{448,9},\phi_{1008,9},\phi_{56,19}   \}$ & 7\\
 $\{ \phi_{1400,8},\phi_{1050,10},\phi_{1575,10},\phi_{175,12},\phi_{350,14} \}$ & 8\\  $\{\phi_{1400,32},\phi_{1050,34},\phi_{1575,34},\phi_{175,36},\phi_{350,38} \}$ & 32\\
 $\{ \phi_{1400,37},\phi_{1344,38},\phi_{448,39},\phi_{1008,39},\phi_{56,49}  \}$ &  37\\
 $\{ \phi_{4480,16},\phi_{7168,17},\phi_{3150,18},\phi_{4200,18},\phi_{4536,18},\phi_{5670,18},$ \\  $\phi_{1344,19},\phi_{2016,19},\phi_{5600,19},\phi_{2688,20},\phi_{420,20},\phi_{1134,20},$\\  $\phi_{1400,20},\phi_{1680,22},\phi_{168,24},\phi_{448,25},\phi_{70,32}  \}$ & 16 \\ \bottomrule
\end{tabularx}
\end{center}
\newpage
\subsubsection{A comment on exceptional orbits}
\label{exceptionalorbits}
The families marked with a $^\spadesuit$ are the only families with just two irreps. There is one such family in $E_7$ and two such families in $E_8$. The orbits for which the associated Orbit representation is one of these are referred to as exceptional orbits. They are known to have somewhat peculiar properties among all nilpotent orbits (See Carter\cite{carter1985finite} Prop 11.3.5 and \cite{benson1972degrees,curtis1975corrections}). The special representations that occur in these families are the only ones which do not give another special representation when tensored with the sign representation. They are also known to posses some special properties from the point of view of the representation theory of Hecke algebras. These are the only cases where $\mathcal{O}_N$ is a special orbit and $Sp[r] \neq \mathcal{O}_H$. Another way to view this anomalous situation would be to say that the natural partial ordering on special representations \footnote{This can be obtained by transferring the closure ordering on the set of Special orbits to the set of Special representation.} of the Weyl group is reversed by a tensoring with sign in all cases except these. There is a version of this inversion map due to Lusztig (denoted earlier in the paper by $i(r)$), which remedies these anomalous cases by assigning the special representation in the family of $\epsilon \otimes r$ to be $i(r)$.

In this context, it is important to note that there are subtler partial orders that are defined by Achar \cite{achar2003order} and Sommers \cite{sommers2006equivalence} which when transferred to Irr(W) may enable the treatment of these cases on a more equal footing with every other instance of duality. From a physical standpoint, it would be interesting to know if these subtler partial orders are related to the partial order implied by the possible Higgsing patterns of the corresponding three dimensional $T[G]$.
\newpage
\subsection{Non-simply laced cases}
\subsubsection{$\mathfrak{g}=B_3$, $\mathfrak{g}^{\vee}=C_3$ and  $\mathfrak{g}=C_3$, $\mathfrak{g}^{\vee}=B_3 $}
$\mid \Lambda^+ \mid = 9$\\
\begin{table}[!h]
\centering 
\caption{Order reversing duality for $\mathfrak{g}=B_3$, $\mathfrak{g}^{\vee}=C_3$} \label{b3c3}
\begin{center}
\begin{tabularx}{\textwidth}{bsbbssb}
\toprule
($\mathcal{O}_N$) & $\tilde{b}$ & $\bar{r}$ & $r$ & $a(f_r)$ & $d$  & $(\mathcal{O}_H$,$C_H)$ \\  
 \midrule 
 \rowcolor{Gray} $[1^7]$  & 9 & $-.[1^3]$ & $[3].-$ & 0 & 9 & $[6]$  \\ 
 $[2^2,1^3]$  & 5 & $-.[2,1]$ & $[2,1].-$ & 1 & 8 & $[4,2]$,$S_2$ \\ 
 \rowcolor{Gray} $[3,1^4]$ & 4 & $[1].[1^2]$ & $[2].[1]$  & 1  & 8 & $[4,2]$ \\ 
  $[3,2^2]$  & 3 & $[1^2].[1]$ & $[1].[2]$  & 2 & 6 & $[3^2]$  \\ 
 \rowcolor{Gray} $[3^2,1]$  & 2 & $-.[3]$ & $[1^3].-$  & 4 & 5 & $[2^2,1^2],S_2$ \\ 
  $[5,1^2]$  &1  & $[2].[1]$  & $[1].[1^2]$ & 4 & 5 & $[2^2,1^2]$ \\ 
 \rowcolor{Gray} $[7]$  & 0 & $[3].-$ & $-.[1^3]$ & 9 & 0 & $[1^6]$ \\ \bottomrule
\end{tabularx}
\end{center}
\end{table}
\begin{table}[!h]
\centering 
\caption{Order reversing duality for $\mathfrak{g}=C_3$, $\mathfrak{g}^{\vee}=B_3$} \label{c3b3}
\begin{center}
\begin{tabularx}{\textwidth}{bsbbssb}
\toprule
($\mathcal{O}_N$) & $\tilde{b}$ & $\bar{r}$ & $r$ & $a(f_r)$ & $d$  & $(\mathcal{O}_H$,$C_H)$ \\  
 \midrule 
 \rowcolor{Gray} $[1^6]$  & 9 & $-.[1^3]$ & $[3].-$ & 0 & 9 & $[7]$  \\ 
 $[2,1^4]$  & 6 & $[1^3].-$ & $-.[3]$ & 1 & 8 & $[5,1^2],S_2$ \\ 
 \rowcolor{Gray} $[2^2,1^2]$ & 4 & $[1].[1^2]$ & $[2].[1]$  & 1  & 8 & $[5,1^2]$ \\ 
  $[2^3]$  & 3 & $[1^2].[1]$ & $[1].[2]$  & 2 & 7 & $[3^2,1]$  \\ 
 \rowcolor{Gray} $[3^2]$  & 2  & $[1].[2]$  & $[1^2].[1]$ & 3 & 6 & $[3,2^2]$ \\ 
 $[4,1^2]$  & 2 & $[2,1].-$ & $-.[2,1]$  & 4 & 5 & $[3,1^4],S_2$ \\ 
 \rowcolor{Gray} $[4,2]$  & 1 & $[2].[1]$ & $[1].[1^2]$ & 4 & 5 & $[3,1^4]$ \\ 
 $[6]$  & 0  & $[3].-$  & $-.[1^3]$ & 9 & 0 & $[1^7]$ \\ \bottomrule
\end{tabularx}
\end{center}
\end{table} 
\subsubsection*{Families with multiple irreps} 
\begin{center}
\begin{tabularx}{\textwidth}{bs}
\toprule
Family $f$ & $a(f)$ \\ \midrule
$ [2].[1],-.[3],[2,1].- $ & 1 \\ 
 $[1].[1^2],[1^3].-,-.[2,1]$ & 4 \\ \bottomrule
\end{tabularx}
\end{center}
\newpage

\subsubsection{$G_2$}
$\mid \Lambda^+ \mid = 6$
\begin{table}[!h]
\centering
\caption{Order reversing duality for $\mathfrak{g}_2$} \label{g2}
\begin{center}
\begin{tabularx}{\textwidth}{bsssssb}
\toprule
($\mathcal{O}_N$) & $\tilde{b}$ & $\bar{r}$ & $r$ & $a(f_r)$ & $d$  & $(\mathcal{O}_H$,$C_H)$ \\ 
 \midrule
 \rowcolor{Gray} 1 & 6 & $\phi_{1,6}$ & $\phi_{1,0}$ & 0 & 6  & $G_2$ \\
 $A_1$ & 3 & $\phi_{1,3}''$ & $\phi_{1,3}''$& 1 & 5  & $(G_2(a_1),S_3)$ \\
 \rowcolor{Gray} $\tilde{A_1}$ & 2 & $\phi_{2,2}$ & $\phi_{2,2}$ & 1 & 5 & $(G_2(a_1),S_2)$ \\
 $G_2(a_1)$ & 1 & $\phi_{2,1}$ & $\phi_{2,1}$ & 1 & 5  & $(G_2(a_1),1)$ \\
 \rowcolor{Gray} $G_2$ & 0 & $\phi_{1,0} $ & $\phi_{1,6}$ & 6 & 0  & 1 \\  \bottomrule
\end{tabularx} 
\end{center}
\label{g2table}
\end{table}
\subsubsection*{Families with multiple irreps}
\begin{center}
\begin{tabularx}{\textwidth}{bs}
\toprule
Family $f$ & $a(f)$ \\ \midrule
 $\{\phi_{2,1}, \phi_{2,2},\phi_{1,3}',\phi_{1,3}'' \}$ & 1 \\ \bottomrule 
\end{tabularx}
\end{center}
\newpage
\subsubsection{$F_4$}
$\mid \Lambda^+ \mid = 24$  \\
\begin{table} [!h]
\centering 
\caption{Order reversing duality for special orbits in $F_4$} \label{f4s}
\end{table}
\begin{center}
\begin{tabularx}{\textwidth}{cssssssb}
\toprule
 $(\mathcal{O}_N)$ & $\tilde{b}$ & $\bar{r}$ & $r$ & $a(f_r)$ & $d$  & $(\mathcal{O}_H)$ \\ 
 \midrule 
 $0$  & 24 & $\phi_{1,24}$ & $\phi_{1,0}$ & 0 & 24   & $F_4$  \\ 
  \rowcolor{Gray}  $\tilde{A_1}$ & 13&$\phi_{4,13}$ & $\phi_{4,1}$& 1 &23 & $F_4(a_1)$ \\
    $A_1 + \tilde{A_1}$ &  10&$\phi_{9,10}$ & $\phi_{9,2}'$& 2 & 22 & $F_4(a_2) $ \\
  \rowcolor{Gray} \footnote{These instances (marked with a $\star$) of the duality map are a bit subtle. Although the Weyl group of the dual is isomorphic in a canonical way to the original, there is an exchange of the long root and the short root. The notation for $\bar{r}$ incorporates this exchange.} $\star A_2$ & 9& $\phi_{8,9}''$ & $\phi_{8,3}''$ & 3 &  21  & $B_3$  \\ 
 $\star \tilde{A_2}$ & 9 &$\phi_{8,9}'$ & $\phi_{8,3}'$& 3 & 21 & $C_3$  \\
 \rowcolor{Gray}  $F_4(a_3)$ & 4 & $\phi_{12,4}$ & $\phi_{12,4}$ & 4 & 20   & $F_4(a_3)$   \\ 
  $\star B_3$ & 3 & $\phi_{8,3}''$ & $\phi_{8,9}''$ & 9 &  15  & $A_2$  \\ 
\rowcolor{Gray} $\star C_3$  & 3 &$\phi_{8,3}'$ &$\phi_{8,9}'$ & 9 & 15 &  $\tilde{A_2}$ \\
 $F_4(a_2)$ &  2&$\phi_{9,2}$  & $\phi_{9,10}$ & 10 & 14   & $A_1 + \tilde{A_1}$   \\ 
  \rowcolor{Gray}  $F_4(a_1)$ & 1 &$\phi_{4,1}$ & $\phi_{4,13}$& 13 & 11 &  $\tilde{A_1}$  \\
 $F_4$ & 0 & $\phi_{1,0}$ & $\phi_{1,24}$ & 24 &  0  & $0$   \\ \bottomrule
\end{tabularx}
\end{center}
\subsubsection*{Families with multiple irreps}
\begin{center}
\begin{tabularx}{\textwidth}{ls}
\toprule
Family $f$ & $a(f)$ \\ \midrule
$\{ \phi_{4,1},\phi_{2,4}',\phi_{2,4} \}$ &  1\\
$\{ \phi_{4,13}, \phi_{2,16}',\phi_{2,16}''  \}$ & 13\\
 $\{ \phi_{12,4},\phi_{16,5},\phi_{6,6}',\phi_{6,6}'',\phi_{9,6}',\phi_{9,6}'',\phi_{4,7}',\phi_{4,7}'',\phi_{4,8},\phi_{1,12}',\phi_{1,12}'' \} $ & 4\\ \bottomrule
\end{tabularx}
\end{center}
\newpage

\begin{table}[!h]
\centering
\caption{Order reversing duality for non trivial special pieces in $F_4$ } \label{f4}
\begin{center}
\begin{tabularx}{\textwidth}{bsssssc}
\toprule
($\mathcal{O}_N$) & $\tilde{b}$ & $\bar{r}$ & $r$ & $a(f_r)$ & $d$  & $(\mathcal{O}_H$,$C_H)$ \\  
 \midrule 
 \rowcolor{Gray} $A_1$  & 16 & $\phi_{2,16}''$  & $\phi_{2,4}'$ & 1 &  23 & ($F_4(a_1),S_2$)\\ 
   $\tilde{A_1}$  & 13 & $\phi_{4,13}$ & $\phi_{4,1}$ & 1 & 23  & $F_4(a_1)$\\ \midrule
     \rowcolor{Gray} $A_2 + \tilde{A_1}$  & 7 & $\phi_{4,7}''$ & $\phi_{4,7}''$ &  4 & 20  & $(F_4(a_3), S_4)$ \\
     $A_1 + \tilde{A_2}$   & 6 & $\phi_{6,6}'$ & $\phi_{6,6}'$ & 4 & 20  & $(F_4(a_3),S_3)$\\
     \rowcolor{Gray} $B_2$  & 6 & $\phi_{9,6}''$ & $\phi_{9,6}''$ & 4 & 20 & $(F_4(a_3),S_2 \times S_2)$ \\
     $C_3(a_1)$  & 5 &  $\phi_{16,5}$  & $\phi_{16,5}$ & 4 & 20 & $(F_4(a_3),S_2)$\\
  \rowcolor{Gray}   $F_4(a_3)$  & 4  & $\phi_{12,4}$ & $\phi_{12,4}$ & 4 & 20  & $F_4(a_3)$ \\
     \bottomrule
\end{tabularx}
\end{center}
\end{table}

\section{Acknowledgements}
It is a pleasure to thank my advisor J. Distler for many discussions. I also thank Andy Neitzke for his comments, David Ben-Zvi \& Jim Humphreys for some kind pointers to the mathematical literature, Anindya Dey and Andy Trimm for discussions and B. Binegar for sharing some unpublished work. G. Lusztig provided helpful advice about some results in \cite{lusztig1984characters} and I thank him for that. Software tools of the ATLAS project\footnote{http://www.liegroups.org/software/} and Jean Michel's developmental version \cite{michel2013development} of the  package CHEVIE \footnote{This is available from http://www.math.jussieu.fr/~jmichel/chevie/chevie.html} were helpful at various stages and I thank the creators and maintainers of these projects. Users of the Q \& A site Mathoverflow\footnote{http://mathoverflow.net/} have indulged my questions at various points and I thank them for that. The material is based upon work supported by the National Science Foundation under Grant Number PHY-1316033.
\newpage
\appendix 
\section{Nilpotent orbits in complex lie algebras}
\label{orbitsappendix}

Nilpotent orbits in the classical cases have a convenient parameterization in terms of certain partitions. For $A_N$, these are just partitions of $N+1$. For the other types $B_N, C_N, D_N$, the orbits are classified by $B-,C-,D-$ type partitions.
The dimension of such an orbit can be expressed in terms of the partition type $[n_i]$ and its transpose $[s_i]$. Let $r_k$ be the number of times the number $k$ appears in the partition $[n_i]$.  Such an orbit will be denoted by $\mathcal{O}_{n_i}$. Its dimension is given by \cite{collingwood1993nilpotent},
\begin{eqnarray}
\text{dim}(\mathcal{O}_{n_i}) &=& \text{dim}(\mathfrak{g}) - \bigg( \sum_i s_i^2 - 1 \bigg)\hspace{1.0 in} \text{for} \hspace{0.5 in}  \mathfrak{g}=A_n \\ 
\text{dim}(\mathcal{O}_{n_i}) &=& \text{dim}(\mathfrak{g}) - \frac{1}{2}\bigg(\sum_i s_i^2 - \sum_{i \in \text{odd}} r_i \bigg)\hspace{0.5 in}\text{for}\hspace{0.5 in} \mathfrak{g}=B_n, D_n \\
\text{dim}(\mathcal{O}_{n_i}) &=& \text{dim}(\mathfrak{g}) - \frac{1}{2}\bigg(\sum_i s_i^2 + \sum_{i \in \text{odd}} r_i  \bigg) \hspace{0.5 in}\text{for} \hspace{0.5 in}\mathfrak{g}=C_n 
\end{eqnarray}
In the exceptional cases, the dimensions of the orbits can be obtained from the tables in \cite{carter1985finite, collingwood1993nilpotent} (also reproduced in \cite{Chacaltana:2012zy}). The closure ordering on the nilpotent orbits plays an important role in many considerations and this is often described by a Hasse diagram. It is often instructive to look at the Hasse diagrams for just the special nilpotent orbits for the order reversing dualities act as an involution on this subset of orbits. In the exceptional cases, such diagrams are available in the Appendices of \cite{Chacaltana:2012zy}. There were originally determined by Spaltenstein in \cite{spaltenstein1982classes}.
\subsection*{Bala-Carter theory}
An efficient classification system for nilpotent orbits that is independent of the existence of partition type classifications was provided in the work of Bala-Carter. Their fundamental insight was to look for distinguished nilpotent orbits in Levi subalgebra $\mathfrak{l}$ of a complex lie algebra $\mathfrak{g}$. Levi subalgebras themselves are classified by subsets of the set of simple roots. By providing a classification of all distinguished nilpotent elements in all Levi subalgebras, Bala-Carter effectively provided a classification scheme for all nilpotent orbits. This complements the classification by partition labels in the classical cases and is somewhat indispensable in the exceptional cases for which there is no partition type classification. When Bala-Carter labels are specified for a nilpotent orbit, the capitalized part of the label identifies a distinguished parabolic subalgebra $\mathfrak{p}$ whose Levi part is Levi subalgebra $\mathfrak{l}$. If there is a further Cartan type label enclosed within parenthesis, this denotes a non-principal nilpotent orbit in that Levi subalgebra. If there is no further label attached, then it is a principal nilpotent orbit in the Levi subalgebra $\mathfrak{l}$. For example, $E_6(a_1)$ and $D_5$ are the BC labels for two different nilpotent orbits in $E_6$. The former is not principal Levi type while the latter is. The BC classification is somewhat indispensable in the exceptional cases since there is no partition type parameterization of the nilpotent orbits.

While it is not absolutely necessary, it is also instructive to assign BC labels to nilpotent orbits in the classical cases. So, it is useful to summarize it here (see \cite{bala1976classes,panyushev1999spherical} for more in this regard). Let $[n_i]$ be the partition describing a classical nilpotent orbit $\rho$ and let $\mathfrak{l}$ be the Bala-Carter Levi \footnote{No relationship is implied here to any of the subalgebras in the main body of the paper. There, Bala-Carter theory is used on both $\mathfrak{g}$ and $\mathfrak{g}^\vee$ sides and the notation for the corresponding Levi subalgebras is introduced therein. }
\begin{itemize}
\item type $A$ : $\mathfrak{l}$ is of Cartan type $A_{n_1-1} + A_{n_2-1} + \ldots $ \\
\item type $B, D$ : If $n_i$ are all distinct and odd, then $\rho$ is distinguished in $\mathfrak{l} = B_n/D_n$. For every pair of $n_i$ that are equal (say to $n$), add a factor of $A_{n-1}$ to $\mathfrak{l}$ and form a reduced partition with the repeating pair removed. Proceed inductively, till the reduced partition is empty. If the final partition is a [3], then add a factor $\tilde{A}_1$. It follows that the principal Levi type orbits have BC labels of the form $A_{i_1} + A_{i_1} + \ldots  + \tilde{A}_1$ or  $A_{i_1} + A_{i_1} + \ldots  + B_n/D_n$.
\item type $C$ : If $n_i$ are all distinct and even, then $\rho$ is distinguished in $\mathfrak{l}=C_n$. For every pair of $n_i$ that are equal (to $n$, say), add a factor of $\tilde{A}_{n-1}$ to $\mathfrak{l}$ and form a reduced partition with the repeating pair removed. Proceed inductively, till the reduced partition is empty. If the final partition is a [2], then add a factor of $A_1$. This implies the principal Levi type orbits have BC labels $\tilde{A}_{i_1} + \tilde{A}_{i_1} + \ldots + A_1$ or $\tilde{A}_{i_1} + \tilde{A}_{i_2} + \ldots + C_n$.
\end{itemize}
Using the above, one can count the number of principal Levi type nilpotent orbits in classical lie algebras. In the exceptional cases, the nilpotent orbits that are principal Levi type are immediately identifiable for they are always written in terms of their BC labels.

\newpage

\section{Representations of Weyl groups}
\label{repsappendix}
Here, the notation that is used in \cite{carter1985finite} to describe irreducible representations of Weyl groups is summarized. In the classical cases, there are certain combinatorial criteria for an irrep to be a special representation and for a set of representation to fall in the same family. These are also reviewed briefly. A general feature obeyed by all Weyl groups is that the trivial representation and the sign representation are special and consequently, they fall into their own families.
\subsection{type $A_{n-1}$}
The irreducible representation of $W[A_n]=S_n$ are given by partitions of $n$. The convention is that $[n]$ corresponds to the trivial representation while $[1^n]$ corresponds to the sign representation. All irreducible representations are special and they occur in separate families.
\subsection{type $B_{n}$ \& $C_{n}$}
The irreducible representations are classified by two partitions $[\alpha].[\beta]$ where $[\alpha]$ and $[\beta]$ are each partitions of $p,q$ such that $p+q=n$. To each such pair of partitions  $[\alpha].[\beta]$, associate a symbol in the following way. 
\begin{itemize}
\item For each ordered pair $[\alpha].[\beta]$, enlarge $\alpha$ or $\beta$ by adding trailing zeros if necessary such $\alpha$ has one part more than $\beta$. 
\item Then consider the following array : \\ 

$\bigg(\begin{array}{cccccccc}
\alpha_1 &   & \alpha_2+1 &  & \ldots & &\alpha_{m+1}+m\\ 
 & \beta_1 &  & \beta_2+1 & \ldots & \beta_m +(m-1) &
\end{array} \bigg)$
\item Apply an equivalence relation on such arrays in the following fashion : \\

$\bigg(\begin{array}{cccccccc}
0 &   & \lambda_1 + 1 & • & \ldots & &\lambda_m+1\\ 
• & 0 &  & \mu_1 +1 & \ldots & \mu_m +1 &
\end{array} \bigg)$ $\sim$ 
$\bigg(\begin{array}{cccccccc}
0 &   & \lambda_1  & • & \ldots & &\lambda_m\\ 
• & 0 &  & \mu_1 & \ldots & \mu_m &
\end{array} \bigg)$
\item Each pair $[\alpha].[\beta]$ then provides a unique equivalence class of arrays. Let a representative for such an array be \\
$\bigg(\begin{array}{cccccccc}
0 &   & \lambda_1  & • & \ldots & &\lambda_m\\ 
• & 0 &  & \mu_1 & \ldots & \mu_m &
\end{array} \bigg)$
\item This is the \textit{symbol} for the corresponding irreducible representation.
\end{itemize}

Two irreps $[\alpha].[\beta]$ and $[\alpha'].[\beta']$ fall in the same family if and only if their symbols are such that their symbols contains the same $\{ \lambda_i , \mu_i\}$ (treated as unordered sets). Within the set of all irreps that fall in a family, there is a unique irrep whose for which the associated symbol satisfies an ordering property :
\begin{equation}
\lambda_1 \leq \mu_1 \leq \lambda_2 \leq \mu_2 \ldots \mu_m \leq \lambda_{m+1}.
\end{equation}
This unique representation within the family is the special representation.
\subsection{type $D_n$}
\label{typeD}
The irreducible representations are classified again by pairs of partitions $[\alpha].[\beta]$, with $\alpha$, $\beta$ being partitions of $p,q$ such that $p+q=n$ but with one additional caveat. If $\alpha=\beta$, then there are two irreducible representations corresponding to this pair $([\alpha].[\alpha])'$ and $([\alpha].[\alpha])''$. Now, associate a symbol to this irrep by the following steps
\begin{itemize}
\item Write $\alpha = (\alpha_1,\alpha_2, \ldots)$, $\beta = (\beta_1,\beta_2,\ldots)$ as non-decreasing strings of integers. Add a few leading zeros if needed such that $\alpha,\beta$ have the same number of parts. Now, consider the array 
$\bigg(\begin{array}{cccc}
\alpha_1 & \alpha_2 +1 & \ldots & \alpha_m + m-1\\ 
\beta_1 & \beta_2 +1 & \ldots & \beta_m + m-1
\end{array} \bigg)$
\item Impose the following equivalence relation on such arrays \\

$\bigg(\begin{array}{ccccc}
0 & \lambda_1+1 & \lambda_2 +1  & \ldots & \lambda_m + 1\\ 
0 & \mu_1 +1 & \mu_2 +1 & \ldots & \mu_m + 1
\end{array} \bigg)$ $\sim$ $\bigg(\begin{array}{cccc}
 \lambda_1 & \lambda_2   & \ldots & \lambda_m \\ 
 \mu_1  & \mu_2  & \ldots & \mu_m 
\end{array} \bigg)$
\item Each $[\alpha].[\beta]$ now determines a unique equivalence class of such arrays. A representative of that equivalence class is the symbol of the irrep. 
\end{itemize}

Two irreps $[\alpha].[\beta]$ and $[\alpha'].[\beta']$ ($\alpha \neq \beta, \alpha' \neq \beta'$) fall in the same family if their symbols are such that the $\lambda_i,\mu_i$ occurring in them are identical (when treated as unordered sets). Within such a family, there is a unique irrep whose symbol satisfies the following ordering property,
\begin{equation}
\lambda_1 \leq \mu_1 \leq \lambda_2 \leq \mu_2 \ldots \lambda_m \leq \mu_m \hspace{0.3 in}\text{or} \hspace{0.3 in} \mu_1 \leq \lambda_1 \leq \mu_2 \leq \lambda_2 \ldots \mu_m \leq \lambda_m.
\end{equation}
This unique irrep  would be the special representations in that family. Irreps corresponding to labels of type $([\alpha].[\alpha])'$ and $([\alpha].[\alpha])''$ are always special and hence occur in their own families.

As an example of the application of the method of symbols, the irreps of $D_4$ and their corresponding symbols are noted in a table.

\begin{table} [!h]
\centering 
\caption{Symbols for irreducible representations of $W(D_4)$}
\begin{center}
\begin{tabular} {c|c}
\toprule
$[\alpha].[\beta]$ & Symbol \\ \midrule 
$[4].[-]$ & $\bigg( \begin{array}{c}
4 \\ 
0
\end{array} \bigg)$  \\
$[3,1].[-]$ & $\bigg( \begin{array}{cc}
1 & 4 \\ 
0 & 1
\end{array} \bigg)$  \\
$[2,2].[-]$ & $\bigg( \begin{array}{cc}
2 & 3 \\ 
0 & 1
\end{array} \bigg)$  \\
$[2,1^2].[-]$ & $\bigg( \begin{array}{ccc}
1 & 2 & 4 \\ 
0 & 1 & 2
\end{array} \bigg)$  \\
$[1^4].[-]$ & $\bigg( \begin{array}{cccc}
1 & 2 & 3 & 4 \\ 
0 & 1 & 2 & 3
\end{array} \bigg)$  \\
$[3].[1]$ & $\bigg( \begin{array}{c}
3 \\ 
1
\end{array} \bigg)$  \\
$[2,1].[1]$ & $\bigg( \begin{array}{cc}
1 & 3 \\ 
0 & 2
\end{array} \bigg)$  \\
$[1^3].[1]$ & $\bigg( \begin{array}{ccc}
1 & 2 & 3 \\ 
0 & 1 & 3
\end{array} \bigg)$  \\
$[2].[2]$ & $\bigg( \begin{array}{cc}
2 \\ 
2
\end{array} \bigg)$  \\
$[2].[1^2]$ & $\bigg( \begin{array}{cc}
0 & 3 \\ 
1 & 2
\end{array} \bigg)$  \\
$[1^2].[1^2]$ & $\bigg( \begin{array}{cc}
1 & 2 \\ 
1& 2
\end{array} \bigg)$  \\
\bottomrule
\end{tabular}
\end{center}
\end{table}

As can be seen from the symbols, the only non-trivial family in the case of $D_4$ is $\{([2,1],[1]),([2^2],-) ,([2],[1^2])\}$. 
\newpage

It is also useful to have the character table of $W(D_4)$ which can be used to compute tensor products with the sign representation. 

\begin{table} [!h]
\centering 
\caption{Character table for $W(D_4)$} \label{d4charactertable}
\begin{center}
\begin{tabularx}{\textwidth}{c|c|c|c|c|c|c|c|c|c|c|c|c|c}
\toprule
 & $c_1$ & $c_2$ & $c_3$ & $c_4$ & $c_5$ & $c_6$  & $c_7$ & $c_8$ & $c_9$ & $c_{10}$ & $c_{11}$ & $c_{12}$ & $c_{13}$ \\ 
 \midrule 
$[-].[1^4]$& 1 & 1 & 1& -1& -1& -1& 1 & 1& 1& 1& 1& -1& -1 \\ \hline
$([11].[11])'$& 3 & -1 & 3 & -1 & 1 & -1& 3& -1& -1 & 0 & 0 & -1 & -1 \\ \hline
$([11].[11])''$& 3 & -1 & 3 &-1 &1 &-1 &-1 &3 &-1 &0 &0 &1 &-1  \\ \hline
$[1].[1^3]$& 4 & 0 & -4& -2& 0 & 2& 0 &0 &0 &1 &-1 &0 &0  \\ \hline
$[1^2].[2]$& 6 & -2 & 6 & 0 & 0& 0 & -2&-2 &2 & 0 & 0 & 0 & 0 \\ \hline
$[1].[21]$& 8& 0  & -8 & 0 & 0 & 0 & 0 & 0 & 0 & -1 & 1& 0& 0 \\ \hline
$[-].[2,1^2]$& 3 & 3 & 3 & -1 & -1 & -1 & -1 & -1 & -1 & 0 & 0& 1& 1 \\ \hline
$[2].[2]$& 3 & -1 & 3& 1& -1& 1& 3& -1& -1& 0& 0& 1& -1 \\ \hline
$[2].[2]$& 3& -1& 3& 1& -1& 1& -1& 3& -1& 0& 0& -1& -1 \\ \hline
$[-].[2^2]$& 2& 2& 2& 0&0 &0 & 2 & 2& 2& -1& -1& 0 & 0  \\ \hline
$[1].[3]$& 4 & 0 & -4 & 2 & 0& -2& 0& 0&0 &1 & -1& 0& 0 \\ \hline
$[-].[1,3]$&3 &3 &3 & 1& 1 &1 & -1 & -1 & -1& 0 & 0 & -1 & -1  \\ \hline
$[-].[4]$&  1 & 1& 1& 1& 1&1 &1 & 1& 1&1 &1 &1 & 1 \\ \bottomrule
\end{tabularx}
\end{center}
\end{table}
where the conjugacy classes $c_i$ are 
\begin{eqnarray*}
c_1 = 1^4.- \\
c_2 = 11.11 \\
c_3 = -.1^4 \\
c_4 = 21^2.-\\
c_5 = 1.21 \\
c_6 = 2.1^2 \\
c_7 = (2^2.-)' \\
c_8 = (2^2,-)'' \\
c_9 = (-.22) \\
c_{10}= 31.-\\
c_{11} = -.31\\
c_{12} = (4.-)'\\
c_{13} = (4.-)''
\end{eqnarray*}

\subsection{Exceptional cases}

The irreps will be denoted by $\phi_{i,j}$, where $i$ is the degree and $j$ is what is usually called the $b$ value of the irreducible representation. In the non-simply laced cases of $G_2$ and $F_4$, there might be more than one representation with same degree and $b$ value. When this occurs, the two representations are distinguished by denoting them as $\phi_{i,j}'$ and $\phi_{i,j}''$ respectively. For example, $G_2$ has $\phi_{1,3}'$ and $\phi_{1,3}''$. Here, note that these two labels will be interchanged if we were to exchange the long root and the short root of $G_2$. The sign and the trivial representation can be identified in this notation as being the ones with the largest $b$ value and zero $b$ value respectively. To give a flavor for this notation in action, here is the character table for $W[G_2]$. The special representation are $\phi_{1,0},\phi_{2,1},\phi_{1,6}$. Every other representation (together with $\phi_{2,1}$) is a member of the only non-trivial family in $W[G_2]$.
\begin{table}[!h]
\centering 
\caption{Character table for $W(G_2)$} \label{g2charactertable}
\begin{center}
\begin{tabular}{c|c|c|c|c|c|c}
\toprule
 & $1$ & $\tilde{A_1}$ & $A_1$ & $G_2$ & $A_2$ & $A_1 + \tilde{A_1}$ \\
\midrule 
$\phi_{1,0}$ & 1 & 1 & 1 & 1 & 1 & 1 \\
$\phi_{1,6}$ & 1 & -1 & -1 & 1 & 1 & 1 \\
$\phi_{1,3}'$ & 1 & 1 & -1 & -1 & 1 & -1 \\
$\phi_{1,3}''$ & 1 & -1 & 1 & -1 & 1  & -1 \\
$\phi_{2,1}$ & 2 & 0 & 0 & 1 & -1 & -2\\
$\phi_{2,2}$ & 2 & 0 & 0 & -1 & -1  & 2\\
\bottomrule
\end{tabular}
\end{center}
\end{table}

There is an interesting duality operation on the set of irreducible representations of the Weyl group. For the most part, this acts as tensoring by the sign representation. An important property of the special representations of a Weyl group is that they are closed under this duality operation (See Section \ref{exceptionalorbits} for more on this duality operation). This can be readily seen to be true by looking at the character tables.

 \newpage
 
 \section{The method of Borel-de Siebenthal }
\label{BoreldeSiebenthal}
The Borel-de Seibenthal algorithm \cite{borel1949sous} can be used to obtain all possible subalgebras that arise as centralizers of semi-simple elements in Lie algebras (See \cite{sommers1998generalization,sommers2001lusztig} and references therein). The algorithm comes down to finding non-conjugate subsystems of the set of extended roots of the Lie algebra. Let $\pi$ denote the set of simple roots and $\Pi$ the corresponding Dynkin diagram. Now, adjoin the lowest root to $\pi$ and form $\tilde{\pi}$, the set of extended roots. Associated to this is the extended Dynkin diagram $\tilde{\Pi}$. The extended Dynkin diagrams formed by this procedure are collected in Fig \ref{extendedroots}. Now, form a sub diagram (possibly disconnected) by removing a node of $\tilde{\Pi}$ and all the lines connecting it. The resulting diagram corresponds to a subalgebra that arises as a centralizer. The Cartan type of the centralizer can be read off directly from the sub diagram. One can proceed by removing more nodes and lines to get all possible centralizers. There is a subset of them whose diagrams can also be obtained by considering just sub diagrams of $\Pi$. These corresponds to the centralizers of semi-simple elements that are also Levi. The more general centralizers are called pseudo-Levi in this paper (following \cite{sommers2001lusztig}). In the body of the paper, pseudo-Levi subalgebras of $\mathfrak{g}^\vee$ play an important role and these are denoted by $\mathfrak{l}^\vee$.  Among the pseudo-Levi subalgebras $\mathfrak{l}^\vee$ that fail to be Levi subalgebras, a particularly interesting class are the ones for which their Langlands dual $\mathfrak{l}$ fails to be a subalgebra of $\mathfrak{g}$ (the Langlands dual of $\mathfrak{g}^\vee$). It follows immediately from the Borel-de Seibenthal procedure that such a scenario can occur only for $\mathfrak{g}$ being non-simply laced. Some examples of these more interesting occurrences are collected here.
\subsection{Centralizer that is not a Levi}
Consider the extended Dynkin diagram for $D_4$ and denote it by $\tilde{\Pi}(D_4)$. There is a sub diagram which is of Cartan type $4A_1$ that does not arise as a sub diagram of $\Pi(D_4)$. This is a pseudo-Levi subalgebra that is not a Levi subalgebra.
\subsection{Pseudo-Levi $\mathfrak{l}^\vee$ such that Langlands dual $\mathfrak{l} \nsubseteq \mathfrak{g}$ }

Consider the extended Dynkin diagram for $\mathfrak{g}^\vee =B_{n+1}$ given by $\tilde{\Pi}(B_{n+1})$. There is a sub diagram which corresponds to a centralizer $\mathfrak{l}^\vee$ of Cartan type $D_n$. Taking Langlands duals, one gets $\mathfrak{g} = C_{n+1}$ and $\mathfrak{l}=D_{n}$. But, $D_n$ is not a subalgebra of $C_{n+1}$. 

\begin{center}

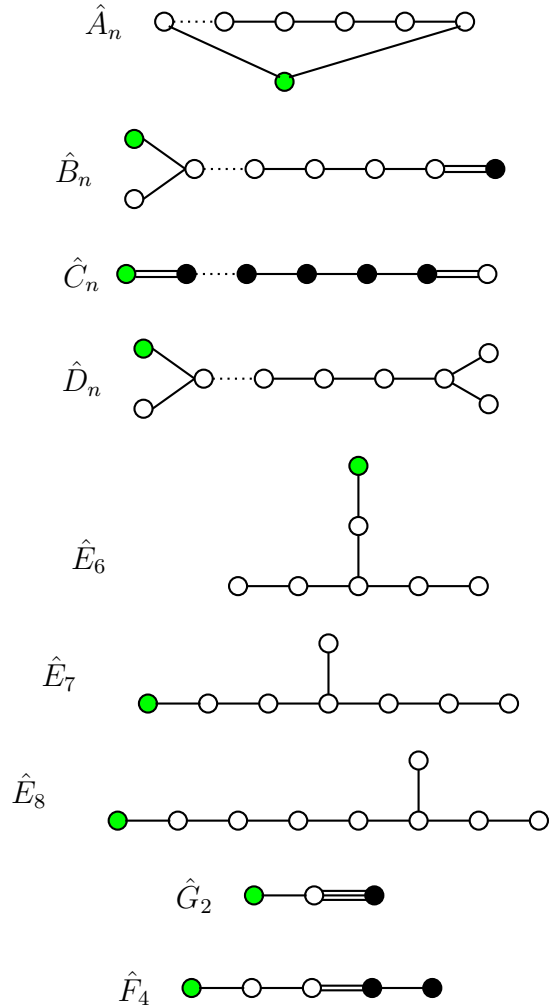
\begin{figure}[!h]

\begin{center}
  \begin{tikzpicture}[scale=.4]
    \draw (-1,0) node[anchor=east]  {$\hat{A}_n$};
    \foreach \x in {0,...,5}
    \draw[xshift=\x cm,thick] (\x cm,0) circle (.3cm);
    \draw[thick, fill=green] (4,-2) circle (.3 cm);
    \draw[thick] (0.15,-0.15) -- (3.85,-1.85);
    \draw[thick] (4.15,-1.85) -- (9.85,-0.15);
    \draw[dotted,thick] (0.3 cm,0) -- +(1.4 cm,0);
    \foreach \y in {1.15,...,4.15}
    \draw[xshift=\y cm,thick] (\y cm,0) -- +(1.4 cm,0);
  \end{tikzpicture}
\end{center}
\begin{center}
  \begin{tikzpicture}[scale=.4]
    \draw (-3,0) node[anchor=east]  {$\hat{B}_n$};
    \foreach \x in {0,...,4}
    \draw[xshift=\x cm,thick] (\x cm,0) circle (.3cm);
    \draw[xshift=5 cm,thick,fill=black] (5 cm, 0) circle (.3 cm);
    \draw[xshift=5 cm,thick] (-7 cm, -1) circle (.3 cm);
    \draw[xshift=5 cm,thick, fill=green] (-7 cm, 1) circle (.3 cm);
    \draw[dotted,thick] (0.3 cm,0) -- +(1.4 cm,0);
    \foreach \y in {1.15,...,3.15}
    \draw[xshift=\y cm,thick] (\y cm,0) -- +(1.4 cm,0);
    \draw[thick] (8.3 cm, 0.1 cm) -- +(1.4 cm,0 cm);
    \draw[thick] (8.3 cm, -0.1 cm) -- +(1.4 cm,-0 cm);
    \draw[thick] (-1.7 cm, 1.0 cm) -- +(1.4 cm,-1.0 cm);
    \draw[thick] (-1.7 cm, -1.0 cm) -- +(1.4 cm,1.0 cm);
  \end{tikzpicture}
\end{center}
\begin{center}
  \begin{tikzpicture}[scale=.4]
    \draw (-2.5,0) node[anchor=east]  {$\hat{C}_n$};
    \foreach \x in {0,...,4}
    \draw[xshift=\x cm,thick,fill=black] (\x cm,0) circle (.3cm);
    \draw[xshift=5 cm,thick] (5 cm, 0) circle (.3 cm);
      \draw[xshift=5 cm,thick, fill=green] (-7 cm, 0) circle (.3 cm);
    \draw[dotted,thick] (0.3 cm,0) -- +(1.4 cm,0);
    \foreach \y in {1.15,...,3.15}
    \draw[xshift=\y cm,thick] (\y cm,0) -- +(1.4 cm,0);
    \draw[thick] (8.3 cm, .1 cm) -- +(1.4 cm,0);
    \draw[thick] (8.3 cm, -.1 cm) -- +(1.4 cm,0);
        \draw[thick] (-1.7 cm, 0.1 cm) -- +(1.7 cm,0 cm);
    \draw[thick] (-1.7 cm, -0.1 cm) -- +(1.7 cm,-0 cm);
  \end{tikzpicture}
\end{center}

\begin{center}
  \begin{tikzpicture}[scale=.4]
    \draw (-3,0) node[anchor=east]  {$\hat{D}_n$};
    \foreach \x in {0,...,4}
    \draw[xshift=\x cm,thick] (\x cm,0) circle (.3cm);
    \draw[xshift=5 cm,thick] (-7 cm, -1) circle (.3 cm);
    \draw[xshift=5 cm,thick, fill=green] (-7 cm, 1) circle (.3 cm);
    \draw[xshift=8 cm,thick] (30: 17 mm) circle (.3cm);
    \draw[xshift=8 cm,thick] (-30: 17 mm) circle (.3cm);
    \draw[dotted,thick] (0.3 cm,0) -- +(1.4 cm,0);
    \foreach \y in {1.15,...,3.15}
    \draw[xshift=\y cm,thick] (\y cm,0) -- +(1.4 cm,0);
    \draw[xshift=8 cm,thick] (30: 3 mm) -- (30: 14 mm);
    \draw[xshift=8 cm,thick] (-30: 3 mm) -- (-30: 14 mm);
     \draw[thick] (-1.7 cm, 1.0 cm) -- +(1.4 cm,-1.0 cm);
    \draw[thick] (-1.7 cm, -1.0 cm) -- +(1.4 cm,1.0 cm);
  \end{tikzpicture}
\end{center}

\begin{center}
  \begin{tikzpicture}[scale=.4]
    \draw (-4,1) node[anchor=east]  {$\hat{E}_6$};
    \draw[xshift=5 cm,thick, fill=green] (-1, 4) circle (.3 cm);
    \foreach \x in {0,...,4}
    \draw[thick,xshift=\x cm] (\x cm,0) circle (3 mm);
    \foreach \y in {0,...,3}
    \draw[thick,xshift=\y cm] (\y cm,0) ++(.3 cm, 0) -- +(14 mm,0);
    \draw[thick] (4 cm,2 cm) circle (3 mm);
    \draw[thick] (4 cm, 3mm) -- +(0, 1.4 cm);
    \draw[thick] (4 cm, 2.3cm) -- +(0, 1.4 cm);
  \end{tikzpicture}
\end{center}

\begin{center}
  \begin{tikzpicture}[scale=.4]
    \draw (-4,1) node[anchor=east]  {$\hat{E}_7$};
    \draw[xshift=5 cm,thick, fill=green] (-7, 0) circle (.3 cm);
    \foreach \x in {0,...,5}
    \draw[thick,xshift=\x cm] (\x cm,0) circle (3 mm);
    \foreach \y in {0,...,4}
    \draw[thick,xshift=\y cm] (\y cm,0) ++(.3 cm, 0) -- +(14 mm,0);
    \draw[thick] (4 cm,2 cm) circle (3 mm);
    \draw[thick] (4 cm, 3mm) -- +(0, 1.4 cm);
    \draw[thick] (-1.75 cm, 0 cm) -- +(1.50 cm,0 cm);
  \end{tikzpicture}
\end{center}
\begin{center}
  \begin{tikzpicture}[scale=.4]
    \draw (-4,1) node[anchor=east]  {$\hat{E}_8$};
    \draw[xshift=5 cm,thick, fill=green] (-7, 0) circle (.3 cm);
    \foreach \x in {0,...,6}
    \draw[thick,xshift=\x cm] (\x cm,0) circle (3 mm);
    \foreach \y in {0,...,5}
    \draw[thick,xshift=\y cm] (\y cm,0) ++(.3 cm, 0) -- +(14 mm,0);
    \draw[thick] (8 cm,2 cm) circle (3 mm);
    \draw[thick] (8 cm, 3mm) -- +(0, 1.4 cm);
     \draw[thick] (-1.75 cm, 0 cm) -- +(1.50 cm,0 cm);
  \end{tikzpicture}
\end{center}

\begin{center}
  \begin{tikzpicture}[scale=.4]
    \draw (-3,0) node[anchor=east]  {$\hat{G}_2$};
    \draw[thick] (0 ,0) circle (.3 cm);
    \draw[thick,fill=black] (2 cm,0) circle (.3 cm);
     \draw[xshift=5 cm,thick, fill=green] (-7 cm, 0) circle (.3 cm);
    \draw[thick] (30: 3mm) -- +(1.5 cm, 0);
    \draw[thick] (0: 3 mm) -- +(1.4 cm, 0);
    \draw[thick] (-30: 3 mm) -- +(1.5 cm, 0);
    \draw[thick] (-1.75 cm, 0 cm) -- +(1.50 cm,0 cm);
  \end{tikzpicture}
\end{center}
\begin{center}
  \begin{tikzpicture}[scale=.4]
    \draw (-5,0) node[anchor=east]  {$\hat{F}_4$};
    \draw[thick] (-2 cm ,0) circle (.3 cm);
    \draw[thick] (0 ,0) circle (.3 cm);
    \draw[xshift=5 cm,thick, fill=green] (-9 cm, 0) circle (.3 cm);
    \draw[thick,fill=black] (2 cm,0) circle (.3 cm);
    \draw[thick,fill=black] (4 cm,0) circle (.3 cm);
    \draw[thick] (15: 3mm) -- +(1.5 cm, 0);
    \draw[xshift=-2 cm,thick] (0: 3 mm) -- +(1.4 cm, 0);
    \draw[thick] (-15: 3 mm) -- +(1.5 cm, 0);
    \draw[xshift=2 cm,thick] (0: 3 mm) -- +(1.4 cm, 0);
     \draw[thick] (-3.75 cm, 0 cm) -- +(1.50 cm,0 cm);
  \end{tikzpicture}
\end{center}
\label{extendedroots}
\caption{Extended Dynkin diagrams}
\end{figure}
\end{center}
\newpage

\section{Macdonald-Lusztig-Spaltenstein (j-) induction}
\label{jinduction}
This is a general procedure that can be used to generate irreducible representations of a Weyl group $W[\mathfrak{g}]$ from irreducible representations of parabolic subgroups $W_p$.  One can use this method to generate a large number of the irreducible representations of $W[\mathfrak{g}]$. In types $A,B,C$, one can actually generate $\textit{all}$ of them by $j$-induction. In other types, there is often quite a few irreducible representations that can't be obtained by $j$ induction. A special case of this method that involves induction only from the sign representation of the parabolic subgroup $W_p$ was developed originally by Macdonald \cite{macdonald1972some}. 

\subsection{Macdonald induction}

Let $W_p$ be a parabolic subgroup of the Weyl group $W[\mathfrak{g}]$. This is equivalent to saying that $W_p$ is the Weyl group of a Levi subalgebra of $\mathfrak{g}$. Then, consider the positive root $e_\alpha$ in the root system corresponding to $W_p$. The positive roots are linear functionals on $\mathfrak{h}$. Form the following rational polynomial,
\begin{equation}
P = \prod _{e_\alpha > 0} e_\alpha.
\end{equation}
Let $w$ be an element of the Weyl group $W[\mathfrak{g}]$. Consider the algebra formed by all polynomials of the form $w(P)$. This is a subalgebra of the symmetric algebra and is naturally a $W[\mathfrak{g}]$ module. In fact, it furnishes an irreducible representation of the Weyl group $W[\mathfrak{g}]$. By choosing different subgroups $W_p$, one obtains different irreps of $W[\mathfrak{g}]$. This is a special case of $j$ induction where one uses the sign representation of the smaller Weyl group to induce from. Within the notation of the more general j-induction, the Macdonald method would correspond to $j_{W_p}^W (sign)$.

\subsection{Macdonald-Lusztig-Spaltenstein induction}

The generalization of the Macdonald method to what is called $j$ induction was provided by Lusztig- Spaltenstein in \cite{lusztig1979induced}. What follows is a very brief review. See \cite{carter1985finite,geck2000characters} for more detailed expositions. 

Let $V$ be a vector space on which $W[\mathfrak{g}]$ acts by reflections. Let $W_r$ now be any reflection subgroup of $W[\mathfrak{g}]$. Let $V^{W_r}$ be the subspace of $V$ fixed by reflections in $W_r$. There is a decomposition $V = \bar{V} \oplus V^{W_r}$. Consider the space of homogeneous polynomial functions on $\bar{V}$ of some degree $d$ and denote it by $P_d(\bar{V})$. Let $r'$ be \textit{any} univalent irrep of $W_r$. This means that $r'$ occurs with multiplicity one in $P_d(\bar{V})$ for some $d$. The $W[\mathfrak{g}]$ module generated by $r'$ is irreducible and univalent and it denoted by $j_{W_r}^W(r')$. When, $r'$ is the sign representation and $W_r$ is the Weyl group of Levi subalgebra (= a parabolic subgroup of the Weyl group), this reduces to the Macdonald method.

The action of $j$ induction is most transparent in type $A$. For types $B,C,D$, it can still be described by suitable combinatorics. However, in practice, it is most convenient to use packages like CHEVIE to calculate $j$ induction. Below, some sample cases are recorded. 
\subsubsection{j-induction in type A}
In type $A$, one can get all irreducible representations using $j$ induction of the sign representation from various parabolic subgroups. The various Levi subalgebras in type A have a natural partition type classification and consequently, so do their Weyl group. Let $W_P$ be a parabolic subgroup of partition type $P$. Let, $P^T$ be the transpose partition. Then, $j_{W_P}^W = P^T$, where $P^T$ is the partition label for the irreducible representations of $S_n$.
\subsubsection{Example : j-induction in $A_3$}
Here is a detailed example of $j$ induction in action for type A. Introduce the following subgroups of the Weyl group $S_4$ by their Deodhar-Dyer labels (which are used in CHEVIE to index reflection subgroups). The label is of the form $[r_1,r_2 \ldots]$ and corresponds to a subset of the set of positive roots (in the ordering used by CHEVIE). By a theorem of Deodhar \& Dyer \cite{deodhar1989note,dyer1990reflection}, this is a characterization of non-conjugate reflection subgroups.
\begin{center}
\begin{tabular}{c|c|c}
Subgroup  & Deodhar-Dyer label & Cartan type of assoc. subalgebra \\ \hline
$W_{[4]}$ & $[r_1,r_2,r_3]$ & $A_3$ \\
$W_{[3,1]}$ & $[r_1,r_2]$ & $A_2$ \\
$W_{[2,2]}$ & $[r_1,r_3]$ & $A_1 + A_1$ \\
$W_{[2,1^2]}$ & $[r_1]$ & $A_1$ \\
$W_{[1^4]}$ & $[\varnothing]$ & $\varnothing$ 
\end{tabular}
\end{center}
Denote the irreducible representation of $W=S_4$ by the usual partition labels ($[1^4]$ is the sign representation while $[4]$ is the identity representation). Applying j-induction using the sign representation in each of the subgroups above, one gets
\begin{eqnarray*}
j_{W_{1,2,3}}^W (sign)  &=&  [1^4] \\
j_{W_{1,2}}^W (sign) &=& [2,1^2] \\
j_{W_{1,3}}^W (sign) &=& [2,2]\\
j_{W_{1}}^W (sign) &=& [3,1] \\
j_{W_{\varnothing}}^W (sign) &=& [4]
\end{eqnarray*}
\subsubsection{Example : j-induction in $D_4$} 
Introduce the following subgroups of $W(D_4)$ using Deodhar-Dyer labels,
\begin{center}
\begin{tabular}{c|c|c}
Subgroup  & Deodhar-Dyer label & Cartan type \\ \hline
$W_{1,2,3,4}$ & $[r_1,r_2,r_3,r_4]$ & $D_4$ \\
$W_{2,3,4}$ & $[r_1,r_3,r_4]$ & $A_3$ \\
$W_{1,3,4}$ & $[r_2,r_3,r_4]$ & $A_3$ \\
$W_{1,2,3}$ & $[r_1,r_2,r_3]$ & $A_3$ \\
$W_{1,2,4,12}$ & $[r_1,r_2,r_4,r_{12}]$ & $4A_1$ \\
$W_{1,3}$ & $[r_1,r_3]$ & $A_2$ \\
$W_{3,10}$ & $[r_3,r_{10}]$ & $2A_1$ \\
$W_{1,12}$ & $[r_1,r_{12}]$ & $2A_1$ \\
$W_{1,2}$ & $[r_1,r_2]$ & $2A_1$ \\
$W_{1}$ & $r_1$ & $A_1$ \\
$W_{\varnothing}$ & $[\varnothing]$ & $\varnothing$ 
\end{tabular}
\end{center}

One obtains the following results useful for j-induction,
\begin{eqnarray*}
j_{W_{1,2,3,4}}^W (sign) &=& [1^4].- \\
j_{W_{1,2,3,4}}^W ([1^3].[1])  &=& [1^3].[1] \\
j_{W_{2,3,4}}^W (sign) &=& ([1^2].[1^2])' \\
j_{W_{1,3,4}}^W (sign)  &=& ([1^2].[1^2])'' \\
j_{W_{1,2,3}}^W (sign)  &=& ([2].[1^2])'' \\
j_{W_{1,2,4,12}}^W (sign)  &=& [2^2].- \\
j_{W_{1,3}}^W (sign)  &=& [2,1].[1] \\
j_{W_{1,2}}^W (sign)  &=& [3,1].- \\
j_{W_{3,10}}^W (sign)  &=& ([2].[2])'\\
j_{W_{1,4}}^W (sign)  &=& ([2].[2])''\\
j_{W_{1}}^W (sign)  &=& [3].[1] \\
j_{W_{\varnothing}}^W (sign)  &=& [4].- \\
\end{eqnarray*}

The choice of the subgroups and the resulting irreps is no accident. The irreducible representations obtained here by $j$ induction are precisely the Orbit representations for $D_4$ and they occur as $\bar{r}$ in Table \ref{d4table}. 

\subsubsection{Example : j-induction in $G_2$}

As a final example of $j$ induction, here are some results for $G_2$ that are relevant for the compiling of Table \ref{g2table}. 
Introduce the following subgroups of $W(G_2)$.
\begin{center}
\begin{tabular}{c|c|c}
Subgroup  & Deodhar-Dyer label & Cartan type \\ \hline
$W_{1,2}$ & $[r_1,r_2]$ & $G_2$ \\ 
$W_{2,3}$ & $[r_2,r_3]$ & $A_2$\\
$W_{2,6}$ & $[r_2,r_6]$ & $A_1 \times A_1$\\
$W_{1}$ & $[r_1]$ & $A_1$ \\ 
$W_{\varnothing}$ & $[\varnothing]$ & $\varnothing$ 
\end{tabular}
\end{center}

With this, one can note the following instances of $j$ induction,
\begin{eqnarray*}
j_{W_{1,2}}^W (sign) = \phi_{1,6} \\
j_{W_{2,3}}^W (sign) = \phi_{1,3}'' \\
j_{W_{2,6}}^W (sign) = \phi_{2,2} \\
j_{W_{1}}^W (sign) = \phi_{2,1} \\
j_{W_{\varnothing}}^W (sign)= \phi_{1,0}
\end{eqnarray*}

The instances of $j$ induction were again chosen such that the result is an Orbit representation of $G_2$. An important observation due to Lusztig is that in any Weyl group, the Orbit representations can always be obtained by $j$ induction.

\newpage

\bibliographystyle{JHEP.bst}
\bibliography{description.bbl}

\end{document}